\documentclass[12pt,preprint]{aastex}

\newcommand\Msun{M$_{\odot}$}

\newcommand\Halpha{H$\alpha$ }

\newcommand\Tmb{$T_{\mathrm{mb}}$}
\newcommand\Tkin{$T_{\mathrm{kin}}$}

\shorttitle{Sub-mm obs. of GMCs in the LMC}
\shortauthors{Minamidani et al.}

\begin{document}


\title{
Sub-millimeter Observations of Giant Molecular Clouds 
in the Large Magellanic Cloud: 
Temperature and Density as Determined 
from $J=3-2$ and $J=1-0$ transitions of CO
}


\author{
Tetsuhiro Minamidani\altaffilmark{1,13}, 
Norikazu Mizuno\altaffilmark{1}, 
Yoji Mizuno\altaffilmark{1}, 
Akiko Kawamura\altaffilmark{1}, 
Toshikazu Onishi\altaffilmark{1}, 
Tetsuo Hasegawa\altaffilmark{2}, 
Ken'ichi Tatematsu\altaffilmark{2}, 
Masafumi Ikeda\altaffilmark{3}, 
Yoshiaki Moriguchi\altaffilmark{1}, 
Nobuyuki Yamaguchi\altaffilmark{2}, 
J\"urgen Ott\altaffilmark{4,*}, 
Tony Wong\altaffilmark{5,14}, 
Erik Muller\altaffilmark{5}, 
Jorge L. Pineda\altaffilmark{6}, 
Annie Hughes\altaffilmark{5,7}, 
Lister Staveley--Smith\altaffilmark{8}, 
Ulrich Klein\altaffilmark{6}, 
Akira Mizuno\altaffilmark{1,15}, 
% Lars E. B. Johansson\altaffilmark{7}, 
Silvana Nikoli\'{c}\altaffilmark{9,16}, 
Roy S. Booth\altaffilmark{9,17}, 
Arto Heikkil\"{a}\altaffilmark{9}, 
Lars--\AA ke Nyman\altaffilmark{10}, 
Mikael Lerner\altaffilmark{10}, 
Guido Garay\altaffilmark{11}, 
Sungeun Kim\altaffilmark{12},
Motosuji Fujishita\altaffilmark{1},
Tokuichi Kawase\altaffilmark{1},
Monic\'{a} Rubio\altaffilmark{11},
\and
Yasuo Fukui\altaffilmark{1}
}

\email{tetsu@a.phys.nagoya-u.ac.jp}


\altaffiltext{1}{Department of Astrophysics, Nagoya University, Furo-cho, Chikusa-ku, Nagoya 464-8602, Japan}
\altaffiltext{2}{National Astronomical Observatory of Japan, Mitaka, Tokyo 181-8588, Japan}
\altaffiltext{3}{Research Center for the Early Univers and Department of Physics, University of Tokyo, Tokyo 113-0033, Japan}
\altaffiltext{4}{National Radio Astronomy Observatory, 520 Edgemont Road, Charlottesville, VA 22903-2475}
\altaffiltext{5}{CSIRO Australia Telescope National Facility, PO Box 76, Epping, NSW 1710, Australia}
\altaffiltext{6}{Radioastronomisches Institut der Universit\"{a}t Bonn, Auf dem H\"{u}gel 71, 53121 Bonn, Germany}
\altaffiltext{7}{Center for Supercomputing and Astrophysics, Swinburne University of Technology, Hawthorn, VIC 3122, Australia}
\altaffiltext{8}{School of Physics, M013, University of Western Australia, 35 Stirling Highway, Crawley, WA 6009, Australia}
\altaffiltext{9}{Onsala Space Observatory, 439-92 Onsala, Sweden}
\altaffiltext{10}{European Southern Observatory, Casilla 19001, Santiago 19, Chile}
\altaffiltext{11}{Departament de Astronomia, Universidad de Chile, Casilla 36-D, Santiago, Chile}
\altaffiltext{12}{Astronomy \& Space Science Department, Sejong University, 98 Kwangjin-gu, Kunja-dong, Seoul, 143-747, Korea}

\altaffiltext{13}{Present address: Department of Physics, Faculty of Science, Hokkaido University, N10W8, Kita-ku, Sapporo 060-0810, Japan}
\altaffiltext{14}{Present address: Department of Astronomy, MC 221, University of Illinois, Urbana, IL 61801}
\altaffiltext{15}{Present address: Solar-Terrestrial Environment Laboratory, Nagoya University, Furo-cho, Chikusa-ku, Nagoya 464-8601, Japan}
\altaffiltext{16}{Present address: Departament de Astronomia, Universidad de Chile, Casilla 36-D, Santiago, Chile}
\altaffiltext{17}{Present address: Hartebeesthoek Radio Astronomy Observatory, P.O.Box 443, Krugersdorp 1740, South Africa}

\altaffiltext{*}{J\"urgen Ott is a Jansky Fellow of the National Radio Astronomy Observatory.}


\begin{abstract}
We have carried out sub-mm $^{12}$CO($J=3-2$) observations of 6 giant molecular clouds (GMCs) in the Large Magellanic Cloud (LMC) with the ASTE 10m sub-mm telescope at a spatial resolution of 5 pc and very high sensitivity. 
We have identified 32 molecular clumps in the GMCs and revealed significant details of the warm and dense molecular gas with $n$(H$_2$) $\sim$ 10$^{3-5}$ cm$^{-3}$ and $T_{\mathrm{kin}}$ $\sim$ 60 K. 
These data are combined with $^{12}$CO($J=1-0$) and $^{13}$CO($J=1-0$) results and compared with LVG calculations. 
%%%%%%%%%%
%\textcolor{blue}{%
The results indicate that clumps we detected are distributed continuously from cool ($\sim$ 10 -- 30 K) to warm ($\sim$ higher than 30 -- 200 K), and warm clumps are distributed from less dense ($\sim$ 10$^3$ cm$^{-3}$) to dense ($\sim$ 10$^{3.5 - 5}$ cm$^{-3}$).%
%}%
%%%%%
We found that the ratio of $^{12}$CO($J=3-2$) to $^{12}$CO($J=1-0$) emission is sensitive to and is well correlated with the local \Halpha flux. 
We interpret that differences of clump propeties represent an evolutionary sequence of GMCs in terms of density increase leading to star formation.Type I and II GMCs (starless GMCs and GMCs with HII regions only, respectively) are at the young phase of star formation where density does not yet become high enough to show active star formation and Type III GMCs (GMCs with HII regions and young star clusters) represents the later phase where the average density is increased and the GMCs are forming massive stars. 
The high kinetic temperature correlated with \Halpha flux suggests that FUV heating is dominant in the molecular gas of the LMC.
\end{abstract}


\keywords{
Magellanic Clouds --- 
galaxies: individual (LMC) ---
ISM: clouds ---
ISM: molecules --- 
radio lines: ISM --- 
submillimeter
}

\section{Introduction}

It is of a fundamental importance in astronomy to understand the evolution of galaxies. Since a major constituent of galaxies is stars, the formation of stars is a fundamental process in galactic evolution. The properties of stars characterize the basic contents of galaxies and their time evolution. We understand from studies of the Milky Way galaxy that giant molecular clouds (GMCs), whose mass ranges from 10$^5$ to 10$^7$ \Msun, are the principal sites of star formation and that it perhaps holds the true in other galaxies as well. We also recognize that the GMC properties (e.g., $L_{\mathrm{CO}}$ -- line width relation, index of mass spectrum) are similar among five galaxies in the Local Group according to the spatially resolved studies \citep{Blitz2006}. This supports the idea that studies of GMCs will be useful in understanding the fundamentals of galactic evolution through the formation and evolution of GMCs and star formation therein.

Observational studies of GMCs have been most effectively made by the mm interstellar carbon monoxide emission line at 2.6 mm which allows us to probe molecular gas whose density is greater than $\sim$ 100 cm$^{-3}$. We note that the most abundant species, molecular hydrogen, does not have appropriate line emissions in the mm and sub-mm region due to its zero permanent electric dipole moment and large separation between the lowest energy levels, which are not excited significantly in the typical physical conditions of molecular clouds.

Recent advances in sub-mm observations have allowed us to determine physical parameters of molecular clouds over much larger ranges than in the mm region by comparing line intensities between different transitions. These sub-mm studies were initiated by the SEST 15m telescope in Chile followed by instruments in Hawaii, Mauna Kea and in the Swiss Alps at an altitude range from 3700 m to 4200 m, including the CSO 10m, JCMT 15m, and KOSMA 3m telescopes, and the AST/RO 1.6m telescope in Antarctica. Subsequently, in the 2000's, the developements of new instruments at an altitude of $\sim$ 5000 m in Atacama in northern Chile resulted in a superior capability because of the high altitude and dry characteristics of the site. The instruments installed in Atacama include the ASTE 10m, APEX 12m and NANTEN2 4m telescopes. All these instruments are beginning to take new molecular data with significantly better quality than before in terms of noise level as well as angular resolution. It is also noteworthy that the current frequency coverage extends as high as the 800GHz band and even the THz region.

Among nearby galaxies we can observe at reasonably high spatial resolutions, the Large and Small Magellanic Clouds offer us a unique opportunity to achieve the highest resolutions due to their unrivaled closeness, 50 -- 60 kpc. In particular, the Large Magellanic Cloud (LMC) is actively forming stars in clusters and is an ideal laboratory for us to study star formation, particularly massive star formation in star cluster. In the LMC, the metallicity is a factor of $\sim$ 3 lower than in the Solar neighborhood \citep{Dufour1982, Dufour1984, Rolleston2002}. Also, the visual extinctions are lower and the FUV field is stronger in the LMC than in the Milky Way \citep{Israel1986}, characterizing the initial conditions of star formation.

The first spatially resolved complete sample of GMCs in a single galaxy has been obtained towards the whole LMC with the NANTEN 4m telescope in 2.6 mm CO emission at 40 pc resolution \citep{Fukui1999, Fukui2001, Fukui2006b, Mizuno2001}. These studies revealed the three types of GMCs in terms of star formation activities; Type I is starless, Type II is with HII regions only, and Type III is associated with active star formation indicated by huge HII regions and young star clusters, where the stars identified are only O stars due to the sensitivity limitation. It also revealed that the lifetime of a GMC is as short as $\sim$ 30 Myrs \citep{Fukui2006a, Kawamura2006, Kawamura2007}. These previous studies naturally place the LMC as one of the prime targets for sub-mm studies to derive the physical parameters of GMCs.

Another aspect which deserves our attention is that very young, rich stellar clusters are forming in the LMC. These are so called populous clusters which are very rare in the Milky Way and resemble globulars formed in the primeval Milky Way. The open clusters forming in the Milky Way are small in the number of stars and loose in spatial distribution. Along with the low metallicity of the LMC, it is an interesting possibility to use molecular data to investigate the formation mechanism of super star clusters at the molecular cloud stage.

%%%%%%%%%%
%In the past, there have been some studies that used sub-mm spectra ($J=3-2$ or $J=4-3$ of CO) of the molecular clouds in the LMC. These studies suggest the molecular gas may be warmer and/or denser than in the Milky Way. 

In the past, there have been some studies that used the higher transitions ($J=2-1$, $J=3-2$, $J=4-3$, $J=7-6$) of CO spectra of the molecular clouds in the LMC \citep[e.g.,][]{Sorai2001,Johansson1998,Heikkila1999,Bolatto2005,Israel2003,Kim2004,Kim2006}. These studies suggest the molecular gas may be warmer and/or denser than in the Milky Way.

%% \citet{Sorai2001} observed 34 points in the LMC with Tokyo-Onsala-ESO-Calan 60 cm radio telescope in the $J=2-1$ transition of CO. They found that the intensity ratio of CO($J=2-1$) to CO($J=1-0$), $R_{2-1/1-0}$, varies point to point, and suggested that the enhancement of $R_{2-1/1-0}$ at $\sim$ 100 pc scale is due to warm and/or dense gas. They also stated that the enhancement of $R_{2-1/1-0}$ in the 30 Dor region and its southern part is not primarily due to the external UV radiation.%
%%%%%
\citet{Johansson1998} used the SEST 15m telescope to observe the central part of the 30 Doradus nebula (rms $\sim$ 0.2 K in 0.5 km s$^{-1}$ velocity resolution for $J=1-0$ and rms $\sim$ 1.0 K in 0.5 km s$^{-1}$ velocity resolution for $J=3-2$), and the southern HII regions N 158C, N 159, and N 160 with a few prominent CO clouds in the $J=2-1$ and $J=3-2$ transitions of CO. They find that the kinetic temperatures are 10 -- 80 K and the highest temperature is towards 30 Dor. The smallest beam size and grid spacing are 15$\arcsec$ and 11$\arcsec$  respectively in the $J=3-2$ emission. 
\citet{Heikkila1999} used SEST to observe the $J=3-2$ transition of CO in N 159 and 30 Doradus, as well as other rarer molecular species. This study aimed at obtaining chemical abundance, while it also provides more information on cloud temperature etc, from CO($J=3-2$) data. The kinetic temperatures they derived are 50 K in 30 Dor-10, 15 K in 30 Dor-27, and 20 -- 25 K in N 159W and N 160.
\citet{Bolatto2005} employed the AST/RO to observe the $^{12}$CO($J=4-3$) transition at 461 GHz with a 109$\arcsec$ beam. They observed 9 regions in the LMC at 6$\arcmin$$\times$6$\arcmin$  field all with HII regions and derived kinetic temperatures from a comparison between the CO($J=4-3$) and ($J=1-0$) transitions. N 48, N 55A, N 79, N 83A, N 113, N 159W, N 167, N 214C, and LIRL 648 are included. They derive temperatures of 100 -- 300 K and note a trend that higher temperatures occur in moderate density regions, 100 -- 1000 cm$^{-3}$, and the lower temperatures in much denser regions 10$^{4-5}$ cm$^{-3}$. These studies were preceded by a suggestion that significant amounts of warm molecular gas may exist in the LMC \citep{Israel2003}. 
\citet{Kim2004} also made similar observations towards an HII region, N 44, and suggest very dense gas of $\sim$ 10$^5$ cm$^{-3}$. Most recently, \citet{Kim2006} derived $T_{\mathrm{kin}}$ = 100 K, $n$ $\sim$ 10$^{4.3}$ cm$^{-3}$ for 30 Dor from the intensity ratios of $^{12}$CO($J=7-6$) to $^{12}$CO($J=4-3$) and $^{12}$CO($J=1-0$) to $^{13}$CO($J=1-0$). 

In the present study, we aim to obtain sub-mm molecular data at better S/N ratios than in the previous studies to make estimates of temperatures and densities over a large sample in the LMC. We will combine the $^{12}$CO($J=3-2$) data obtained with the ASTE telescope and CO($J=1-0$) data obtained with the SEST and Mopra telescopes. In order to make reasonable comparisons between the two transitions, $J=3-2$ and $J=1-0$, we shall smooth the ASTE results (22$\arcsec$  beam) to the same resolution as the SEST data (45$\arcsec$) and use LVG calculations to estimate density and temperature. We shall also employ the $^{13}$CO($J=1-0$) data where available to place constraints on the physical parameters.  

This paper is organized as follows: Section 2 describes the observations. Section 3 and 4 show the results and data analysis, respectively. In section 5, we discuss the physical properties of clumps and evolutional sequence of GMCs. Finally, we present a summary in section 6.

\section{Observations}

\subsection{Selection of GMCs}

The present targets were chosen from the NANTEN catalog of $^{12}$CO($J=1-0$) GMCs compiled by \citet{Fukui2006b}. This catalog is based on the 2nd survey, with a factor of $\sim$ 2 higher sensitivity than the 1st survey \citep{Mizuno2001}. The catalog includes 272 CO clouds 230 of which are detected at three or more observed positions and they are classified into the three Types; 56 (24.3 \%) Type I (starless) GMCs, 120 (52.2 \%) Type II GMCs (those with HII regions only), and 54 (23.5 \%) Type III GMCs (those with HII regions and young star clusters), where "stars" refer only to O stars due to the limited sensitivity of existing observations \citep{Kawamura2007, Fukui2006a, Kawamura2006, Blitz2006}.

Among these GMCs, we mainly focus on Type III GMCs whose $^{12}$CO($J=3-2$) intensities are expected to be strong due to the highly excited conditions. In addition, relatively high resolution $^{12}$CO and $^{13}$CO $J=1-0$ data, observed with the SEST 15m or Mopra 22m telescopes, are collected to derive physical properties of molecular clouds through the LVG analysis at a 10 pc scale.

\begin{figure}
\epsscale{.80}
\plotone{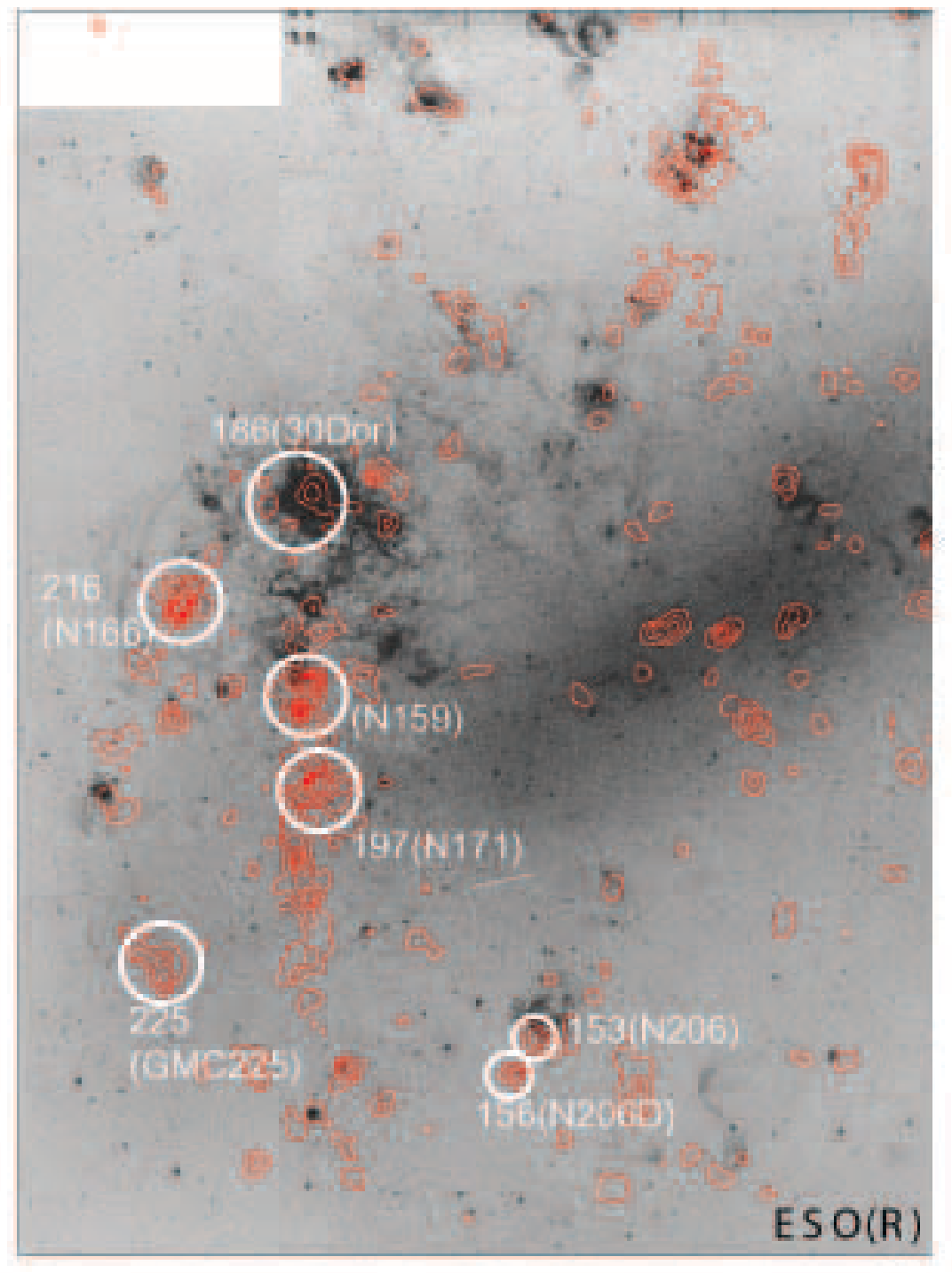}
\caption{CO velocity integrated intensity map \citep{Fukui2006b, Fukui2001} overlaid on ESO(R) image. The observed 7 regions are indicated by white circle. \label{fig01}}
\end{figure}

In the present study, we observe GMCs in the south-east region of the LMC, which contains 30 Doradus -- the largest and most massive HII region in the Local Group. We observe the molecular ridge extending southward from 30 Doradus, and "CO Arc" along the southeastern optical edge \citep{Fukui1999}. 3 Type III GMCs, LMC N J0538-6904 (the 30 Dor region), LMC N J0540-7008 (the N 159 region and the N 171 region), and LMC N J0530-7106 (the N 206 region), are selected as the principal targets, and two Type II GMCs, LMC N J0544-6923 (the N 166 region) and LMC N J0532-7114 (the N 206D region), and a Type I GMC, LMC N J0547-7041 (the GMC 225 region), are included for reference. The locations of the observed GMCs and regions are shown in Figure \ref{fig01}, and their coordinates and the data used in this paper are summarized in Table \ref{tbl01}. Hereafter, the region names, which are in the parenthesis above or column 4 of Table \ref{tbl01}, are used to identify the regions.

\subsection{$^{12}$CO($J=3-2$)}

	Observations of the $^{12}$CO($J=3-2$) transition at 345GHz were made with the ASTE (Atacama Submillimeter Telescope Experiment) telescope at Pampa la Bola in Chile \citep{Ezawa2004} in October 2004. The half power beam width was measured to be 22$\arcsec$ at 345 GHz by observing the planets. This corresponds to 5.3 pc at the distance of the LMC, 50 kpc. The telescope was equipped with a single "cartridge-type" SIS receiver, sensitive from 324 to 384 GHz, which is of a similar design to that for ALMA \citep{Kohno2005}. The spectrometer was an XF-type digital auto-correlator \citep{Sorai2000}, and was used in the wideband mode, which has a bandwidth of 512 MHz with 1024 channels. The spectrometer provided a velocity coverage and resolution of 450 km s$^{-1}$ and 0.44 km s$^{-1}$, respectively, at 345 GHz. We observed 6 GMCs (7 regions) in the Large Magellanic Cloud as shown in Figure \ref{fig01} and listed in Table \ref{tbl01}. These observations were carried out by position switching at a grid spacing of 20$\arcsec$ or 30$\arcsec$ for the entire clouds, and of 10$\arcsec$ or 15$\arcsec$ for the regions around the local peaks of the integrated intensity. The pointing error was measured to be within 7$\arcsec$ in peak to peak by observing CO point sources R Dor or o Cet every 2 hours during this observing term. The spectral intensities were calibrated by employing the standard room-temperature chopper-wheel technique. We observed Ori-KL once a day, and N 159W every 2 hours to check the stability of intensity calibration, and the intensity variation during these observations was less than 13 \%. We use 0.7 for the main-beam efficiency at 345 GHz, which was measured by observing Jupiter. The system noise temperature was typically 300 K in double-sideband (DSB) including the atmosphere towards the zenith. The typical r.m.s. noise fluctuations were $\sim$ 0.25 K at a velocity resolution of 0.44 km s$^{-1}$ for a 1 minute integration for an on-position. In total, about 1400 points were observed in Equatorial coordinates (B1950). Velocities were relative to the Local Standard of Rest (LSR). These observations were made remotely from an ASTE operation room of San Pedro de Atacama, Chile, using the network observation system N-COSMOS3 developed by NAOJ \citep{Kamazaki2005}.

\subsection{$^{12}$CO($J=1-0$) and $^{13}$CO($J=1-0$)}

\subsubsection{Mopra observations}

	A 20$\arcmin$ $\times$ 120$\arcmin$ region, the prominent molecular ridge extending from 30 Doradus southward, was mapped in the $J=1-0$ transition of $^{12}$CO at a frequency of 115 GHz with the 22m ATNF Mopra telescope, in 5 runs from May to September 2005. This region contains the 30 Dor, N 159 and N 171 regions. The newly implemented on-the-fly (OTF) mode was used, in which the telescope takes data continuously while moving across the sky. Spectra were taken at a 6$\arcsec$ spacing so that the 33$\arcsec$ Mopra beam would be well oversampled in the scanning direction; the row spacing was 8$\arcsec$, also assuring oversampling. The typical system noise temperature, $T_{\mathrm{sys}}$, was 500 K in the single side band (SSB) towards the zenith. The pointing was checked on the SiO maser R Dor every 2 hours; typical pointing error was less than 5$\arcsec$ rms. The digital correlator was configured to output 1024 channels across 64 MHz in each of two orthogonal polarizations. The velocity resolution and coverage were 0.16 and 160 km s$^{-1}$, respectively at 115 GHz. The observing time was about 100 minutes per field (5$\arcmin$$\times$5$\arcmin$), providing rms noise fluctuations of $\sim$ 0.34 K at a velocity resolution of 0.65 km s$^{-1}$. Initial spectral processing (baseline fitting and calibration onto a $T_\mathrm{A}^*$ scale) was performed using the livedata task in AIPS++, and the spectra were gridded into data cubes using the AIPS++ gridzilla task. During the gridding, a Gaussian smoothing kernel with a FWHM taken at a half of the beam size was convolved with the data, so the effective resolution of the output cubes was 36$\arcsec$. The cubes were then rescaled onto a $T_{\mathrm{mb}}$ scale using an "extended beam" efficiency of 0.55 \citep{Ladd2005}. This takes into account that sources larger than about 80$\arcsec$ in diameter will couple to both the main beam and the inner error beam of the telescope. The cubes were 2 dimensional Gaussian smoothed to a 45$\arcsec$ beam, which is the beam size of SEST at 115 GHz.

\subsubsection{SEST observations}

	Observations towards N 206, N 206D and GMC 225 regions in the $^{12}$CO($J=1-0$) line (115 GHz) were made in August, 2001 and February, 2002, using the SEST 15m telescope at La Silla, Chile. The HPBW was 45$\arcsec$ at 115 GHz, the front end was the IRAM 115 SIS receiver and the spectrometer was a high-resolution AOS with 2048 channels. The typical system noise temperature was 550 K (SSB). The velocity resolution and coverage were 0.2 and 216 km s$^{-1}$ respectively at 115 GHz. We mapped these 3 regions in position switching with a grid spacing of 40$\arcsec$ or 20$\arcsec$. The typical integration time was 1 minute for an on-position, providing rms noise fluctuations of $\sim$ 0.18 K at a velocity resolution of 0.2 km s$^{-1}$.

	Observations towards N 206, N 206D and GMC 225 regions in the $^{13}$CO($J=1-0$) line (110 GHz) were made in February and December, 2002, also using the SEST 15m telescope. We mapped peak positions of $^{12}$CO($J=1-0$) in position switching with a grid spacing of 20$\arcsec$. The system noise temperature was typically 230 K (SSB). The typical integration time was 4 minutes for an on-position, providing rms noise fluctuations of $\sim$ 0.04 K at a velocity resolution of 0.2 km s$^{-1}$.

	The pointing accuracy was 5$\arcsec$ rms. We checked this by observing SiO maser toward R Dor every 2 hours during this observing term. N 159W was observed periodically for pointing checks and intensity calibration. We use 0.8 for the main-beam efficiency at 115 GHz by assuming main-beam temperature $T_{\mathrm{mb}}$ of N 159W to be $\sim$ 6.5 -- 6.9 K to keep consistency with former publications \citep{Johansson1994, Johansson1998}.

	Observations toward N 166 region are described separately by \citet{Garay2002}.

\section{Results}

We first present the $^{12}$CO($J=3-2$) images of the clouds at 5 pc resolution and make an empirical comparison of them with the $^{12}$CO($J=1-0$) distribution (section 3.1.). Next, we define molecular clumps and estimate the physical parameters of each clump (section 3.2.).

\subsection{Distributions of the $^{12}$CO($J=3-2$) emission}

\begin{figure}
\epsscale{.80}
\plotone{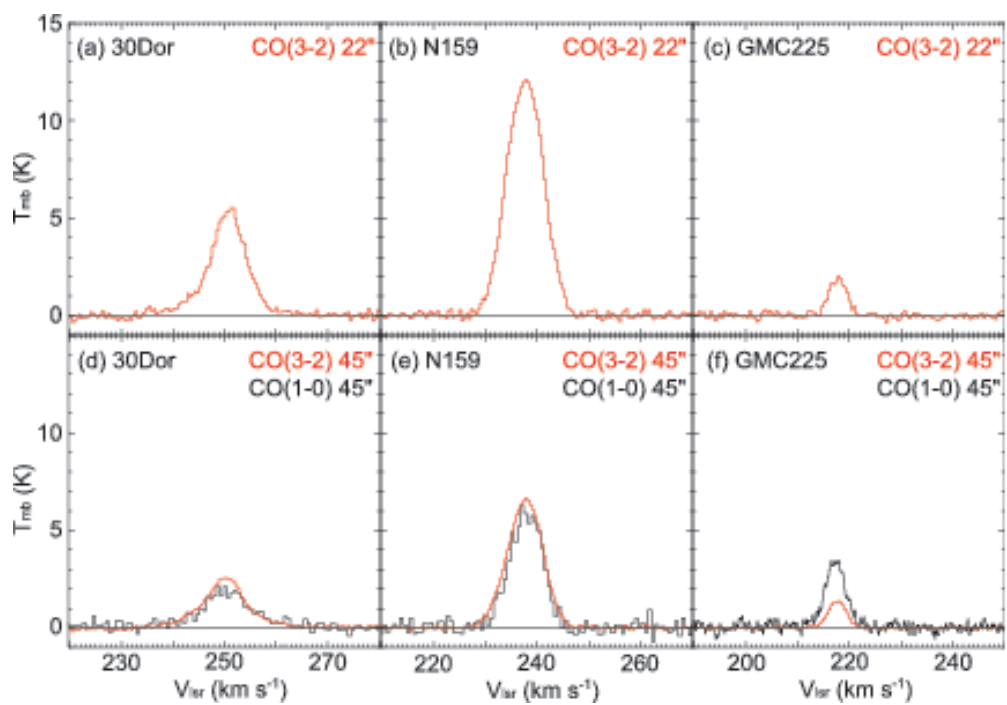}
\caption{$^{12}$CO($J=3-2$) and $^{12}$CO($J=1-0$) profiles at selected points. Note that the vertical scale is main beam temperature, \Tmb. The velocities (abscissae) are relative to the LSR. (a)$^{12}$CO($J=3-2$) profile of 30 Dor. (b)$^{12}$CO($J=3-2$) profile of N 159. (c)$^{12}$CO($J=3-2$) profile of GMC 225. (d) $^{12}$CO($J=1-0$) and smoothed $^{12}$CO($J=3-2$) profiles of 30 Dor. (e) $^{12}$CO($J=1-0$) and smoothed $^{12}$CO($J=3-2$) profiles of N 159. (f) $^{12}$CO($J=1-0$) and smoothed $^{12}$CO($J=3-2$) profiles of GMC 225.\label{fig02}}
\end{figure}

In Figure \ref{fig02}, typical $^{12}$CO($J=3-2$) and $^{12}$CO($J=1-0$) profiles of the 30 Dor, N 159, and GMC 225 regions are presented. The upper panels, Figures \ref{fig02}(a) -- (c), show $^{12}$CO($J=3-2$) profiles. These illustrate the low noise levels of the present data, typically, $\sim$ 0.25 K rms at 0.44 km s$^{-1}$ velocity resolution. Among the present observed positions, the $^{12}$CO($J=3-2$) intensity is strongest at $T_{\mathrm{mb}}$ $\sim$ 12.3 K towards N 159W (Figure \ref{fig02}(b) ). The lower panels, Figures \ref{fig02}(d) -- (f), show $^{12}$CO($J=1-0$) profiles towards the same positions, where the red line indicates $^{12}$CO($J=3-2$) profiles convolved to a 45$\arcsec$ Gaussian beam following the method described in Section 4. The $^{12}$CO($J=3-2$) intensities from the 30 Dor and N 159 regions are a little stronger than the $^{12}$CO($J=1-0$) intensity when convolved to the same resolution. Only towards GMC 225 is the $^{12}$CO($J=3-2$) intensity about 50\% weaker than the $^{12}$CO($J=1-0$) intensity. The peak velocity and line width are nearly the same between the two transitions in these regions.

The distributions of the integrated intensities of $^{12}$CO($J=3-2$) and $^{12}$CO($J=1-0$) are shown in Figures \ref{fig03} -- \ref{fig09}. Detailed descriptions of each region are presented in the following.

\begin{figure}
\epsscale{.80}
\plotone{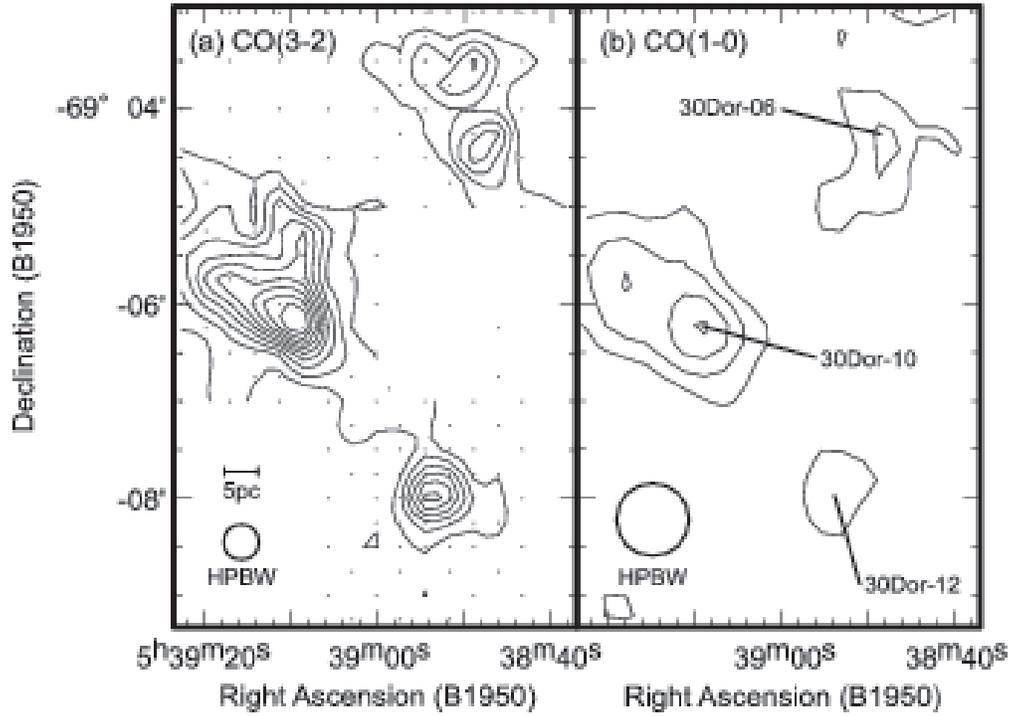}
\caption{(a) Contour map of $^{12}$CO($J=3-2$) integrated intensity in the 30 Doradus region. The contour levels are 5, 10, 15, 20, 25, 30, 35, 40, and 45 K km s$^{-1}$. Observed points are indicated by dots. (b) Contour map of $^{12}$CO($J=1-0$) integrated intensity in the 30 Doradus region. The contour levels are 10, 15, 20, and 25 K km s$^{-1}$. \label{fig03}}
\end{figure}

\subsubsection{30 Dor (Figure \ref{fig03})}

Figures \ref{fig03}(a) and (b) show the distributions of the $^{12}$CO($J=3-2$) and $^{12}$CO($J=1-0$) integrated intensities in the 30 Dor region. We see a general trend that the $J=3-2$ distribution shows more details which are not obvious in the $J=1-0$ distribution owing to the higher angular resolution and possibly due to the more compact distribution of warmer and denser gas in $J=3-2$ than in $J=1-0$. We see three peaks corresponding to the $^{12}$CO($J=1-0$) peaks, 30 Dor-10, 30 Dor-6, and 30 Dor-12, reported in \citet{Johansson1998}. One of them in the north, 30 Dor-06, which is singly peaked in $J=1-0$ appears to be resolved into two peaks with the present beam.

\begin{figure}
\epsscale{.80}
\plotone{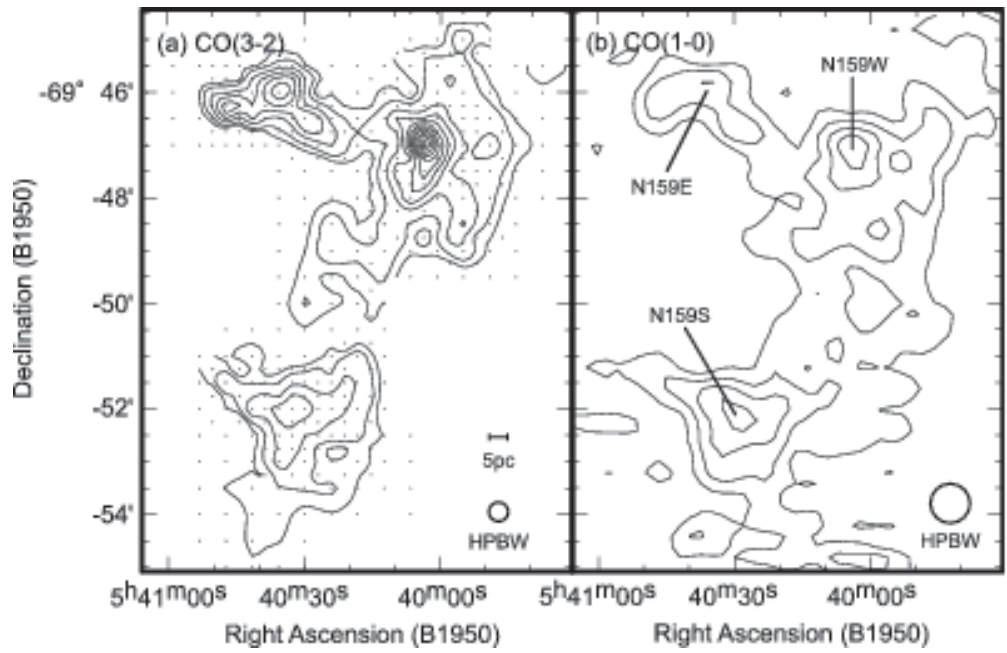}
\caption{(a) Contour map of $^{12}$CO($J=3-2$) integrated intensity in the N 159 region. The contour levels are 5, 10, 20, 30, 40, 50, 60, 70, 80, 90 and 100 K km s$^{-1}$. Observed points are indicated by dots. (b) Contour map of $^{12}$CO($J=1-0$) integrated intensity in the N 159 region. The contour levels are 10, 20, 30, and 40 K km s$^{-1}$. \label{fig04}}
\end{figure}

\subsubsection{N 159 (Figure \ref{fig04})}

Figures \ref{fig04}(a) and (b) show the $^{12}$CO($J=3-2$) and $^{12}$CO($J=1-0$) distributions in N 159. We note that N 159W shows the strongest intensity among the present clouds as well as a very compact peak which is not sufficiently resolved with the present beam.  Its radius is estimated to be a few pc after de-convolution. N 159E also shows a compact distribution with a hint of a sub peak, while N 159S shows similar distributions both in $J=1-0$ and $J=3-2$. The east-west elongation of N 159S may be due to the scanning effect of the OTF mapping, and needs to be confirmed.

\begin{figure}
\epsscale{.80}
\plotone{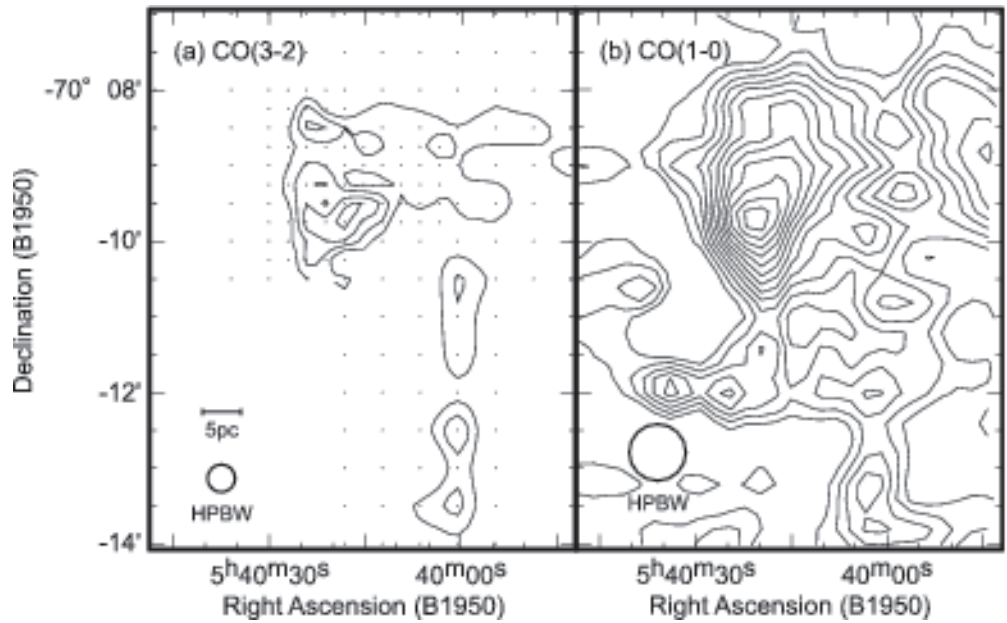}
\caption{(a) Contour map of $^{12}$CO($J=3-2$) integrated intensity in the N 171 region. The contour levels are 8, 12, 16, and 20 K km s$^{-1}$. Observed points are indicated by dots. (b) Contour map of $^{12}$CO($J=1-0$) integrated intensity in the N 171 region. The contour levels are 8, 12, 16, 20, 24, 28, 32, 36, 40, 44, 48, and 52 K km s$^{-1}$. \label{fig05}}
\end{figure}

\subsubsection{N 171 (Figure \ref{fig05})}

Figures \ref{fig05}(a) and (b) show the $^{12}$CO($J=3-2$) and $^{12}$CO($J=1-0$) distributions in N 171. We note that the $J=3-2$ emission is weaker than the $J=1-0$ emission. There are multiple velocity components, at $V_{\mathrm{LSR}}$ = 225 km s$^{-1}$, 230 km s$^{-1}$, and 240 km s$^{-1}$ in both $J=3-2$ and $J=1-0$ as indicated in Table \ref{tbl02} and Table 1 of \citet{Kutner1997}.

\begin{figure}
\epsscale{.80}
\plotone{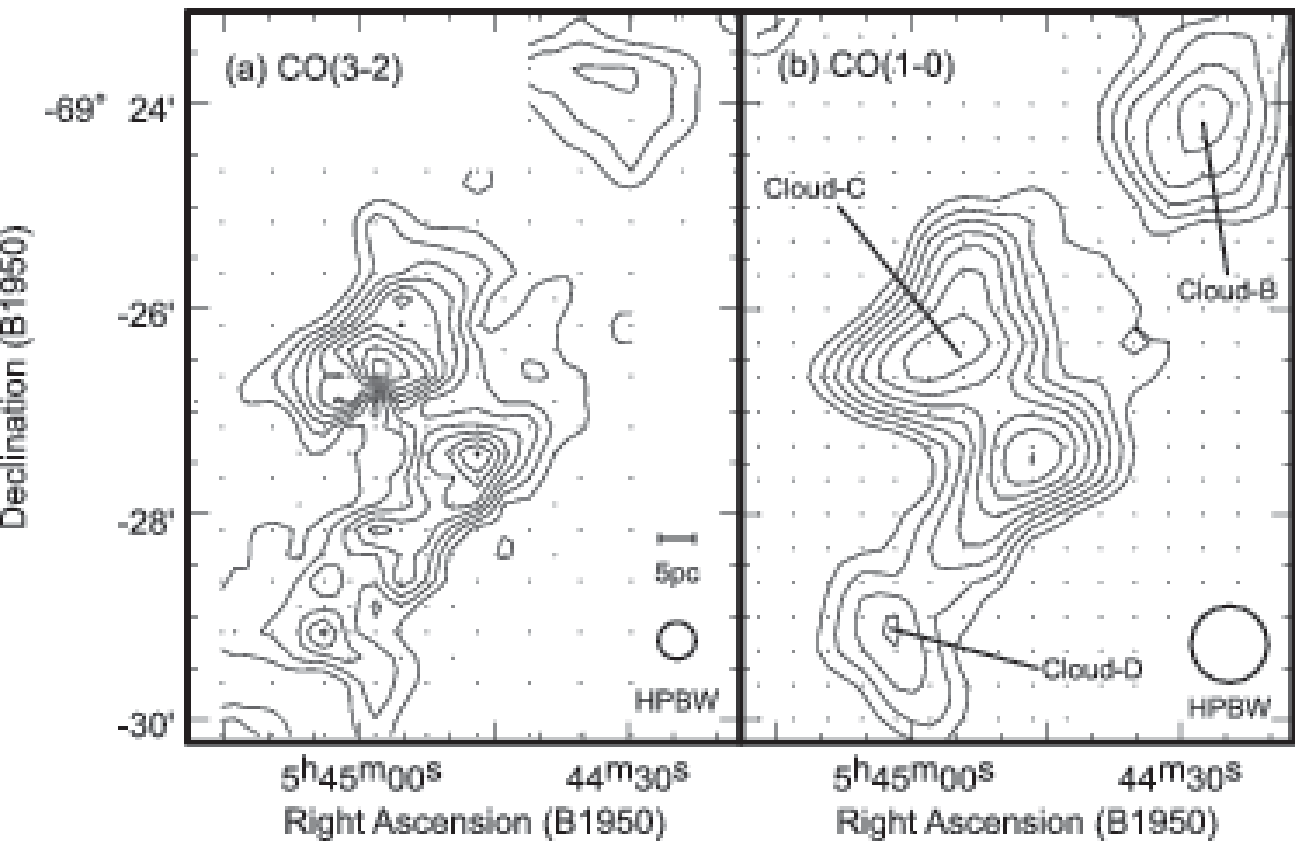}
\caption{(a) Contour map of $^{12}$CO($J=3-2$) integrated intensity in the N 166 region. The contour levels are 6, 8, 10, 12, 14, 16, 18, 20, and 22 K km s$^{-1}$. Observed points are indicated by dots. (b) Contour map of $^{12}$CO($J=1-0$) integrated intensity in the N 166 region. The contour levels are 8, 10, 12, 14, 16, 18, 20, and 22 K km s$^{-1}$. Observed points are indicated by dots. \label{fig06}}
\end{figure}

\subsubsection{N 166 (Figure \ref{fig06})}

Figures \ref{fig06}(a) and (b) show the $^{12}$CO($J=3-2$) and $^{12}$CO($J=1-0$) distributions in N 166. The $J=1-0$ data themselves have already been published by Garay et al. (2002). We see four peaks in both $J=3-2$ and $J=1-0$. Of these peaks, three peaks are named Cloud-B, Cloud-C, and Cloud-D, respectively, as reported by \citet{Garay2002}, although Cloud-C is resolved into two peaks with the present beam and observing grid.

\begin{figure}
\epsscale{.80}
\plotone{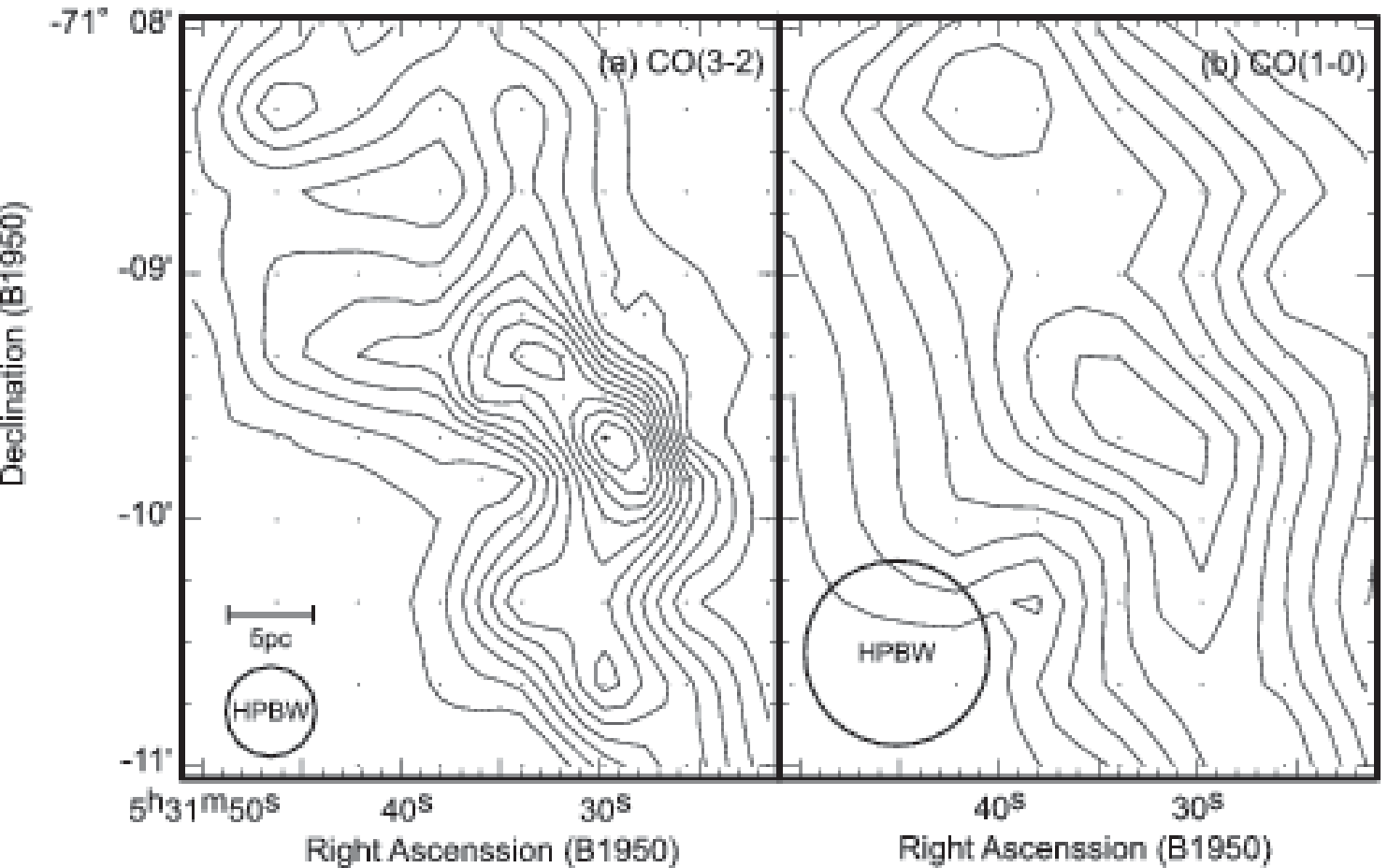}
\caption{(a) Contour map of $^{12}$CO($J=3-2$) integrated intensity in the N 206 region. The contour levels are 4, 6, 8, 10, 12, 14, 16, 18, 20, 22, 24, and 26 K km s$^{-1}$. Observed points are indicated by dots. (b) Contour map of $^{12}$CO($J=1-0$) integrated intensity in the N 206 region. The contour levels are 4, 6, 8, 12, 14, 16, and 18 K km s$^{-1}$. Observed points are indicated by dots. \label{fig07}}
\end{figure}

\subsubsection{N 206 (Figure \ref{fig07})}

Figures \ref{fig07}(a) and (b) show the $^{12}$CO($J=3-2$) and $^{12}$CO($J=1-0$) distributions in N 206. A $J=1-0$ peak appears to be resolved into two sub peaks and a north-south filamentary structure with the higher angular resolution of the $J=3-2$ line.

\begin{figure}
\epsscale{.80}
\plotone{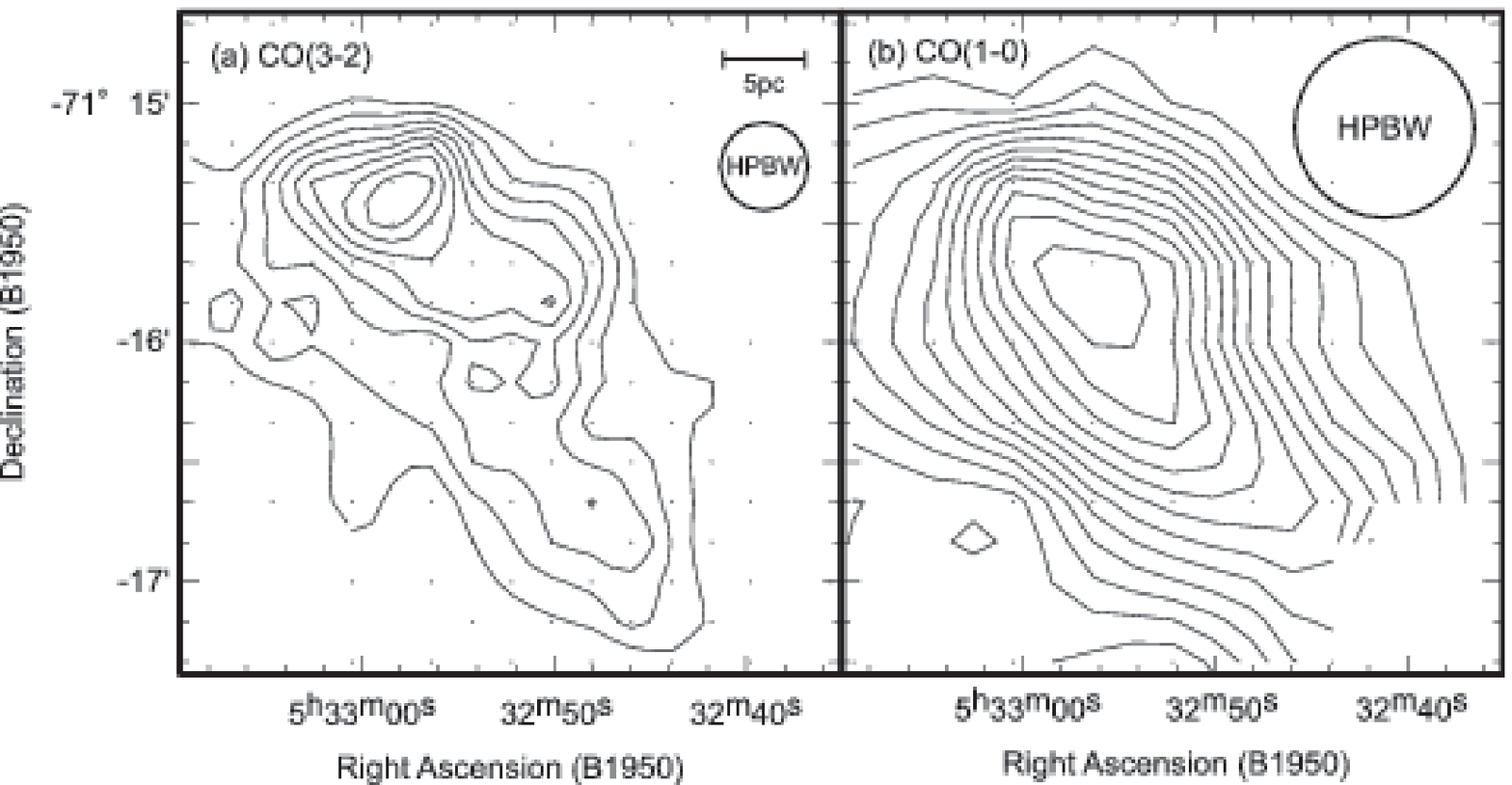}
\caption{(a) Contour map of $^{12}$CO($J=3-2$) integrated intensity in the N 206D region. The contour levels are 4, 6, 8, 10, 12, 14, 16, and 18 K km s$^{-1}$. Observed points are indicated by dots. (b) Contour map of $^{12}$CO($J=1-0$) integrated intensity in the N 206D region. The contour levels are 4, 6, 8, 10, 12, 14, 16, 18, 20, 22, 24, and 26 K km s$^{-1}$. Observed points are indicated by dots. \label{fig08}}
\end{figure}

\subsubsection{N 206D (Figure \ref{fig08})}

Figures \ref{fig08}(a) and (b) show the $^{12}$CO($J=3-2$) and $^{12}$CO($J=1-0$) distributions in N 206D. We can see a head-tail structure in $J=3-2$, although it appears more rounded in $J=1-0$.

\begin{figure}
\epsscale{.80}
\plotone{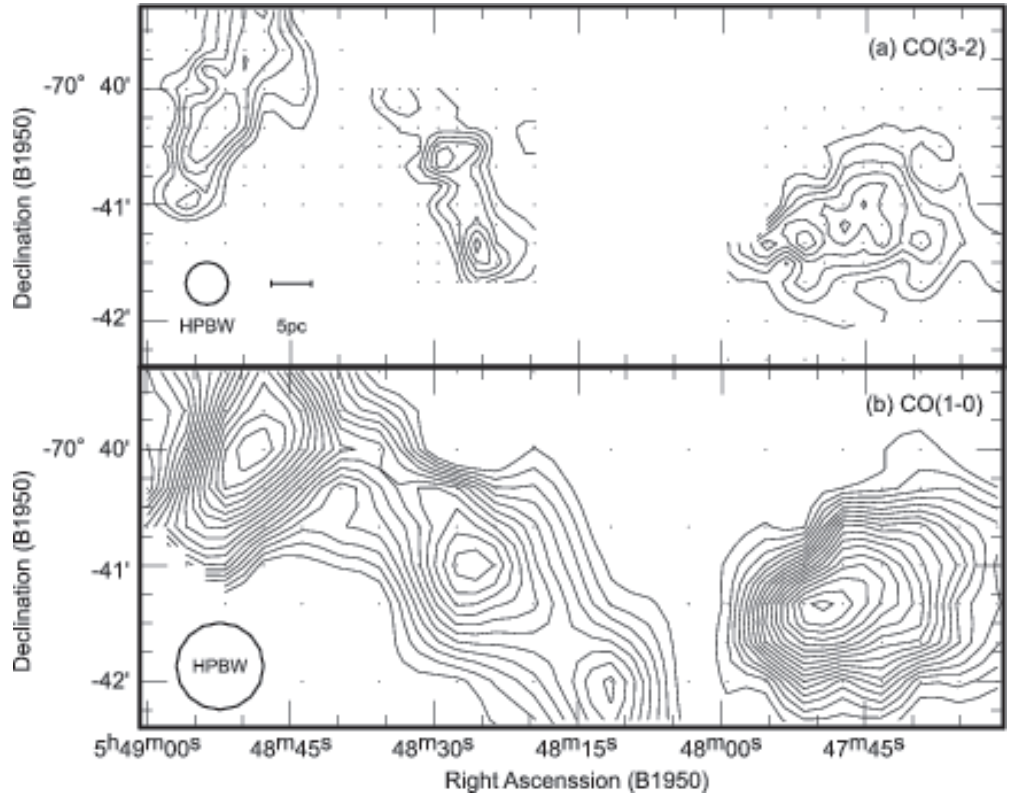}
\caption{(a) Contour map of $^{12}$CO($J=3-2$) integrated intensity in the GMC 225 region. The contour levels are 3, 4, 5, 6, 7, and 8 K km s$^{-1}$. Observed points are indicated by dots. (b) Contour map of $^{12}$CO($J=1-0$) integrated intensity in the GMC 225 region. The contour levels are 3, 4, 5, 6, 7, 8, 9, 10, 11, 12, 13, 14, 15, 16, 17, 18, and 19 K km s$^{-1}$. Observed points are indicated by dots. \label{fig09}}
\end{figure}

\subsubsection{GMC 225 (Figure \ref{fig09})}

Figures \ref{fig09}(a) and (b) show the $^{12}$CO($J=3-2$) and $^{12}$CO($J=1-0$) distributions in GMC 225. We note that $J=3-2$ emissions are weaker than $J=1-0$.

\subsection{Properties of the clumps}

We identified clumps according to the following way in the $J=3-2$ distributions shown in Figures \ref{fig03} -- \ref{fig09}; 
1) Pick up local peaks using the integrated intensity. 
2) Draw a contour at half of the peak integrated intensity level and identify it as a clump unless it contains other local peaks. 
3) When there are other local peaks inside the contour, draw new contours at the 70 \% level of each integrated intensity peak. Then, identify clumps separately if their contours do not contain another local peak (the boundary is taken at the "valley" between clumps), or else identify a clump by using the highest contour as a clump boundary. 
4) If a spectrum has multiple velocity components with a separation of more than 6 km s$^{-1}$, identify these components to be associated with different clumps.

As a result, 32 clumps have been identified; 5 clumps in 30 Dor region, 10 clumps in N 159 region, 6 clumps in N 171 region, 5 clumps in N 166 region, 2 clumps in N 206 region, 1 clump in N 206D region, and 3 clumps in GMC 225 region. Their line parameters at the peak positions are shown in Table \ref{tbl02}, and their physical properties are listed in Table \ref{tbl03}. The clump size, line width, and virial mass range 1.1 -- 12.4 pc, 4.0 -- 12.8 km s$^{-1}$, and 4.6$\times$10$^3$ -- 2.2$\times$10$^5$ \Msun, respectively. 
The smallest clump identified by the procedure above is detected with 2 observing points and its size is 1.1 pc. There are several other weak emissions (below 6 $\sigma$), and they are detected with 1 observing point. These emissions are not identified as clumps. When we assume such small clumps ($R$ $\sim$ 1 pc, d$V$ $\sim$ 7 km s$^{-1}$) are exist, their virial mass is 9.3 $\times$ 10$^3$ \Msun, and this corresponds to detection limit.
There are 18 local peaks above the 6 $\sigma$ noise level which are not identified as clumps, because the stronger peaks are located near these local peaks. Their virial mass is 6.5 $\times$ 10$^4$ \Msun, if their size and line width are similar to identified clumps ($R$ $\sim$ 7 pc, d$V$ $\sim$ 7 km s$^{-1}$). This seems to correspond to completeness limit.

\begin{figure}
\epsscale{.40}
\plotone{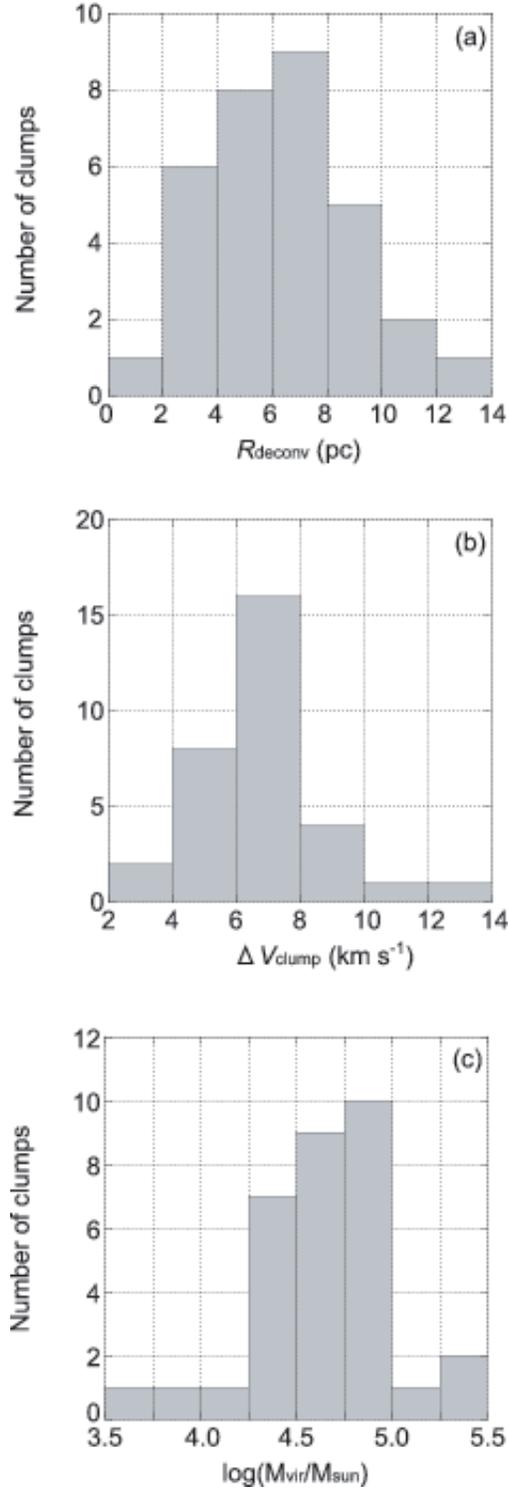}
\caption{Histograms of physical properties of $^{12}$CO($J=3-2$) clumps: (a)size, (b)line width, and (c)virial mass.\label{fig10}}
\end{figure}

%%%%%%%%%%%
%\textcolor{red}{%
If the clumps are identified with the 70 \% level of the peak integrated intensity level instead of the 50 \%, the clump size changes to a half of the original one, althogh the line width does not change. Their virial mass also changes to about a half of the original one.
%}%
%%%%%
Histograms of their physical properties are presented in Figure \ref{fig10}. Typical values are 7 pc, 7 km s$^{-1}$, and 6$\times$10$^4$ \Msun, in size, line width, and virial mass, respectively. Their line widths are larger than those of the GMCs in our galaxy \citep[e.g.,][]{Williams1998, Ikeda1999, Sun2006}. We note a trend that the masses of these clumps are relatively large compared with those of high mass cloud cores in the Milky Way which are of the order of 10$^3$ \Msun \citep[e.g.,][]{Burton2005, Yonekura2005}.
%%%%%%%%%%%
%\textcolor{red}{%
Precise comparisons will be necessary using same tracers and resolution in the future, because those galactic studies above are based on optically thin mm dust continuum and C$^{18}$O emission.
%}%
%%%%%
The clumps in the LMC are also fairly compact, with sizes of several pc or less. Hereafter, the region names and the numbers of clumps are used to identify clumps (e.g., "30 Dor No.1").

\section{Data Analysis}
\subsection{Derivation of line intensity ratios}

The spatial resolution of the present CO data varies depending on the telescope and frequency. We convolved and regridded these data into the same resolution and position using 2 dimensional Gaussian smoothing in order to compare them to derive reliable peak intensity ratios of $^{12}$CO($J=3-2$) to $^{12}$CO($J=1-0$) (hereafter, $R_{3-2/1-0}$). 

We made Gaussian fits to each of the $^{12}$CO($J=3-2$) and $^{12}$CO($J=1-0$) spectra having a single peak in most cases. We derived peak intensities and FWHM line widths through these fittings. The ratios of the two transitions were then derived as the ratio between peak \Tmb s. The distributions of these ratios are shown along with the $^{12}$CO($J=1-0$) distributions in Figures \ref{fig11} -- \ref{fig17} (a). The youngest stellar clusters SWB0, whose ages are estimated to be less than 10 Myrs, are also shown by red circles \citep{Bica1996}. 

Averaged $R_{3-2/1-0}$ over each clump (hereafter, $R_{3-2/1-0,clump}$) was also derived from the averaged $^{12}$CO($J=3-2$) and $^{12}$CO($J=1-0$) spectra over the each clump. A summary of the $R_{3-2/1-0,clump}$ for each clump is shown in Table \ref{tbl04}. The histogram in Figure \ref{fig18} shows that the $R_{3-2/1-0,clump}$ ranges from 0.2 to 1.6. These ratios will be compared with numerical calculations of radiative transfer in the LVG approximation to derive constraints on density and temperature.

\subsection{LVG analysis}

\subsubsection{Calculations of LVG model}

To estimate the physical properties of the molecular gas in the LMC, we performed an LVG analysis \citep{Goldreich1974} of the CO rotational transitions. The LVG radiative transfer code simulates a spherically symmetric cloud of constant density and temperature with a spherically symmetric velocity distribution proportional to the radius, and employs a Castor's escape probability formalism \citep{Castor1970}. It solves the equations of statistical equilibrium for the fractional population of CO rotational levels at each density and temperature. It includes the lowest 40 rotational levels of the ground vibrational level and uses the Einstein's $A$ and H$_2$ impact rate coefficients of \citet{Schoier2005}.

The present calculations incorporate the lowest 40 rotational levels of CO in the ground vibrational state over a kinetic temperature range of $T_{\mathrm{kin}}$ = 5 -- 200 K and a density range of $n$(H$_2$) = 10 -- 10$^6$ cm$^{-3}$. We did not include higher energy levels in the present study, which requires including populations in the lower vibrationally excited states. Therefore, the present work does not cover kinetic temperatures above 200 K, which should be dealt with in the future for analyses of higher sub-mm transitions above $J=4-3$. This imposes a limit of $T_{\mathrm{kin}}$ $\sim$ 200 K in the present study and even higher temperature is not excluded in general below.

We performed calculations for 3 different CO fractional abundances; $X$(CO) = [CO]/[H$_2$] = 1$\times$10$^{-6}$, 3$\times$10$^{-6}$, and 1$\times$10$^{-5}$, and 3 different $^{12}$CO/$^{13}$CO abundance ratios of 20, 25, and 30 (Heikkila et al. 1999).

\subsubsection{Results from $^{12}$CO($J=3-2$) and $^{12}$CO($J=1-0$) data for 32 clumps}

First, we assume that $X$(CO) is uniform among the clumps and derive density and temperature using the LVG results. 

Figure \ref{fig19}(a) illustrates that we obtain the following lower limits for kinetic temperature and density.
For the 8 clumps with $R_{3-2/1-0, clump}$ $\ge$ 1 (30 Dor No.1, 2, 3, 4, N 159 No.1, 2, 6, 8), we estimate $T_{\mathrm{kin}}$ \textgreater  30 K and $n$(H$_2$) \textgreater  10$^3$ cm$^{-3}$. For the 24 clumps with $R_{3-2/1-0, clump}$ \textless  1 (30 Dor No.5, N 159 No.3, 4, 5, 7, 9, 10, N 171 No.1, 2, 3, 4, 5, 6, N 166 No.1, 2, 3, 4, 5, N 206 No.1, 2, N 206D No.1, GMC 225 No.1, 2, 3), we estimate $T_{\mathrm{kin}}$ \textgreater  several K and $n$(H$_2$) \textgreater  several$\times$10$^2$ cm$^{-3}$.
In either case, the lower limits tend to increase with the $R_{3-2/1-0,clump}$.

\subsubsection{Results from $^{12}$CO($J=3-2$), $^{12}$CO($J=1-0$), and $^{13}$CO($J=1-0$) data for 13 clumps}

We can better constrain these physical parameters when $^{13}$CO($J=1-0$) data are available. Figure \ref{fig19} shows the general behavior of the loci of constant $R_{3-2/1-0}$ and constant $R_{12/13}$ (peak intensity ratio of $^{12}$CO($J=1-0$) to $^{13}$CO($J=1-0$)) in the density-temperature plane. It is recognized that the combination of the two will allow us to determine the parameters relatively well since the two lines are nearly "orthogonal" in the plane, except for densities higher than $\sim$ 10$^4$ cm$^{-3}$ (Figure \ref{fig19}). 

Of the 32 clumps, $^{13}$CO($J=1-0$) data are available for 13 clumps, including 4 clumps of $R_{3-2/1-0, clump}$ $\ge$ 1 (30 Dor No.1, 4, N 159 No.1, 2) and 9 clumps with $R_{3-2/1-0, clump}$ \textless 1 (N 159 No.4, N 166 No.1, 3, 4, N 206 No.1, 2, N 206D No.1, GMC 225 No.1, 3). For these 13 clumps, we made a detailed analysis using the $R_{12/13}$ at peaks of $^{12}$CO($J=1-0$) and have obtained the best constraints.

We summarize the input parameters for the 13 clumps in columns 3-5 of Table \ref{tbl05}. All the data refer to the $^{12}$CO($J=1-0$) beam size, 45$\arcsec$; the higher-transition data has been Gaussian smoothed as described in section 4.1. Clump averaged d$v$/d$r$ were used for the calculations. %
$R_{3-2/1-0,clump}$ and $R_{12/13}$ at the peak of $^{12}$CO($J=1-0$) were used, and the errors of these ratios are both estimated as $\pm$20 \%, which are derived from errors of absolute intensity calibration. These correspond 27 $\sigma$ and 7 $\sigma$ noise levels of $^{12}$CO($J=3-2$) and $^{12}$CO($J=1-0$), respectively. This indicates the errors of intensity ratios are dominated by the error of absolute intensity calibration.
%
%%%%%% $R_{3-2/1-0,clump}$ were used, and the error of this ratio is estimated as $\pm$ 20 \% (27 $\sigma$ of $^{12}$CO($J=3-2$) intensity at velocity resolution of 0.44 km s$^{-1}$), which is dominated by the error in the absolute calibration. $R_{12/13}$ at the peak of $^{12}$CO($J=1-0$) were used, and the error of this ratio is estimated to be $\pm$20\% (7 $\sigma$ of $^{12}$CO($J=1-0$) intensity at velocity resolution of 0.5 km s$^{-1}$), which is again dominated by the error in the absolute calibration. %%%%%%
%%%%%%%%%%
%\textcolor{red}{%
The way of the clump definition does not change $R_{3-2/1-0,clump}$, but changes d$v$/d$r$, and d$v$/d$r$ is about 2 times as large as the original one, when the clumps are identified with the 70 \% level of the peak integrated intensity level. This, however, does not affect the results of the LVG calculations.
%}%
%%%%%

We describe three typical cases; (a) 30 Dor No.1, (b) N 206 No.1, and (c) GMC 225 No.1 are shown in Figure \ref{fig22}. A fractional CO abundance of $X$(CO) of 3$\times$10$^{-6}$ was used throughout. The horizontal axis is molecular hydrogen density ($n$(H$_2$)), and the vertical axis is the gas kinetic temperature (\Tkin). Solid lines are $R_{3-2/1-0,clump}$ and dashed lines are $R_{12/13}$ at $^{12}$CO($J=1-0$) peaks. Hatched areas indicate the overlap regions of these two ratios within the errors which is allowed from the observed ratios. It also includes the uncertainty due to a possible variation of the $^{12}$CO/$^{13}$CO abundance ratio from 20 to 30. Hereafter, we shall call clumps above 30 K "warm" and those below "cold". Clumps with densities greater than 10$^{3.5}$ cm$^{-3}$ are referred to as "dense" and those at lower densities as "less dense".

30 Dor No.1 exhibits a $R_{3-2/1-0,clump}$ of $\sim$ 1.4, suggesting that it is a warm (\Tkin = 50 -- 200 K) and dense ($n$ = 10$^{3-5}$ cm$^{-3}$) clump (Figure \ref{fig20}(a)). N 206 No.1(Figure \ref{fig20}(b)) and GMC 225 No.1(Figure \ref{fig20}(c)) show of $R_{3-2/1-0, clump}$ $\sim$ 0.7 and $\sim$0.4, respectively. Figures \ref{fig20}(b) and (c) indicate that they are warm ($T_{\mathrm{kin}}$ = 30 -- 200 K) and less dense clumps ($n$ = 10$^3$ cm$^{-3}$) and cold ($T_{\mathrm{kin}}$ = 15 -- 40 K) and less dense ($n$ = 10$^3$ cm$^{-3}$) clumps, respectively. We find that $R_{12/13}$ is useful to discriminate the temperature difference except in the case of 30 Dor No.1.

All the figures for the 13 clumps for all different fractional CO abundances (1$\times$10$^{-6}$, 3$\times$10$^{-6}$, and 1$\times$10$^{-5}$) are given in Appendix A. In Table \ref{tbl05} we present the results of for the 13 clumps for a fixed fractional CO abundance of $X$(CO) = 3$\times$10$^{-6}$. The results of previous studies are also summarized in Table \ref{tbl05} for comparison. We find that the present results are the most extensive among these studies in terms of the number of samples, while they are basically consistent with previous study for the individual clumps or clouds. There are relatively large differences in both density and temperature in N 159 No.4 (N 159S) only, and this is because of the high density tracers are used in \citet{Heikkila1999}.

Figure \ref{fig21} summarizes the calculated densities and temperatures for the 13 clumps. We see first that the temperature ranges from 10 K up to more than 200 K and density from 10$^3$ cm$^{-3}$ to 10$^5$ cm$^{-3}$. %
%%%%%%%%%%
%\textcolor{blue}{%
Clumps we detected are distributed continuously from cool ($\sim$ 10 -- 30 K) to warm ($\sim$ higher than 30 -- 200 K), and warm clumps are distributed from less dense ($\sim$ 10$^3$ cm$^{-3}$) to dense ($\sim$ 10$^{3.5 - 5}$ cm$^{-3}$), although cool clumps are all less dense. The three cases shown in Figure \ref{fig20} represent typical cases, "warm and dense" ($T_{\mathrm{kin}}$ $\sim$ higher than 30 -- 200 K, $n$(H$_2$) $\sim$ 10$^{3.5 - 5}$ cm$^{-3}$), "warm and less dense" ($T_{\mathrm{kin}}$ $\sim$ higher than 30 -- 200 K, $n$(H$_2$) $\sim$ 10$^3$ cm$^{-3}$), and "cool and less dense" ($T_{\mathrm{kin}}$ $\sim$ 10 -- 30 K, $n$(H$_2$) $\sim$ 10$^3$ cm$^{-3}$).%
%}%
%%%%%

\subsubsection{Effects of $X$(CO)}

We will now discuss the possible effect of changing $X$(CO), as summarized in Table \ref{tbl06}. If we adopt $X$(CO) = 1$\times$10$^{-6}$, the contours of $R_{3-2/1-0}$ shift to lower temperature and the contours of $R_{12/13}$ shift to higher density. Accordingly, the solution shifts to lower temperature, higher density. If we adopt $X$(CO) = 1$\times$10$^{-5}$, the $R_{3-2/1-0}$ contours shift to higher temperature and the $R_{12/13}$ contours shift to lower density. Accordingly, the solution shifts to higher temperature, lower density (Table \ref{tbl06}).

Next, we vary $X$(CO) from clump to clump. \citet{Heikkila1999} estimate $X$(CO) in 30 Dor No.1 (30 Dor-10), N 159 No.1 (N 159W), and N 159 No.4 (N 159S) to be 1$\times$10$^{-6}$, 1$\times$10$^{-5}$, and 3$\times$10$^{-6}$, respectively. Table \ref{tbl06} indicates that 30 Dor No.1 shows similar values to \citet{Heikkila1999}, whereas N 159 No.1 and No.4 become warmer and lower in density than in \citet{Heikkila1999}. This discrepancy may be due to the fact that the high density tracers used by \citet{Heikkila1999} are not used in the present study. To summarize, the assumption of uniform fractional abundance, $X$(CO), is fairly good and the present results do not show significant difference even if we adopt the different fractional abundances used by previous authors.

\subsection{Comparisons to \Halpha flux}

\subsubsection{Relation between $R_{3-2/1-0}$ and \Halpha flux}

\begin{figure}
\epsscale{.80}
\plotone{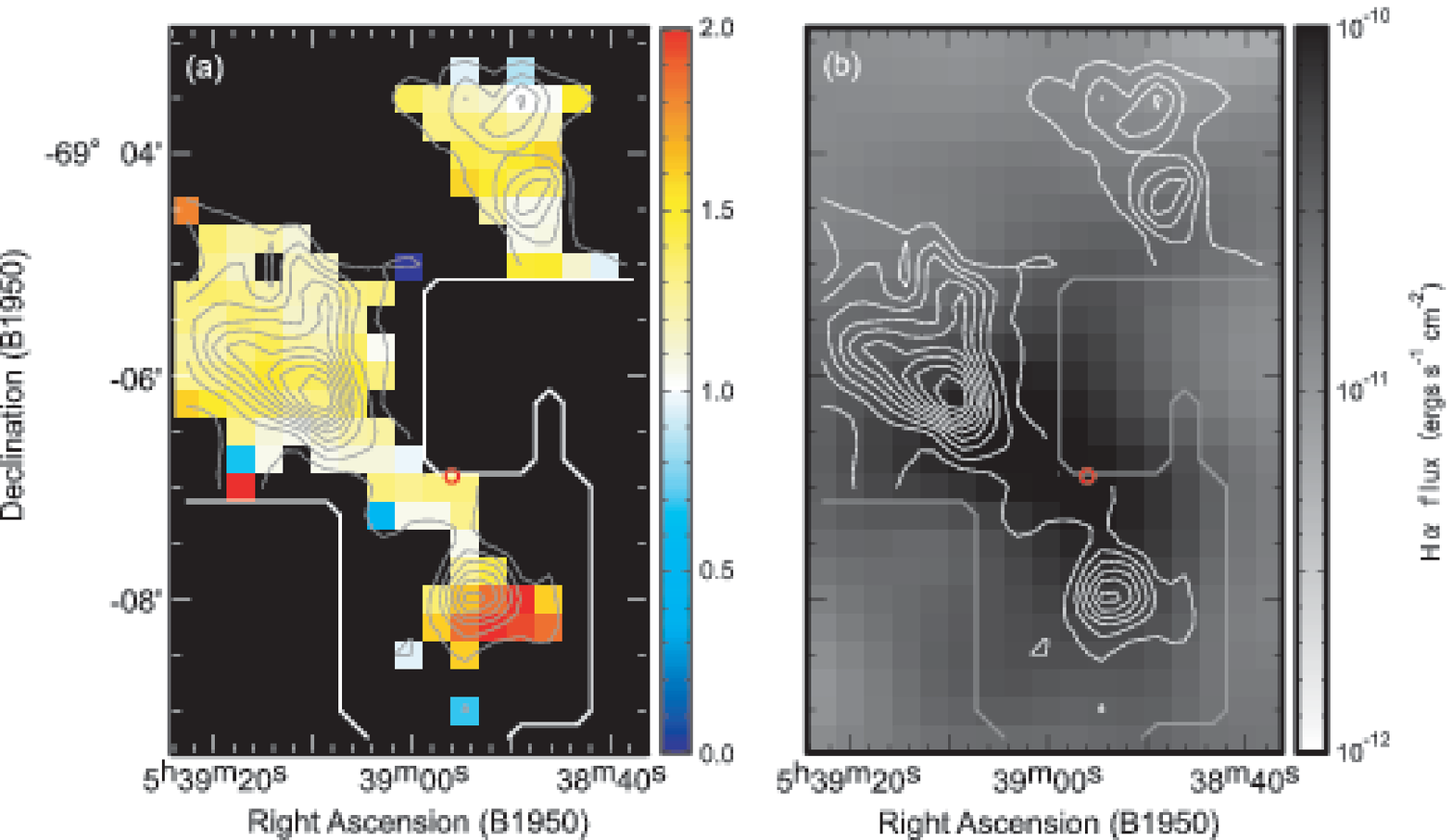}
\caption{(a) Color map of $R_{3-2/1-0}$, and (b) H$\alpha$ flux image of the 30 Doradus region. Contours are $^{12}$CO($J=3-2$) integrated intensity. Contour levels are the same as Figure 3a. Thick lines indicate the observed area in $^{12}$CO($J=3-2$), and the red circle indicates the position of young cluster (\textless 10 Myrs; SWB0).\label{fig11}}
\end{figure}

\begin{figure}
\epsscale{.80}
\plotone{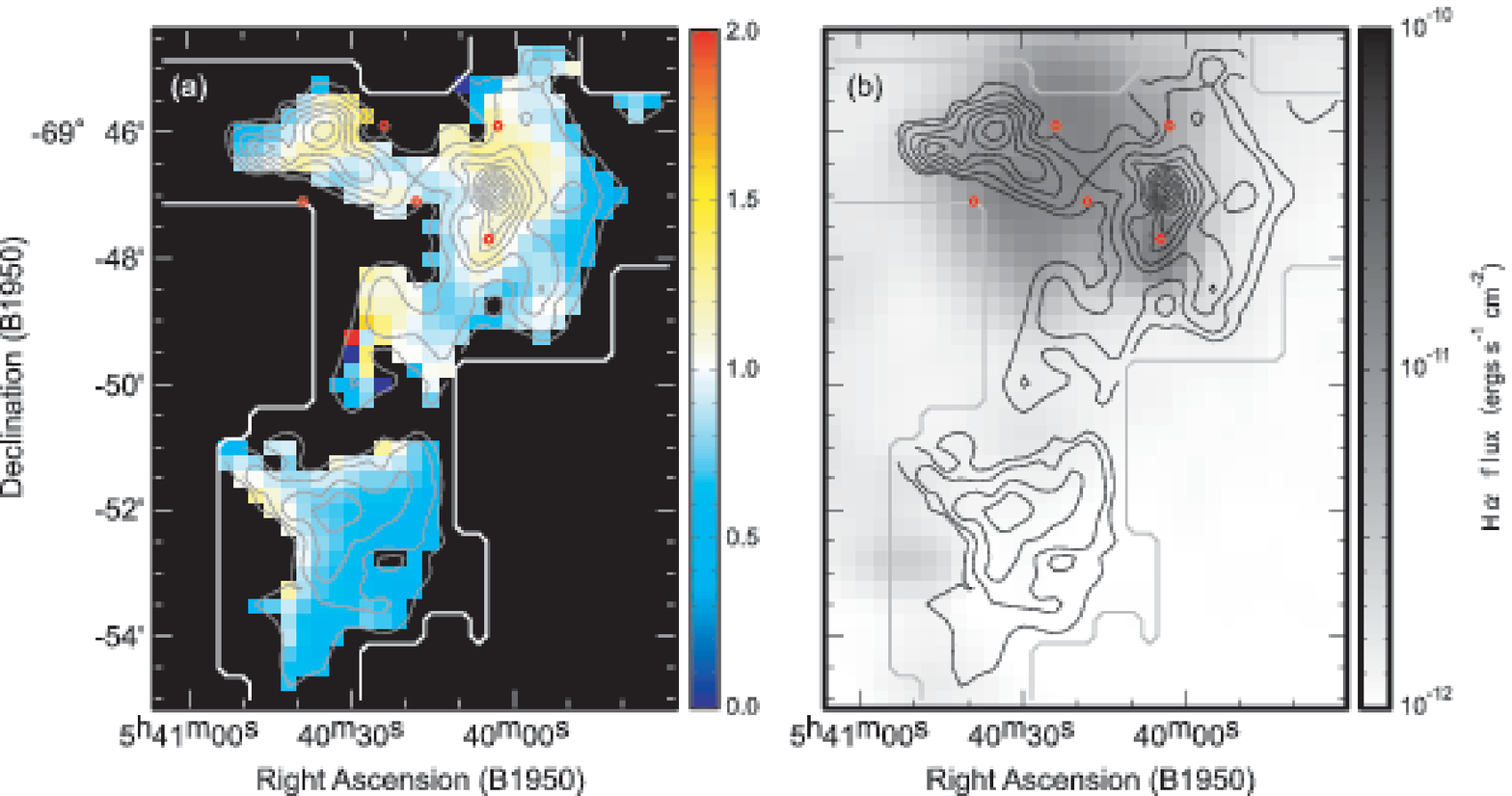}
\caption{(a) Color map of $R_{3-2/1-0}$, and (b) H$\alpha$ flux image of the N 159 region. Contours are $^{12}$CO($J=3-2$) integrated intensity. Contour levels are the same as Figure 4a. Thick lines indicate observed area of $^{12}$CO($J=3-2$), and the red circles indicate the position of young clusters (\textless 10 Myrs; SWB0).\label{fig12}}
\end{figure}

\begin{figure}
\epsscale{.80}
\plotone{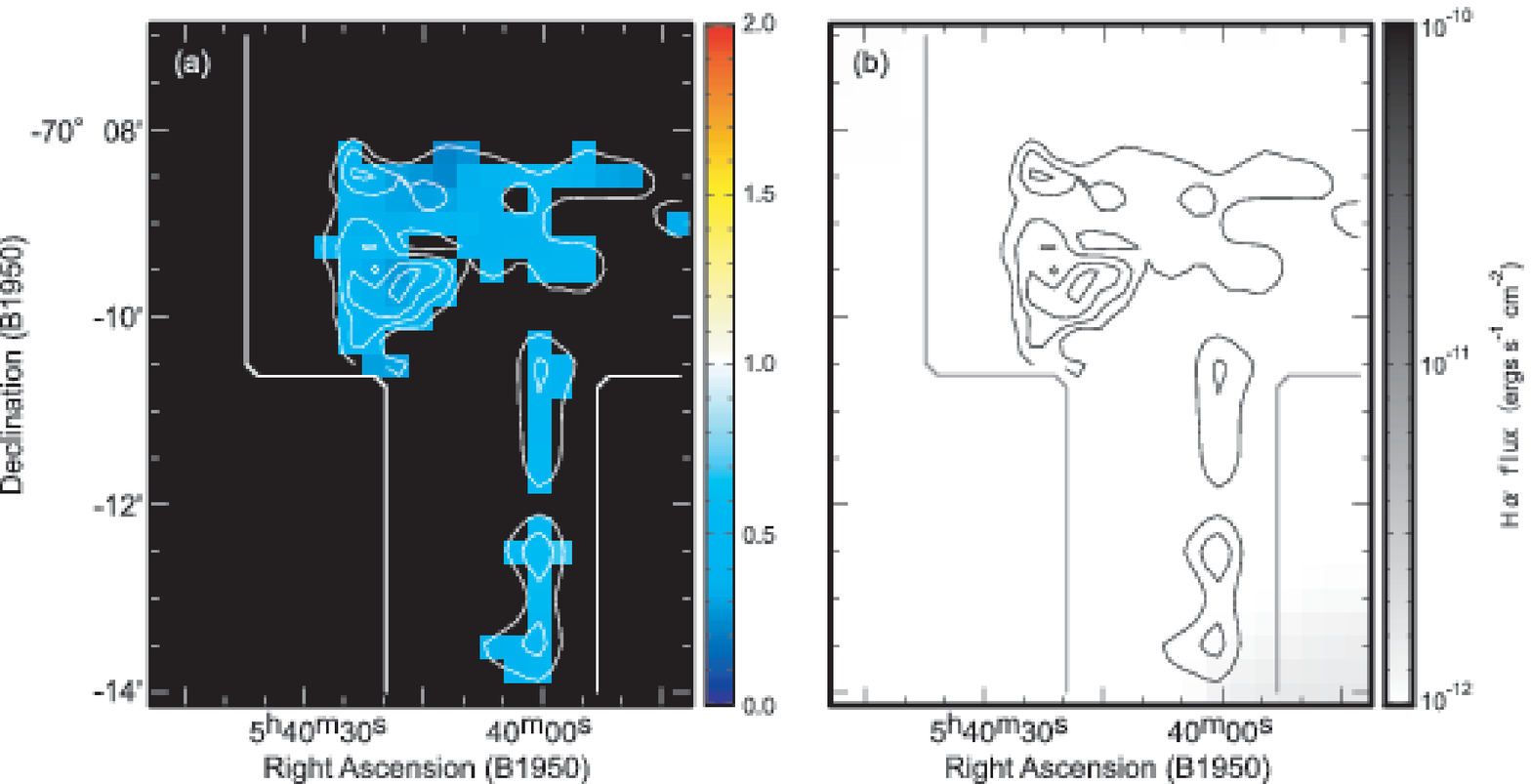}
\caption{(a) Color map of $R_{3-2/1-0}$, and (b) H$\alpha$ flux image of the N 171 region. Contours are $^{12}$CO($J=3-2$) integrated intensity. Contour levels are the same as Figure 5a. Thick lines indicate observed area of $^{12}$CO($J=3-2$).\label{fig13}}
\end{figure}

\begin{figure}
\epsscale{.80}
\plotone{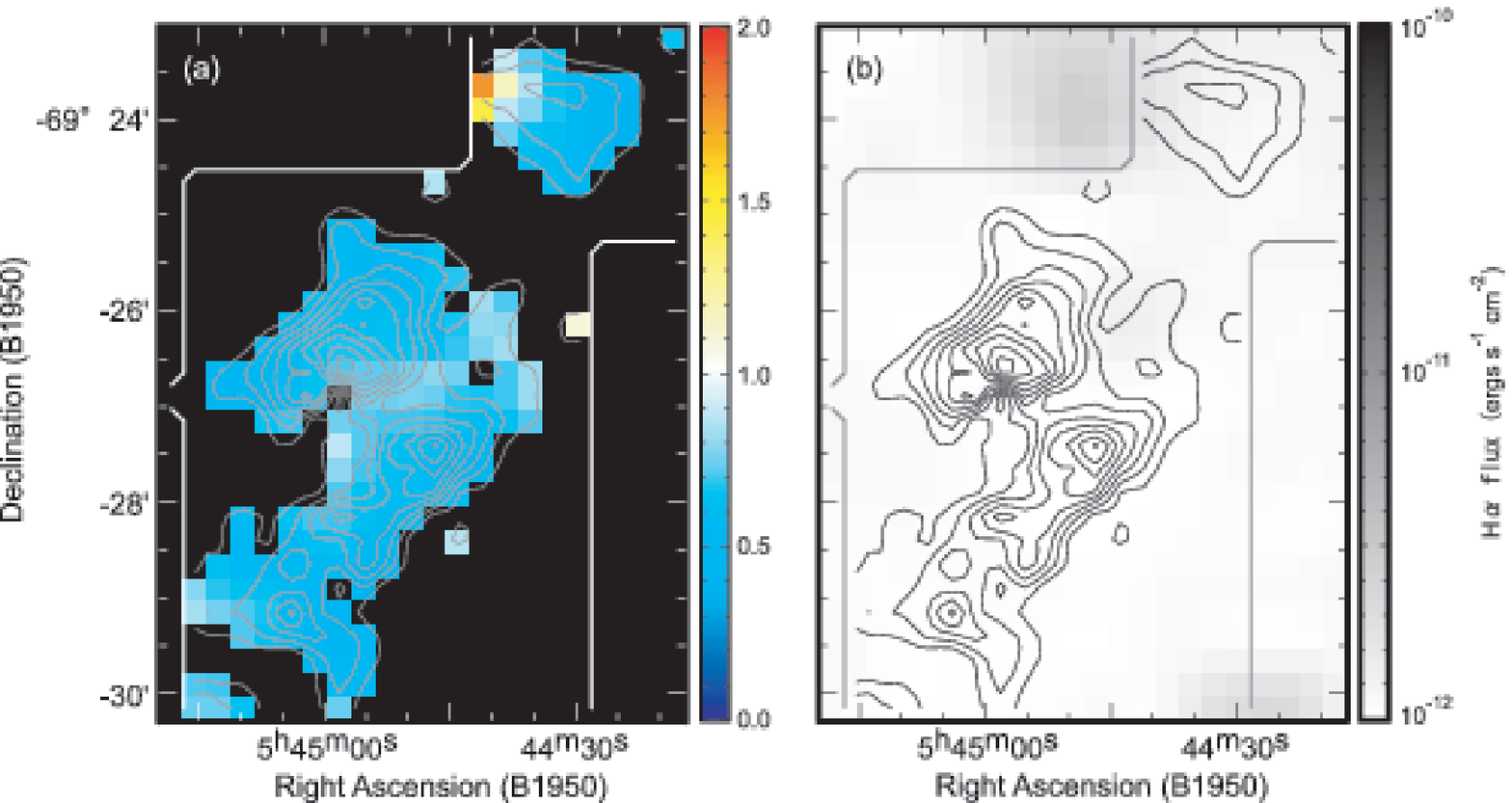}
\caption{(a) Color map of $R_{3-2/1-0}$, and (b) H$\alpha$ flux image of the N 166 region. Contours are $^{12}$CO($J=3-2$) integrated intensity. Contour levels are the same as Figure 6a. Thick lines indicate observed area of $^{12}$CO($J=3-2$).\label{fig14}}
\end{figure}

\begin{figure}
\epsscale{.80}
\plotone{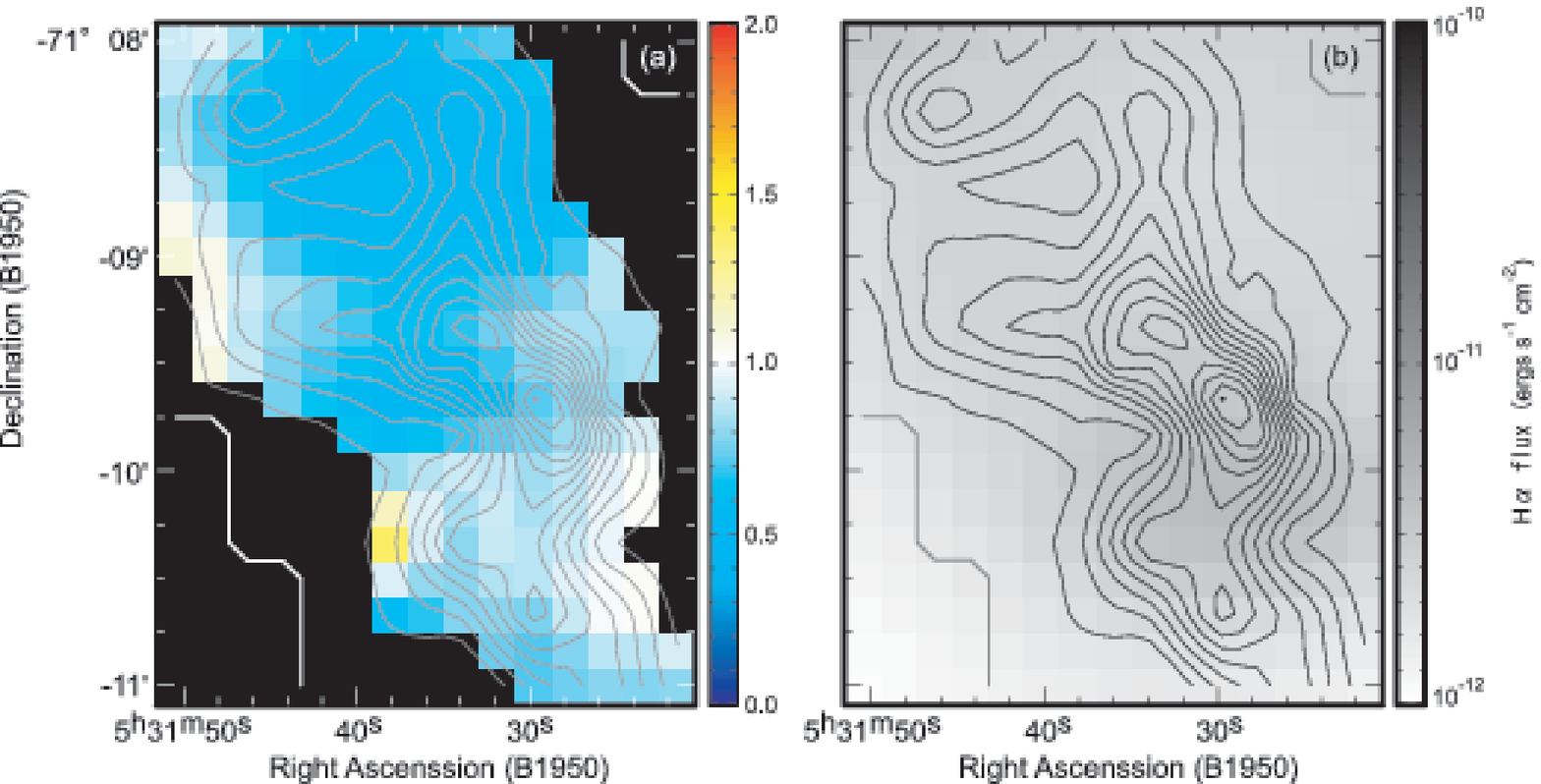}
\caption{(a) Color map of $R_{3-2/1-0}$, and (b) H$\alpha$ flux image of the N 206 region. Contours are $^{12}$CO($J=3-2$) integrated intensity. Contour levels are the same as Figure 7a. Thick lines indicate observed area of $^{12}$CO($J=3-2$).\label{fig15}}
\end{figure}

\begin{figure}
\epsscale{.80}
\plotone{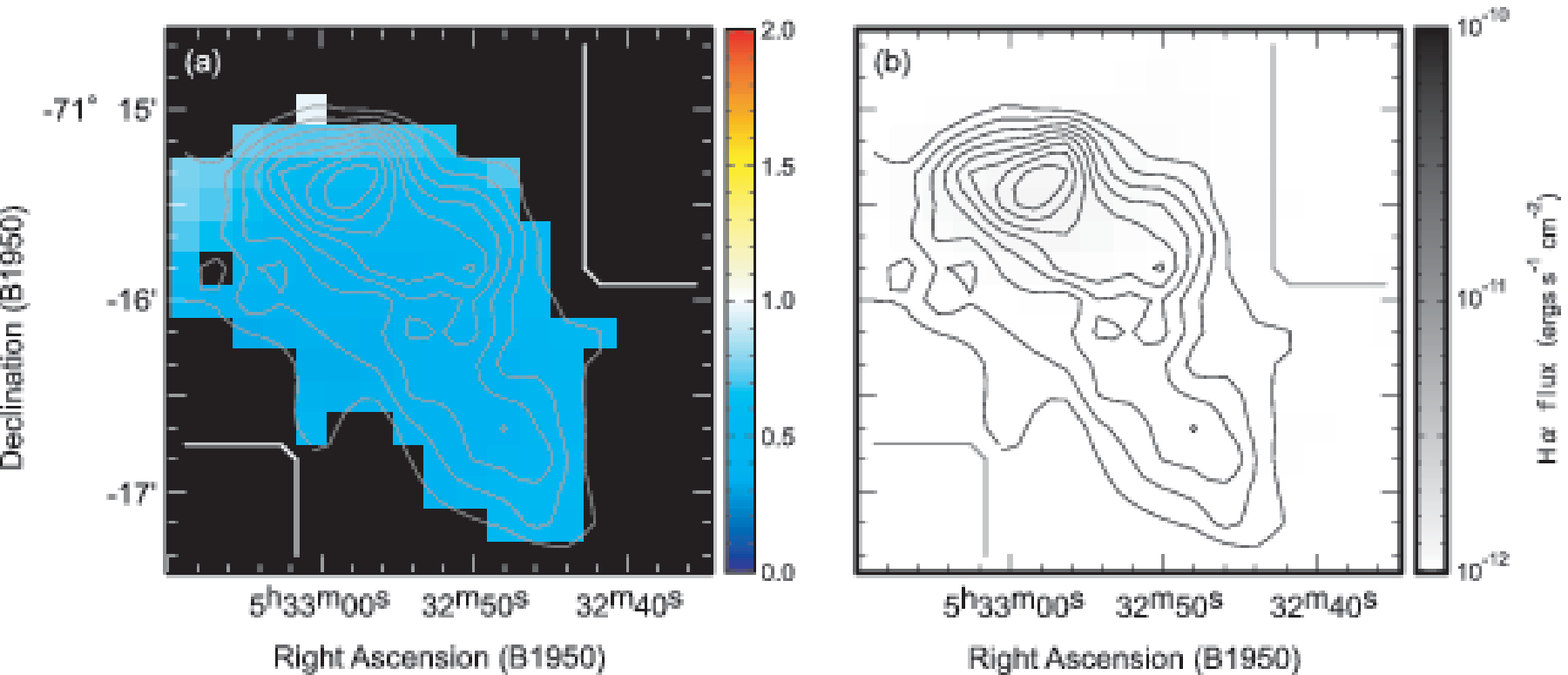}
\caption{(a) Color map of $R_{3-2/1-0}$, and (b) H$\alpha$ flux image of the N 206D region. Contours are $^{12}$CO($J=3-2$) integrated intensity. Contour levels are the same as Figure 8a. Thick lines indicate observed area of $^{12}$CO($J=3-2$).\label{fig16}}
\end{figure}

\begin{figure}
\epsscale{.80}
\plotone{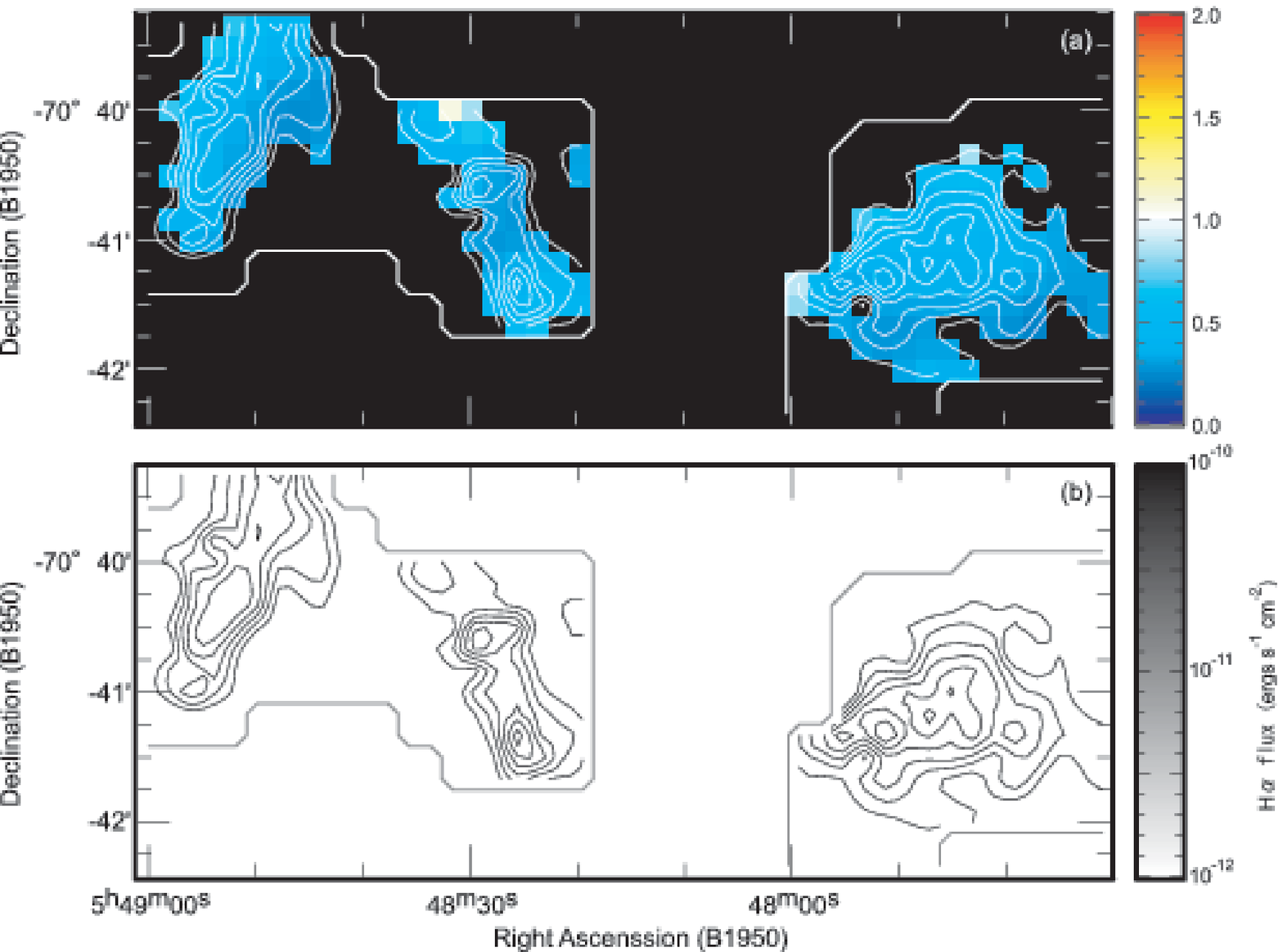}
\caption{(a) Color map of $R_{3-2/1-0}$, and (b) H$\alpha$ flux image of the GMC 225 region. Contours are $^{12}$CO($J=3-2$) integrated intensity. Contour levels are the same as Figure 9a. Thick lines indicate observed area of $^{12}$CO($J=3-2$).\label{fig17}}
\end{figure}

\begin{figure}
\epsscale{.80}
\plotone{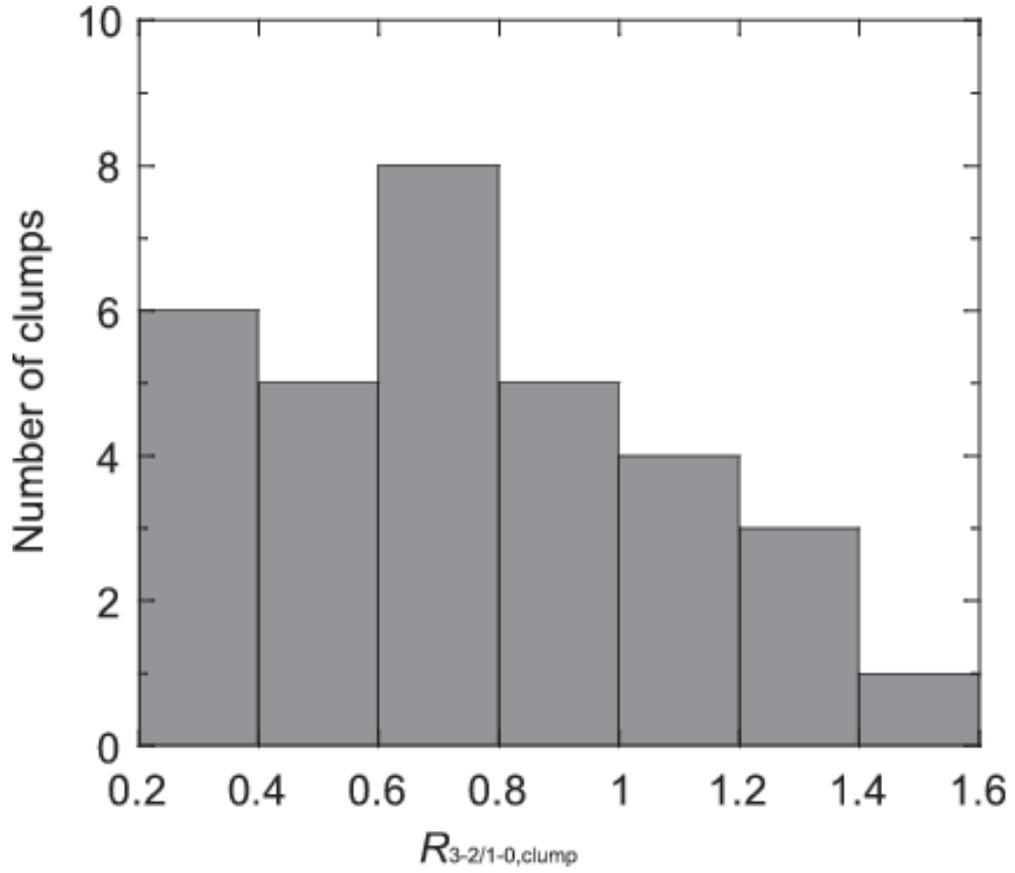}
\caption{Histgram of clump averaged peak intensity ratio of $^{12}$CO($J=3-2$) to $^{12}$CO($J=1-0$) ($R_{3-2/1-0,clump}$).\label{fig18}}
\end{figure}

\begin{figure}
\epsscale{.80}
\plotone{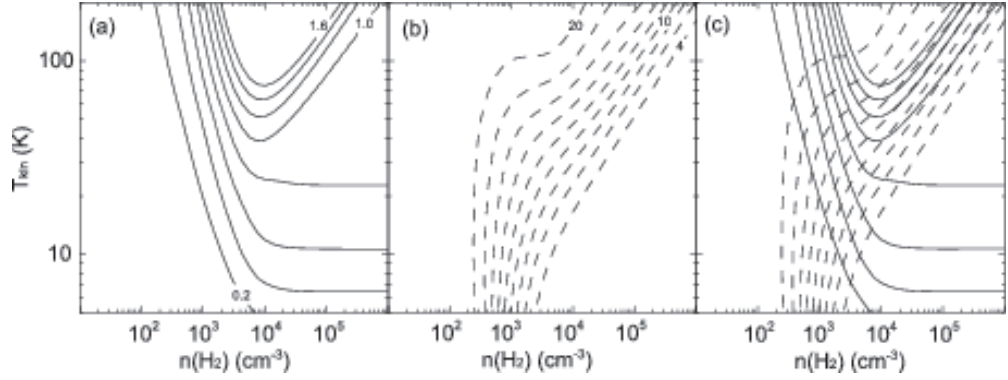}
\caption{Contour plots of LVG analysis for reference. Contours are (a) $R_{3-2/1-0}$, (b) $R_{12/13}$, and (c) a+b. $X$(CO)=3$\times$10$^{-6}$, d$v$/d$r$=1.0km s$^{-1}$ pc$^{-1}$, and abundance ratio of $^{12}$CO/$^{13}$CO is 25. \label{fig19}}
\end{figure}

\begin{figure}
\epsscale{.80}
\plotone{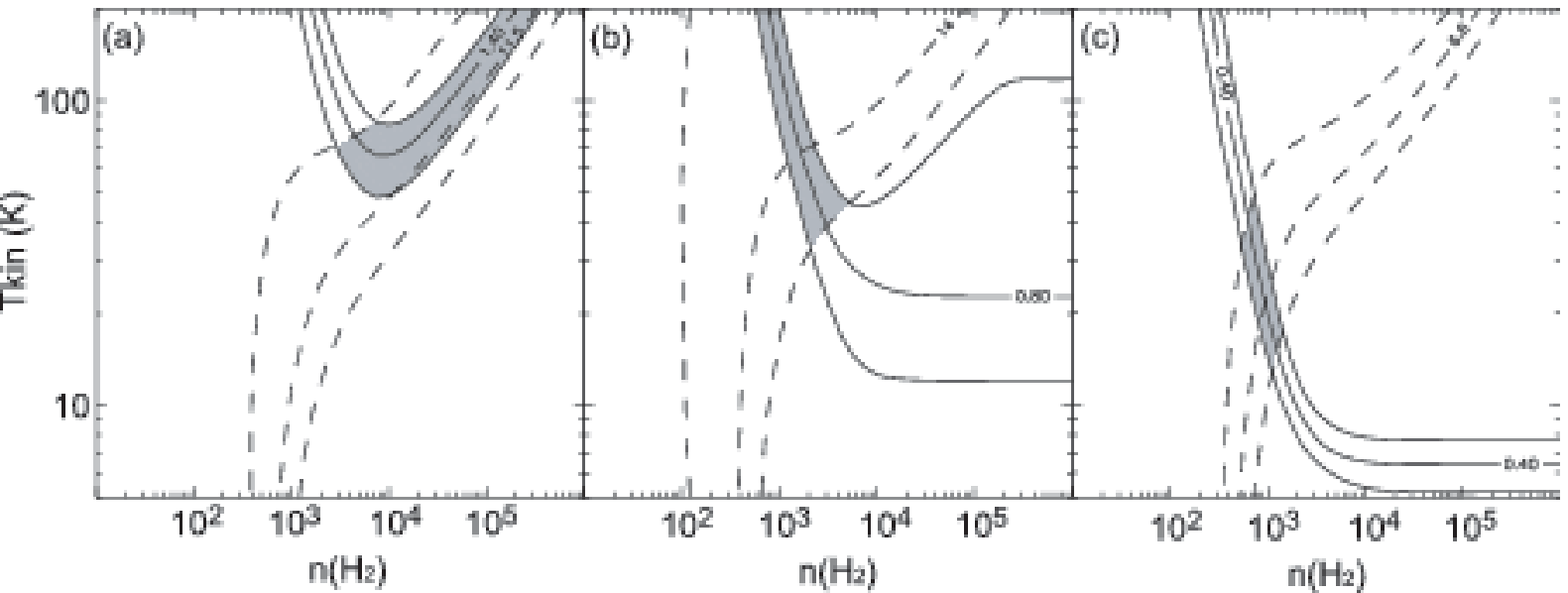}
\caption{Contour plots of LVG analysis of 3 clumps: (a)30 Dor No.1, (b)N 206 No.1, and  (c)GMC 225 No.1. The vertical axis is kinetic temperature \Tkin, and the horizontal axis is molecular hydrogen density $n$(H$_2$). Solid lines are $R_{3-2/1-0, clump}$, and dashed lines are $R_{12/13}$. Hatched areas are the regions in which these two ratios overlap within intensity calibration errors of 20\% and uncertainty due to a possible variation of $^{12}$CO/$^{13}$CO abandance ratio from 20 to 30. \label{fig20}}
\end{figure}

\begin{figure}
\epsscale{.80}
\plotone{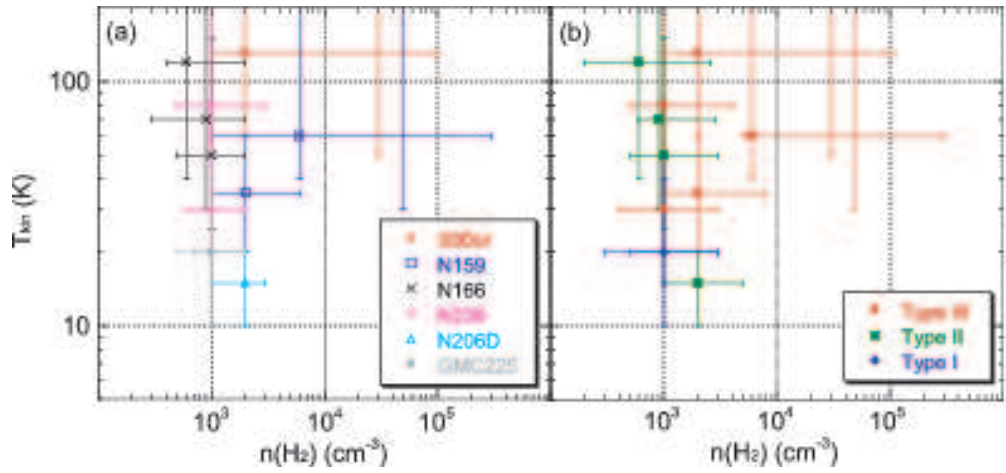}
\caption{Plot of LVG results. The vertical axis is kinetic temperature, \Tkin, and the horizontal axis is molecular hydrogen density $n$(H$_2$).\label{fig21}}
\end{figure}

\begin{figure}
\epsscale{.80}
\plotone{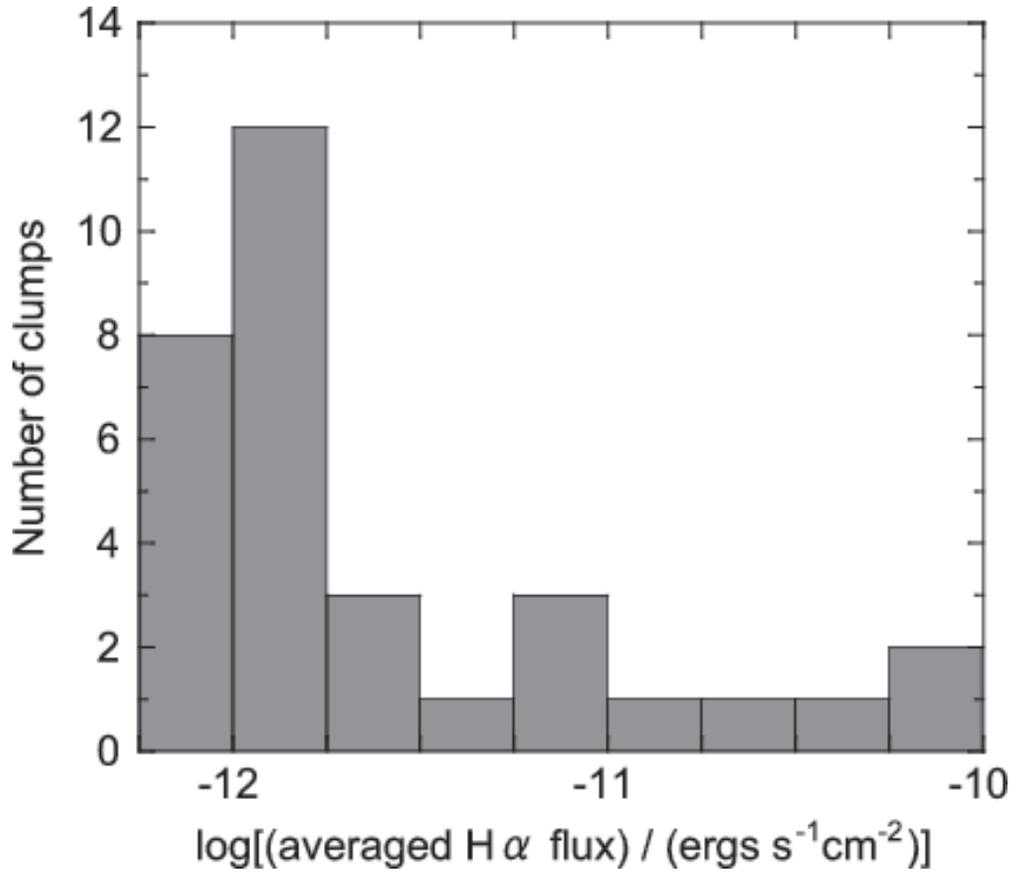}
\caption{Histgram of clump averaged H$\alpha$ flux. Background level is nearly 10$^{-12}$ ergs s$^{-1}$ cm$^{-2}$.\label{fig22}}
\end{figure}

We converted the \Halpha data \citep{Kim1999} towards the present clouds using the method given in the Appendix B. The typical background level of the \Halpha flux is $\sim$ 10$^{-12}$ ergs s$^{-1}$ cm$^{-2}$ at the 40$\arcsec$ scale and ranges up to 10$^{-10}$ ergs s$^{-1}$ cm$^{-2}$ towards strong HII regions (Figure \ref{fig22}). These data were re-gridded into the $^{12}$CO($J=3-2$) data grids. \Halpha flux images with the youngest stellar clusters SWB0 \citep[younger than 10 Myrs]{Bica1996} are shown in Figures \ref{fig11} -- \ref{fig17} (b) and the $^{12}$CO($J=3-2$) contours from Figures \ref{fig03} -- \ref{fig09} are overlayed for comparison in these figures.

Figures \ref{fig11} -- \ref{fig17} indicate a clear trend that the $R_{3-2/1-0}$ is enhanced to 1.0 -- 1.5 towards HII regions or clouds with young clusters, as in 30 Dor and N 159 regions. On the other hand, the ratio is low, less than 1.0, towards clumps with neither HII regions nor clusters, as in GMC 225. We also note that the ratio is enhanced towards the regions where \Halpha is intense or towards the interfaces between clouds and HII regions. A summary of the averaged \Halpha flux over each clumps is shown in Table \ref{tbl04}. It is not fully guaranteed that the \Halpha emission is actually in contact with the molecular gas and some of the apparent coincidence could be fortuitous. Nonetheless, previous studies comparing CO and \Halpha flux indicate a strong correlation between them and lend a support to the assumption that almost all the coincidences indicate actual physical association \citep{Fukui1999, Yamaguchi2001c, Kawamura2007}.

\begin{figure}
\epsscale{.80}
\plotone{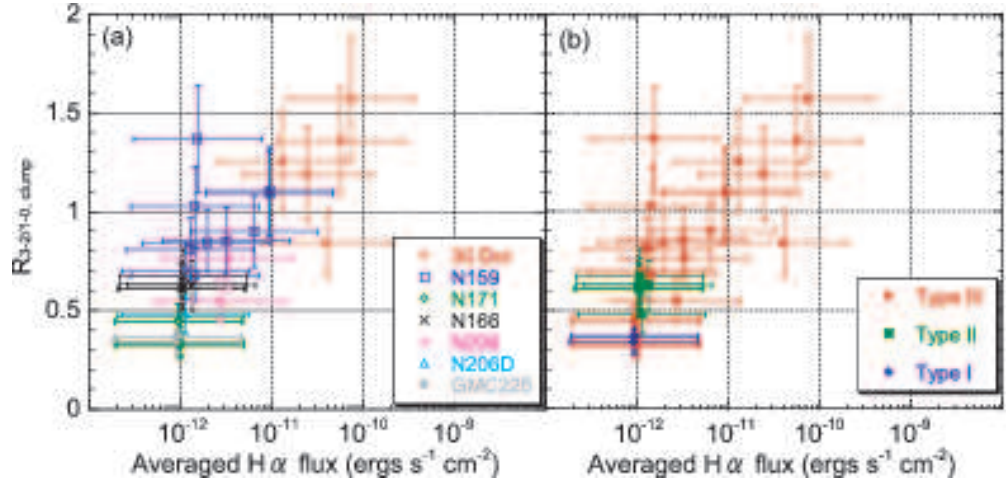}
\caption{Plots of $R_{3-2/1-0, clump}$ as a function of clump averaged H$\alpha$ flux by (a)region and (b)GMC type, respectively. \label{fig23}}
\end{figure}

Figure \ref{fig23} shows the correlation between $R_{3-2/1-0,clump}$ and the averaged \Halpha flux over each clump. It is clear that $R_{3-2/1-0,clump}$ is well correlated with the averaged \Halpha flux, with a correlation coefficient of $\sim$ 0.79. This is a fairly good empirical relationship and should be tested to see if it holds true in other galaxies. The relation suggests that higher $R_{3-2/1-0}$ reflect higher temperatures or higher densities. It is notable that the clumps with the averaged \Halpha flux greater than 10$^{-11}$ ergs s$^{-1}$ cm$^{-2}$ correspond to high $R_{3-2/1-0,clump}$ of 1.0 -- 1.5. In the barred spiral galaxy M83, the CO($J=3-2$)/($J=1-0$) integrated intensity ratio exceeds unity at the nucleus, whereas the ratio gradually decreases to 0.6 -- 0.7 with distance from center. The ratio is constant through the disk region \citep{Muraoka2007}.

\subsubsection{Comparison with physical properties and \Halpha emisson}

\begin{figure}
\epsscale{.80}
\plotone{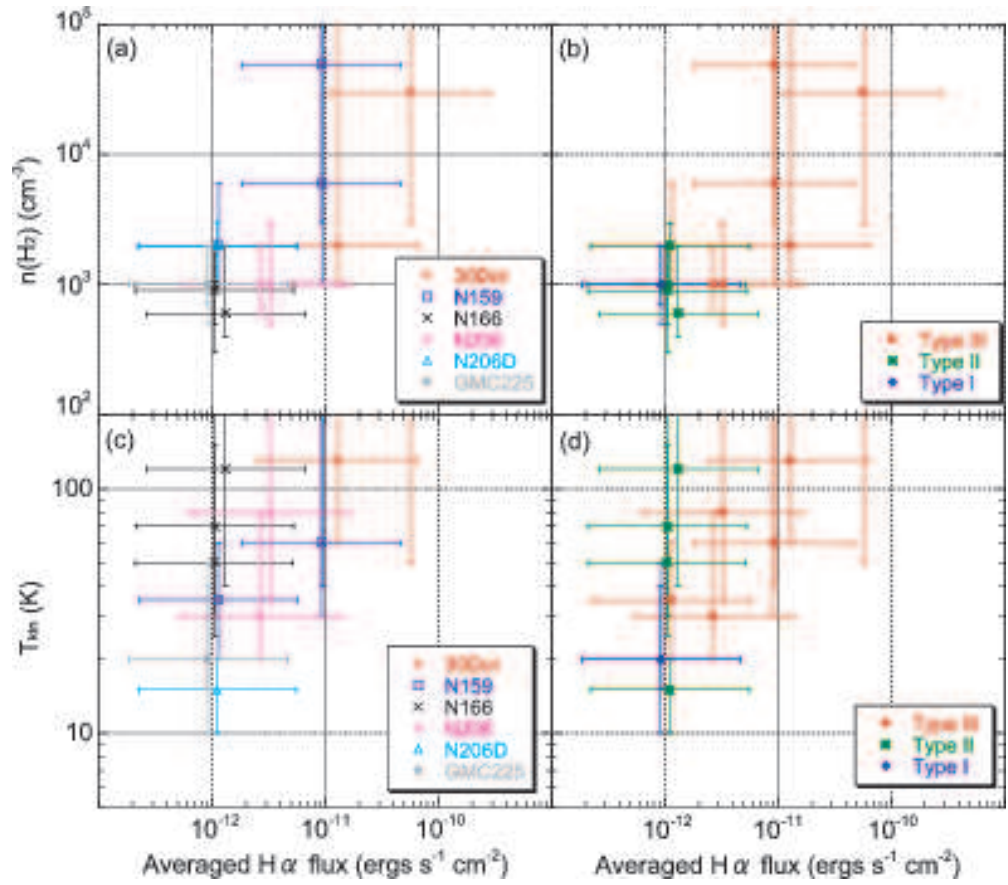}
\caption{Plots of physical properties as a function of clump averaged H$\alpha$ flux. (a) $n$(H$_2$) by region. (b) $n$(H$_2$) by GMC type. (c) $T_{\mathrm{kin}}$ by region. (d) $T_{\mathrm{kin}}$ by GMC type.\label{fig24}}
\end{figure}

Figure \ref{fig24} shows plots of density and temperature as functions of averaged \Halpha flux, and gives us another insight into these properties. Where averaged \Halpha is strong (10$^{-11}$ergs s$^{-1}$ cm$^{-2}$) the clumps are always warm at around $T_{\mathrm{kin}}$ = 100 K or more. On the other hand, when averaged \Halpha is weak, density is always low but temperature can be either high or low. 

The high density clumps show high $R_{3-2/1-0,clump}$ of 1.0 -- 1.5 and are associated with strong \Halpha flux, while the low density clumps show low $R_{3-2/1-0, clump}$ of 0.5 -- 1.0 with weak \Halpha flux. 

Since we are averaging the \Halpha intensity in each clump, we may be diluting the localized low \Halpha flux towards some of the clumps in Type III and II GMCs. The effects of \Halpha emission or nearby clusters on the molecular gas are perhaps local phenomena as indicated by the comparison in Figures \ref{fig11} -- \ref{fig17} for the individual clumps.

\section{Discussions}

\subsection{Dense and compact clumps as candidates for proto-cluster condensations}

We have carried out sub-mm $^{12}$CO($J=3-2$) observations of GMCs in the LMC which are most extensive and highly sensitive compared to the previous studies. Six GMCs were selected based on the NANTEN CO survey of the LMC, including 3 Type III GMCs actively forming O stars in addition to 3 Type I/II GMCs which are quiet in O-star formation or cluster formation, although the formation of low to intermediate mass star is not excluded. The spatial resolution of $\sim$ 5 pc and the high sensitivity allowed us to identify 32 molecular clumps in these GMCs and to reveal significant details of the warm and dense molecular gas with $n$(H$_2$) $\sim$ 10$^{3-5}$ cm$^{-3}$ and $T_{\mathrm{kin}}$ $\sim$ 10 -- 200 K.

The typical mass of the molecular clumps is large, in the range of 5$\times$10$^3$ -- 2$\times$10$^5$ M$_{\odot}$ with radii of 1 -- 12 pc. Of all of our objects, N 159 No.1 or -W shows the strongest concentration of mass of $\sim$ 7$\times$10$^4$ M$_{\odot}$ within a radius of $\sim$ 5 pc. The masses seem to be larger than those of typical Milky Way GMCs such as those in the eta Car region \citep[e.g.,][]{Yonekura2005}%
%%%%%%%%%%%
%\textcolor{red}{%
, although the propeties of these galactic GMCs are based on optically thin C$^{18}$O data.
%}%
%%%%%
We suggest that these are good candidates for the precursors of rich super clusters like R136 in 30 Dor which includes more than 10$^4$ stars in a small volume with a radius of $\sim$ 1 pc. It is of particular interest to look for even denser gas towards them in higher excitation transitions of the sub-mm region.

\subsection{Density and temperature of the clumps and implications}

The results of our LVG analysis indicate that clumps are distributed from cool to warm in temperature and from less dense to dense in density. These differences of clump properties in density and temperature show good correspondence with the GMC Types based on the star formation activity, as well as with the \Halpha emission of ionized gas associated with each clump.
Clumps in Type III GMCs are warm ($T_{\mathrm{kin}}$ $\sim$ 30 -- 200 K) and are either dense ($n$(H$_2$) $\sim$ 10$^{3.5-5}$ cm$^{-3}$) or less dense ($n$(H$_2$) $\sim$ 10$^3$cm$^{-3}$). Clumps in Type II GMCs are either warm ($T_{\mathrm{kin}}$ $\sim$ 30 -- 200 K) or cool ($T_{\mathrm{kin}}$ $\sim$ 10 -- 30 K) and less dense ($n$(H$_2$) $\sim$ 10$^3$ cm$^{-3}$). Clumps in Type I GMC are cool ($T_{\mathrm{kin}}$ $\sim$ 10 -- 30 K) and less dense ($n$(H$_2$) $\sim$ 10$^3$cm$^{-3}$). The physical parameters of clumps are generally correlated with the star formation activity of GMCs and can perhaps be interpreted in terms of evolutionary effects.

%%%%%%%%%%
%\textcolor{blue}{%
Our interpretation is that defferences of clump density and temperature represent an evolutionary sequence of GMCs in terms of density increase leading to star formation; Type I/II GMCs are at a young phase of star formation where density has not yet reached high enough values to cause active massive star formation, and Type III GMCs represent the later phase where the average density is higher, including both high and low density sub-types. The high density clumps in Type III GMCs show high $R_{3-2/1-0,clump}$ of 1.0 -- 1.5 and are associated with strong \Halpha flux while the low density clumps in Type III GMCs show low $R_{3-2/1-0,clump}$ of 0.5 -- 1.0 with weak \Halpha flux. 
%}%
%%%%%

We suggest two alternative ideas to explain %
%%%%%%%%%%
%\textcolor{blue}{%
the density difference of the clumps %
%}% 
%%%%%
in Type III GMCs; one is that density is being enhanced by shock compression driven by HII regions and the other is that gravitational condensation of each clump plays a role in the density increase. The former may be difficult because the shock front may occupy a small volume which is likely missed with the present 5 pc beam. It seems thus favorable that the latter scenario is working mainly to enhance density. 

The present study, which resolved the smaller clumps in GMCs at 5 -- 10 pc scales, indicates that the clumps may have physical properties affected by local properties such as the \Halpha distribution. It should be interesting to investigate the variations among these internal clumps and their relation to star formation.

\subsection{FUV Heating of the molecular gas in the LMC}

The present findings that the $R_{3-2/1-0,clump}$ is well correlated with \Halpha flux suggests that the heating of molecular gas by far-ultraviolet (FUV) photons may be effective in the LMC where the dust opacity is lower and the FUV intensity is higher than in the Milky Way. 
The molecular gas in the Milky Way is mainly heated by cosmic ray protons of $\sim$ 100 MeV as discussed by a number of authors, although the surface layers of molecular clouds with small visual extinctions at $A$v $\sim$ a few mag or less may be dominated by the FUV heating \citep[e.g.,][]{Kaufman1999}.
Some authors have made detailed calculations of gas heating and cooling under the effects of FUV radiation fileds \citep{Kaufman1999}. We shall try to present a picture that can be applied to the present results below.

First, the gas temperature is determined through the balance between the cooling and heating. According to Table 4 in \citet{Goldsmith1978}, the total cooling rate is 6.8$\times$10$^{-27}T^{2.2}$ ergs cm$^{-3}$ s$^{-1}$ for $X$(CO)/(d$v$/d$r$) = 4$\times$10$^{-5}$ and $n$(H$_2$) = 10$^3$ cm$^{-3}$. In 30 Dor region, since $X$(CO)/(d$v$/d$r$) = 3$\times$10$^{-6}$ / 0.8 = 3.75$\times$10$^{-6}$ and this value is 10 times lower than the value used in \citet{Goldsmith1978}, $n$(H$_2$) = 10$^4$ cm$^{-3}$ can be read 10$^3$ cm$^{-3}$, then the cooling rate is estimated as 1.7$\times$10$^{-22}$ ergs cm$^{-3}$ s$^{-1}$ for $T$ = 100 K.
This value is a factor 2 -- 3 smaller than that of Galactic clouds with $n$(H$_2$) = 10$^4$ cm$^{-3}$ and $T$ = 50 K.

We shall assume that the heating by cosmic ray electrons is not important in the LMC. This assumption is not directly confirmed, but it is consistent with the low non-thermal fraction of the LMC's radio continuun emission \citep[e.g.,][]{Hughes2006} and studies that suggest a significant fraction of cosmic ray electrons are able to escape from low luminosity galaxies \citep[e.g.,][]{Bell2003, Skillman1988}

The FUV flux ($G_0$) is estimated as 3500 for 30 Doradus \citep{Bolatto1999, Poglitsch1995, Werner1978, Israel1979}, and 300 for N 159 \citep{Bolatto1999, Israel1996, Israel1979}. In Orion, it is estimated as 25 \citep{Bolatto1999, Stacey1993}. The FUV flux in the LMC is larger than that in the Milky Way.

PDR models are calculated by \citet{Kaufman1999} which incorporate the chemical and physical processes that form and destruct atoms or molecules, as well as ionization effects. Figuer 1 of \citet{Kaufman1999} shows the kinetic temperature for a molecular gas layer with density of $n$ (cm$^{-3}$) under FUV flux of $G_0$ at the surface. PDR surface temperatures are estimated as listed in Table \ref{tbl07}.
These indicate that temperature becomes as high as 100 -- 300 K on the PDR surface under the conditions of the clumps in Type III GMCs in the LMC. These temperatures are basically consistent with the temperatures of the warm clumps in the present sample.

Generally speaking, at a scale of $\sim$ 10 pc, $T_{\mathrm{kin}}$ $\sim$ 100 K seems to be higher than the kinetic temperatures typical in Milky Way GMCs, where the Milky Way values are usually derived from the $^{12}$CO($J=1-0$) emission only \citep[e.g., eta Car $T_{\mathrm{kin}}$ $\sim$ 50 K by][]{Yonekura2005}. This suggests that the heating of molecular clouds may be stronger in the LMC than in the Milky Way and the molecular temperature may be higher. If this is correct, the lower metallicity, resulting in lower extinction, is the basic cause for the higher temperature in addition to the stronger FUV field in the LMC. We shall note in the end that this higher temperature in the molecular gas possibly leads to an increase of the Jeans mass of molecular clumps, which may favor the formation of rich super clusters in the LMC. This is consistent with the higher mass of the molecular clumps which may represent precursors of the clusters.

The present work has undertaken to sample 6 GMCs (7 regions) to have a uniform determination of the density and temperature in the LMC. The number of GMCs is still limited to 6 among $\sim$ 300 detected with NANTEN. We should make more efforts to collect appropriate data sets in the sub-mm wavelengths to improve our understanding of the cloud properties. NANTEN2, ASTE and others will certainly be powerful in achieving this goal.

\section{Summary}

We summarize the results as follows.

1) We have used the ASTE 10m telescope to obtain the distribution of $^{12}$CO($J=3-2$) emission at 345 GHz towards 6 GMCs (7 regions) in the LMC. We have identified 32 clumps in these GMCs at $\sim$ 5pc resolution. The radius, line width and virial mass are estimated as 1.1 -- 12.4 pc (7 pc), 4.0 -- 12.8 km s$^{-1}$ (7 km s$^{-1}$), and 4.6$\times$10$^3$ -- 2.2$\times$10$^5$ \Msun (6$\times$10$^4$ \Msun), respectively, with the average values in the parenthesis.

%%%%%%%%%%
%\textcolor{blue}{%
2) We have compared the present results with LVG radiative line transfer calculations in order to obtain the density and temperature estimated for the 13 clumps using $R_{3-2/1-0,clump}$ and $R_{12/13}$ at 45$\arcsec$ resolution. The clumps are distributed from cool ($\sim$ 10 -- 30 K) to warm (more than $\sim$ 30 -- 200 K) and from less dense ($n$(H$_2$) $\sim$ 10$^3$ cm$^{-3}$) to dense ($n$(H$_2$) $\sim$ 10$^{3.5 - 5}$ cm$^{-3}$)
%}%
%%%%%

3) The \Halpha flux towards these clumps is well correlated with the $^{12}$CO($J=3-2$)/$^{12}$CO($J=1-0$) ratio, $R_{3-2/1-0,clump}$, and clumps with \Halpha fluxes greater than 10$^{-11}$ ergs s$^{-1}$ cm$^{-2}$ have large $R_{3-2/1-0,clump}$ of $\sim$ 1.5. The $^{12}$CO($J=1-0$) data were taken with the SEST 15m and Mopra 22m telescopes.

4) The typical mass of the molecular clumps ranges 5$\times$10$^3$ -- 2$\times$10$^5$ M$_{\odot}$ with radii of 1 -- 12 pc. Of all of our objects, N 159 No.1 or -W shows the strongest concentration of mass of $\sim$ 7$\times$10$^4$ M$_{\odot}$ within a radius of $\sim$ 5 pc. We suggest that these are good candidates for the precursors of rich super clusters like R136 in 30 Dor.

%%%%%%%%%%
%\textcolor{blue}{%
5) We suggest that differences of clump properties represent an evolutionary seqeunce of GMCs in terms of density increase leading to star formations. Type I/II GMCs are at a young phase of star formation where density has not yet reached high enough values to cause active massive star formation, and Type III GMCs represent the later phase where the average density is higher, including both high and low density sub-types.
%}%
%%%%%

6) The high kinetic temperature correlated with \Halpha flux suggests that FUV heating is dominant in the molecular gas of the LMC. The low fraction of non-thermal radio continuum emission and calculations of PDR models support this suggestion. Furthermore, the high temperature in the molecular gas possibly leads to an increase of the Jeans mass of molecular clumps, which may favor the formation of rich super clusters.


%% Included in this acknowledgments section are examples of the
%% AASTeX hypertext markup commands. Use \url without the optional [HREF]
%% argument when you want to print the url directly in the text. Otherwise,
%% use either \url or \anchor, with the HREF as the first argument and the
%% text to be printed in the second.

\acknowledgments

A part of this study was financially supported by MEXT Grant-in-Aid for Scientific Research on Priority Area (No. 15071202 and No. 15071203) and by JSPS (No. 14102003, core-to-core program 17004, and No. 18684003).
T.M. is supported by JSPS Research Fellowships for Young Scientists.
M.R. is supported by the Chilean {\sl Center for Astrophysics} FONDAP No. 15010003. SK was supported in part by Korea Science and Engineering Foundation
(KOSEF) under a cooperative agreement with the Astrophysical Research
Center of the Structure and Evolution of the Cosmos (ARCSEC).

ASTE is a joint project between Japan and Chile. The telescope is operated by the ASTE team, including NAOJ, University of Tokyo, Nagoya University, Osaka Prefecture University, and Universidad de Chile. We are grateful to all the members of ASTE team.


%% in parentheses, after the keyword, but they will not be verified.

%% {\it Facilities:} \facility{ASTE}, \facility{MOPRA}, \facility{SEST}.



\appendix

\section{Complete LVG results of all clumps}

\begin{figure}
% \figurenum{E1}
\epsscale{.80}
\plotone{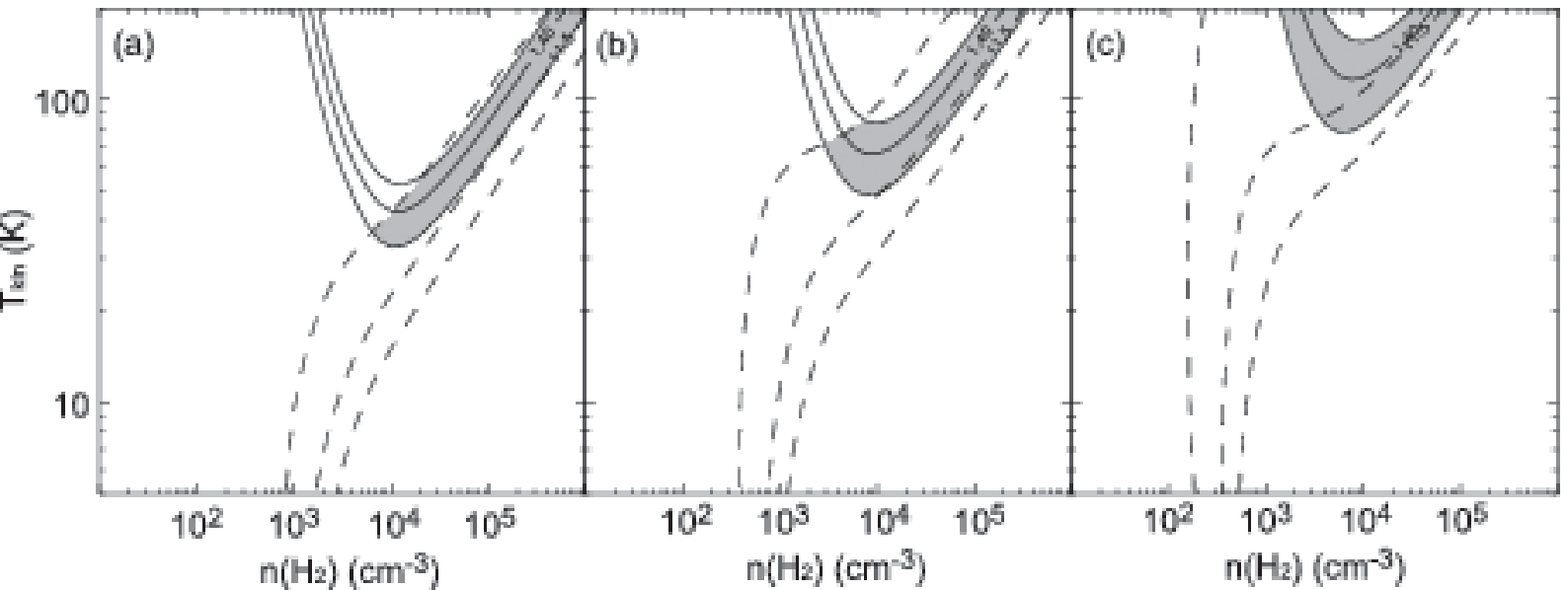}
\caption{Contour plots of LVG analysis of 30 Dor No.1. The vertical axis is kinetic temperature, \Tkin, and the horizontal axis is molecular hydrogen density, $n$(H$_2$). Solid lines are $R_{3-2/1-0, clump}$, and dashed lines are $R_{12/13}$. Hatched areas are the regions in which these two ratios oberlap within intensity calibration errors and uncertainty due to a possible variation of $^{12}$CO/$^{13}$CO abandance ratio. $X$(CO) = (a)1$\times$10$^{-6}$, (b)3$\times$10$^{-6}$, and (c)1$\times$10$^{-5}$. \label{fig25}}
\end{figure}

\begin{figure}
% \figurenum{E2}
\epsscale{.80}
\plotone{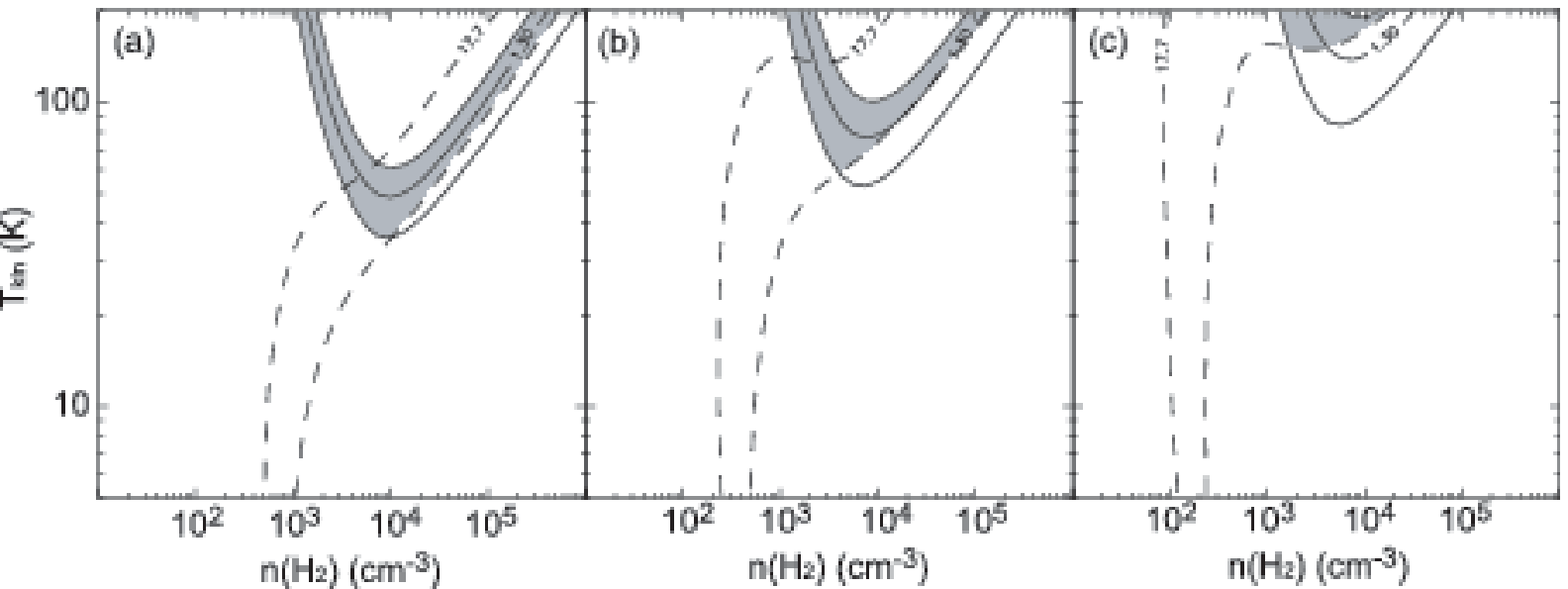}
\caption{Contour plots of LVG analysis of 30 Dor No.4. The vertical axis is kinetic temperature, \Tkin, and the horizontal axis is molecular hydrogen density, $n$(H$_2$). Solid lines are $R_{3-2/1-0, clump}$, and dashed lines are $R_{12/13}$. Hatched areas are the regions in which these two ratios oberlap within intensity calibration errors and uncertainty due to a possible variation of $^{12}$CO/$^{13}$CO abandance ratio. $X$(CO) = (a)1$\times$10$^{-6}$, (b)3$\times$10$^{-6}$, and (c)1$\times$10$^{-5}$. \label{fig26}}
\end{figure}

\begin{figure}
% \figurenum{E3}
\epsscale{.80}
\plotone{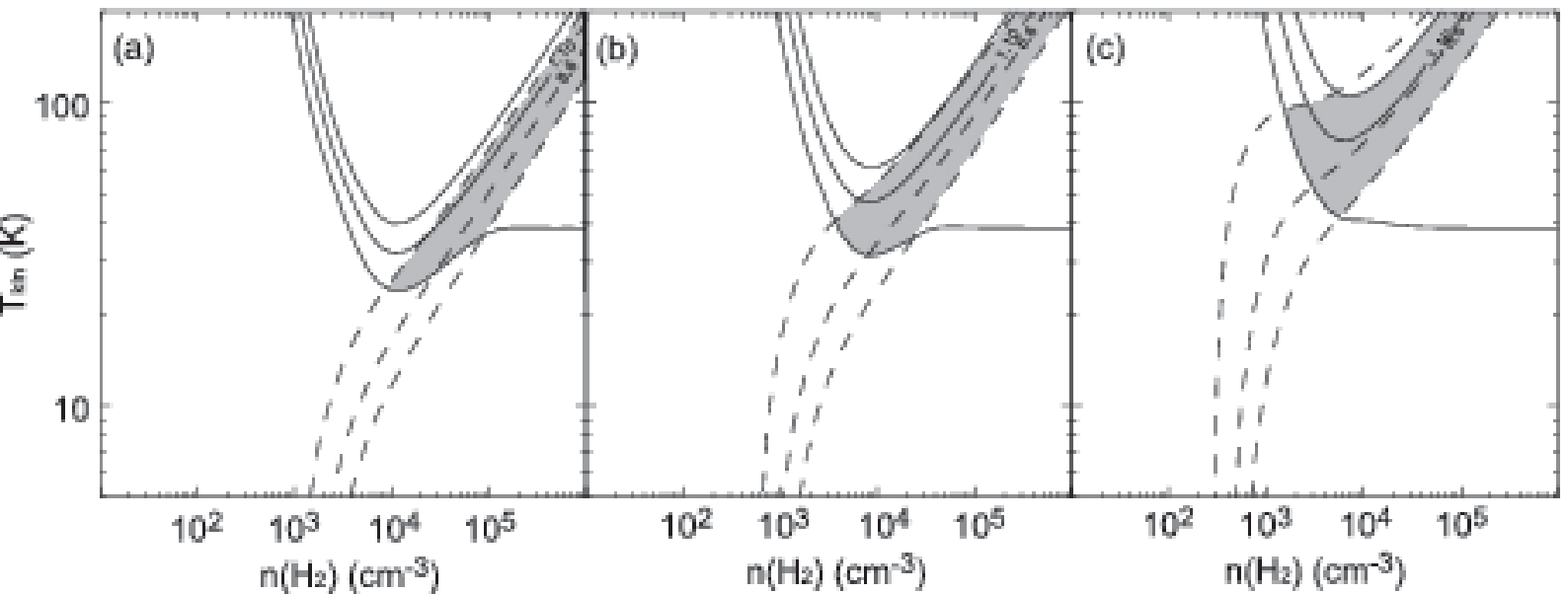}
\caption{Contour plots of LVG analysis of N 159 No.1. The vertical axis is kinetic temperature, \Tkin, and the horizontal axis is molecular hydrogen density, $n$(H$_2$). Solid lines are $R_{3-2/1-0, clump}$, and dashed lines are $R_{12/13}$. Hatched areas are the regions in which these two ratios oberlap within intensity calibration errors and uncertainty due to a possible variation of $^{12}$CO/$^{13}$CO abandance ratio. $X$(CO) = (a)1$\times$10$^{-6}$, (b)3$\times$10$^{-6}$, and (c)1$\times$10$^{-5}$. \label{fig27}}
\end{figure}

\begin{figure}
% \figurenum{E4}
\epsscale{.80}
\plotone{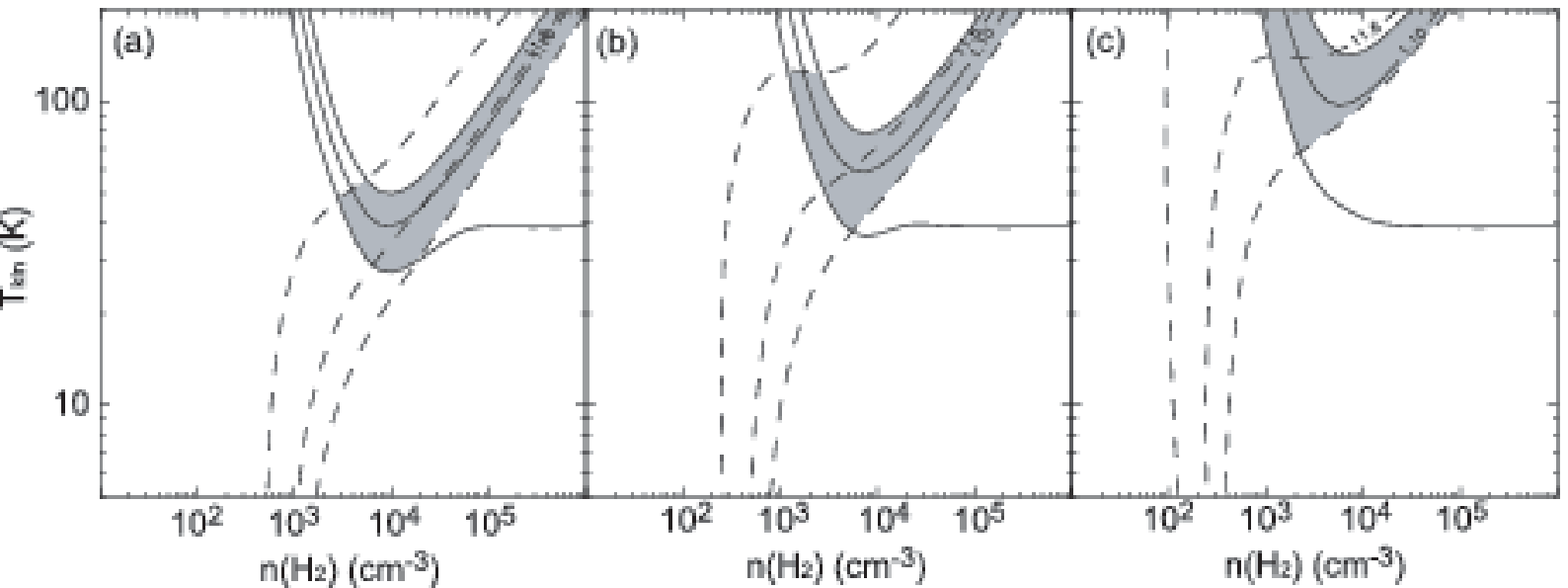}
\caption{Contour plots of LVG analysis of N 159 No.2. The vertical axis is kinetic temperature, \Tkin, and the horizontal axis is molecular hydrogen density, $n$(H$_2$). Solid lines are $R_{3-2/1-0, clump}$, and dashed lines are $R_{12/13}$. Hatched areas are the regions in which these two ratios oberlap within intensity calibration errors and uncertainty due to a possible variation of $^{12}$CO/$^{13}$CO abandance ratio. $X$(CO) = (a)1$\times$10$^{-6}$, (b)3$\times$10$^{-6}$, and (c)1$\times$10$^{-5}$. \label{fig28}}
\end{figure}

\begin{figure}
% \figurenum{E5}
\epsscale{.80}
\plotone{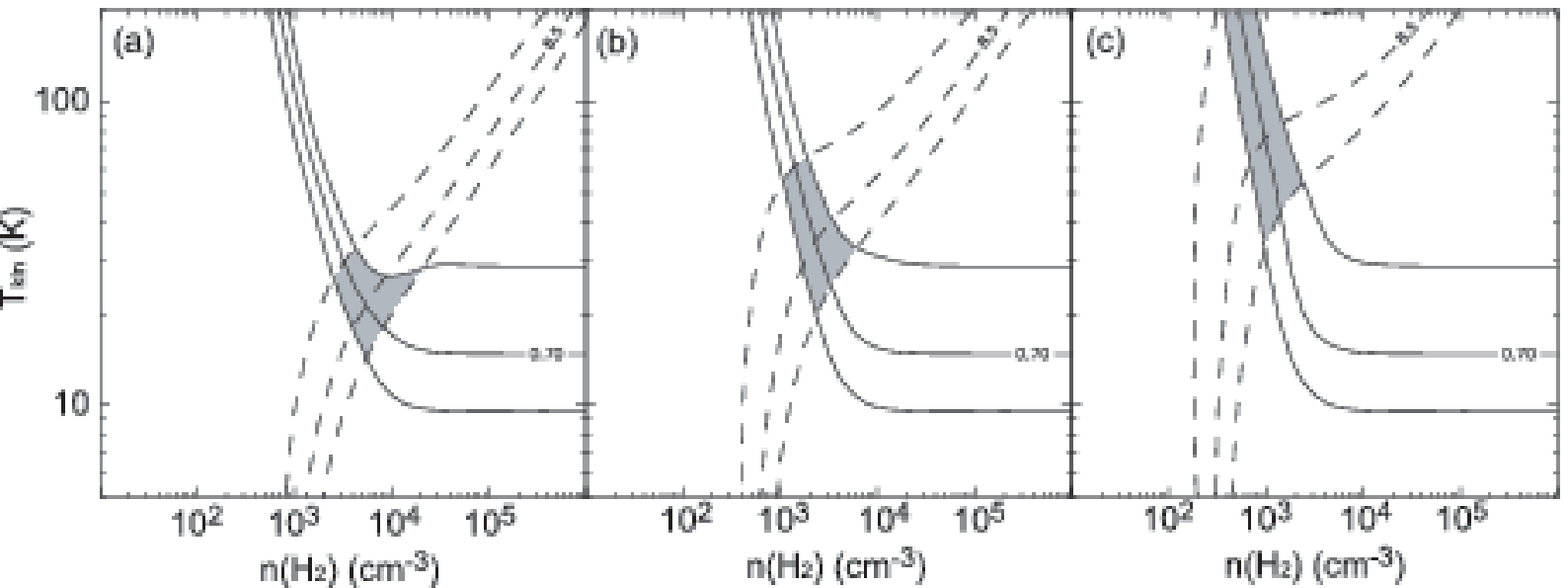}
\caption{Contour plots of LVG analysis of N 159 No.4. The vertical axis is kinetic temperature, \Tkin, and the horizontal axis is molecular hydrogen density, $n$(H$_2$). Solid lines are $R_{3-2/1-0, clump}$, and dashed lines are $R_{12/13}$. Hatched areas are the regions in which these two ratios oberlap within intensity calibration errors and uncertainty due to a possible variation of $^{12}$CO/$^{13}$CO abandance ratio. $X$(CO) = (a)1$\times$10$^{-6}$, (b)3$\times$10$^{-6}$, and (c)1$\times$10$^{-5}$. \label{fig29}}
\end{figure}

\begin{figure}
% \figurenum{E6}
\epsscale{.80}
\plotone{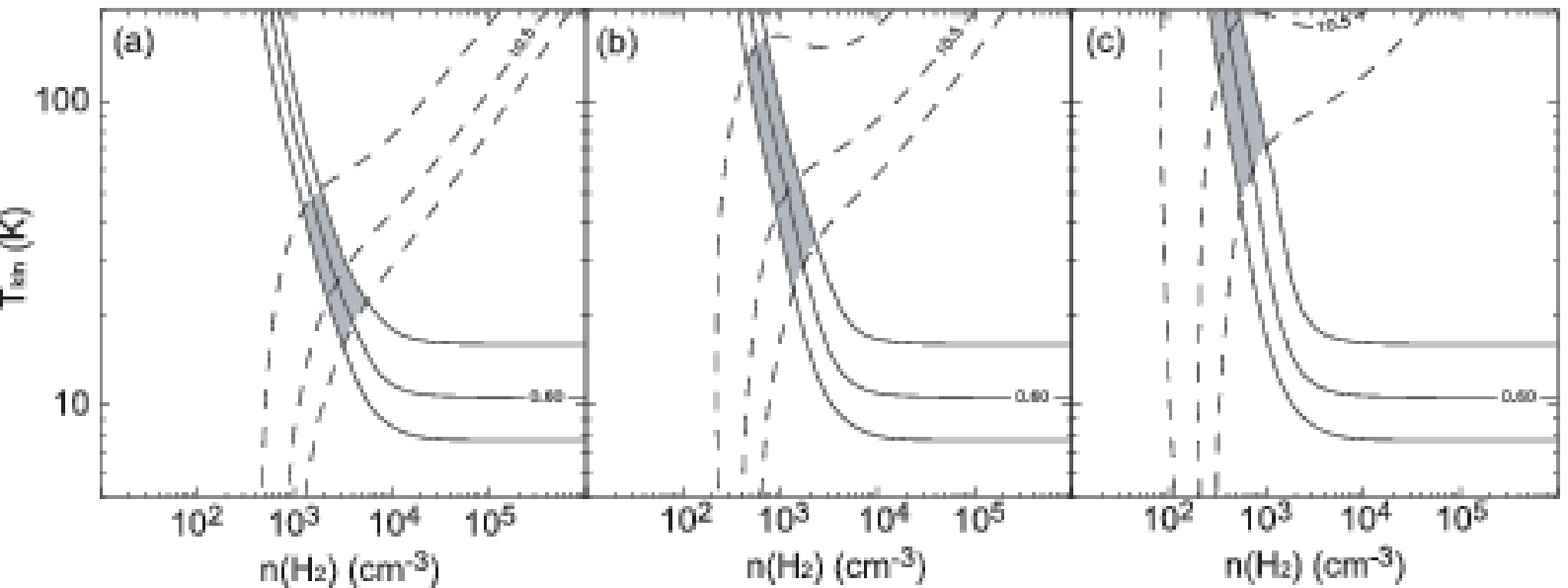}
\caption{Contour plots of LVG analysis of N 166 No.1. The vertical axis is kinetic temperature, \Tkin, and the horizontal axis is molecular hydrogen density, $n$(H$_2$). Solid lines are $R_{3-2/1-0, clump}$, and dashed lines are $R_{12/13}$. Hatched areas are the regions in which these two ratios oberlap within intensity calibration errors and uncertainty due to a possible variation of $^{12}$CO/$^{13}$CO abandance ratio. $X$(CO) = (a)1$\times$10$^{-6}$, (b)3$\times$10$^{-6}$, and (c)1$\times$10$^{-5}$. \label{fig30}}
\end{figure}

\begin{figure}
% \figurenum{E7}
\epsscale{.80}
\plotone{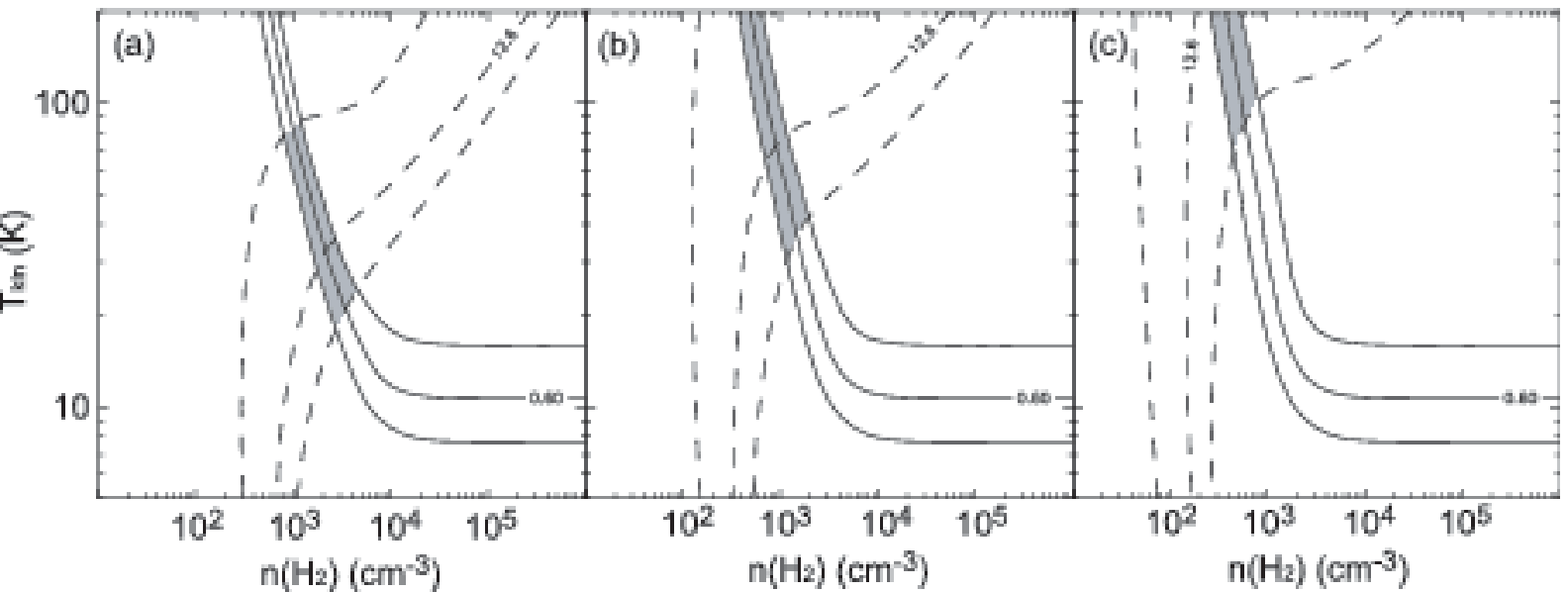}
\caption{Contour plots of LVG analysis of N 166 No.3. The vertical axis is kinetic temperature, \Tkin, and the horizontal axis is molecular hydrogen density, $n$(H$_2$). Solid lines are $R_{3-2/1-0, clump}$, and dashed lines are $R_{12/13}$. Hatched areas are the regions in which these two ratios oberlap within intensity calibration errors and uncertainty due to a possible variation of $^{12}$CO/$^{13}$CO abandance ratio. $X$(CO) = (a)1$\times$10$^{-6}$, (b)3$\times$10$^{-6}$, and (c)1$\times$10$^{-5}$. \label{fig31}}
\end{figure}

\begin{figure}
% \figurenum{E8}
\epsscale{.80}
\plotone{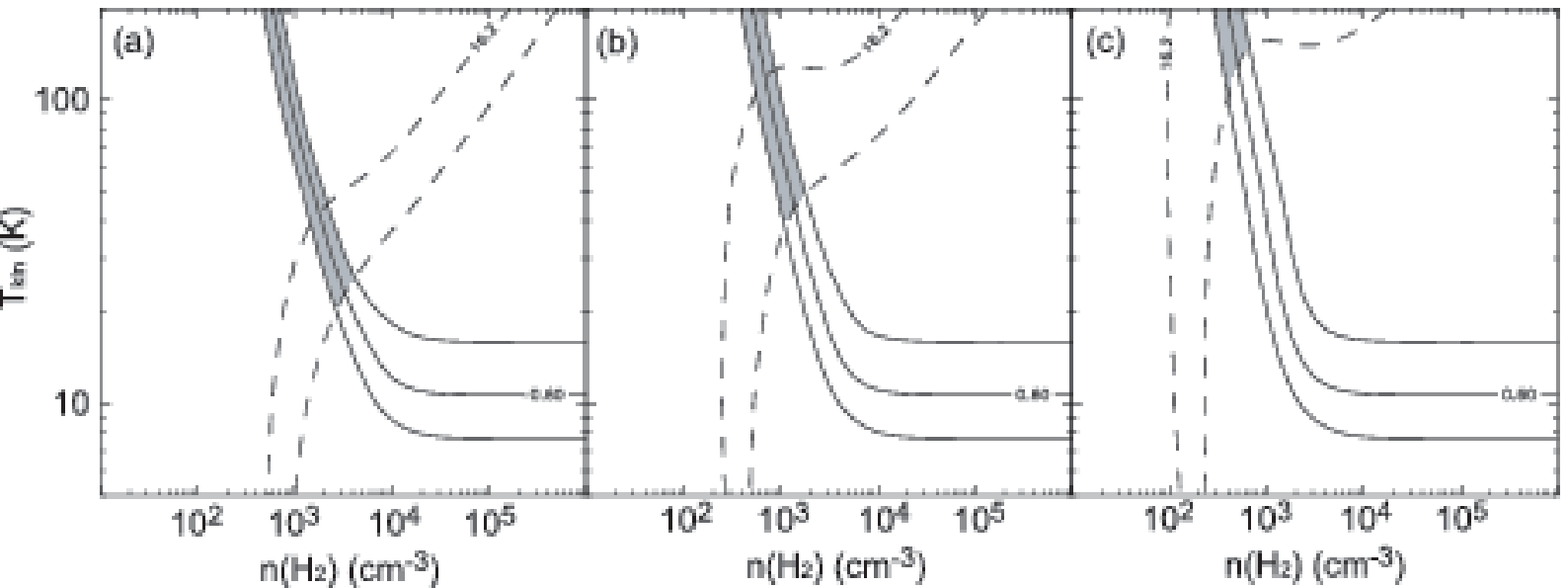}
\caption{Contour plots of LVG analysis of N 166 No.4. The vertical axis is kinetic temperature, \Tkin, and the horizontal axis is molecular hydrogen density, $n$(H$_2$). Solid lines are $R_{3-2/1-0, clump}$, and dashed lines are $R_{12/13}$. Hatched areas are the regions in which these two ratios oberlap within intensity calibration errors and uncertainty due to a possible variation of $^{12}$CO/$^{13}$CO abandance ratio. $X$(CO) = (a)1$\times$10$^{-6}$, (b)3$\times$10$^{-6}$, and (c)1$\times$10$^{-5}$. \label{fig32}}
\end{figure}

\begin{figure}
% \figurenum{E9}
\epsscale{.80}
\plotone{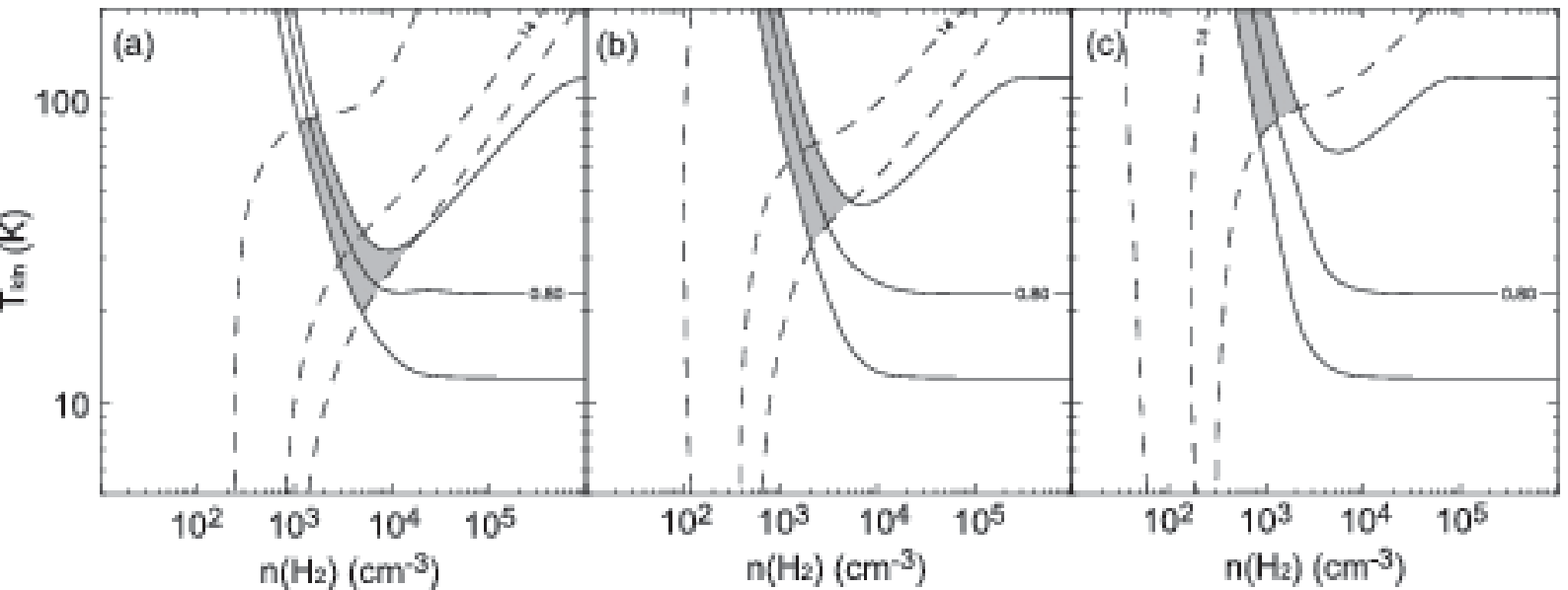}
\caption{Contour plots of LVG analysis of N 206 No.1. The vertical axis is kinetic temperature, \Tkin, and the horizontal axis is molecular hydrogen density, $n$(H$_2$). Solid lines are $R_{3-2/1-0, clump}$, and dashed lines are $R_{12/13}$. Hatched areas are the regions in which these two ratios oberlap within intensity calibration errors and uncertainty due to a possible variation of $^{12}$CO/$^{13}$CO abandance ratio. $X$(CO) = (a)1$\times$10$^{-6}$, (b)3$\times$10$^{-6}$, and (c)1$\times$10$^{-5}$. \label{fig33}}
\end{figure}

\begin{figure}
% \figurenum{E10}
\epsscale{.80}
\plotone{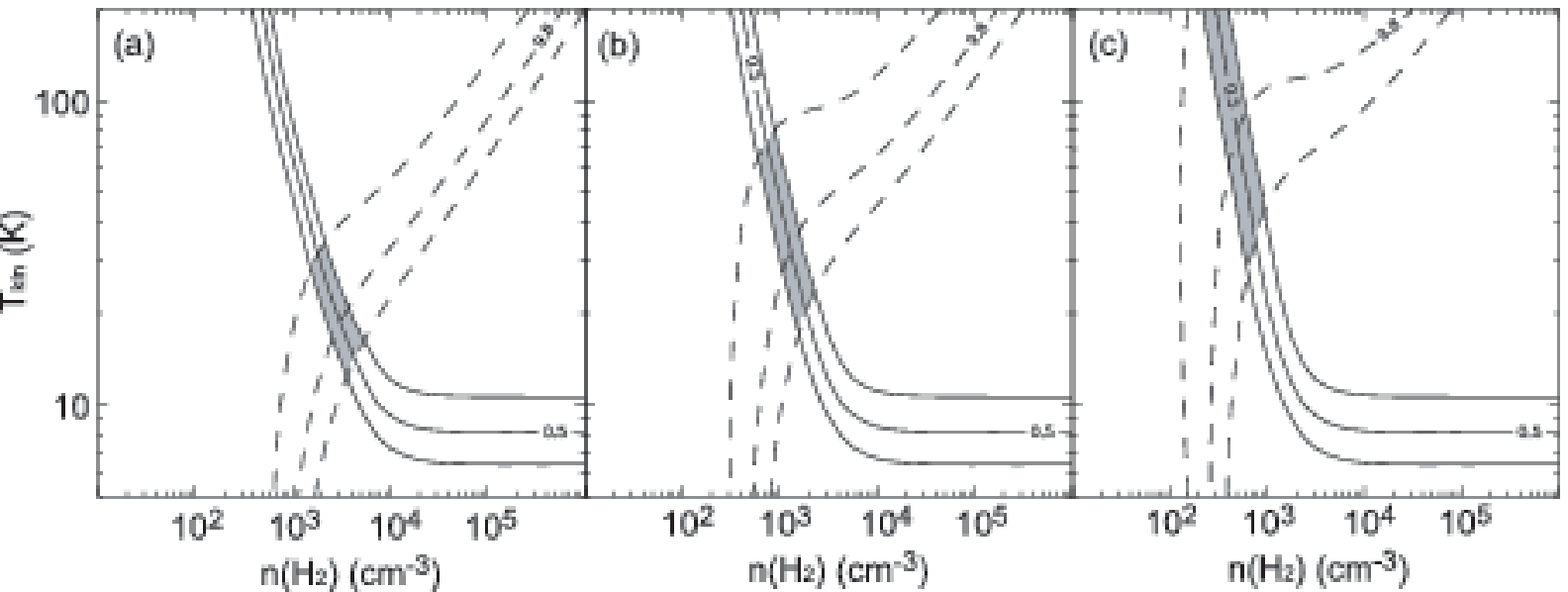}
\caption{Contour plots of LVG analysis of N 206 No.2. The vertical axis is kinetic temperature, \Tkin, and the horizontal axis is molecular hydrogen density, $n$(H$_2$). Solid lines are $R_{3-2/1-0, clump}$, and dashed lines are $R_{12/13}$. Hatched areas are the regions in which these two ratios oberlap within intensity calibration errors and uncertainty due to a possible variation of $^{12}$CO/$^{13}$CO abandance ratio. $X$(CO) = (a)1$\times$10$^{-6}$, (b)3$\times$10$^{-6}$, and (c)1$\times$10$^{-5}$. \label{fig34}}
\end{figure}

\begin{figure}
% \figurenum{E11}
\epsscale{.80}
\plotone{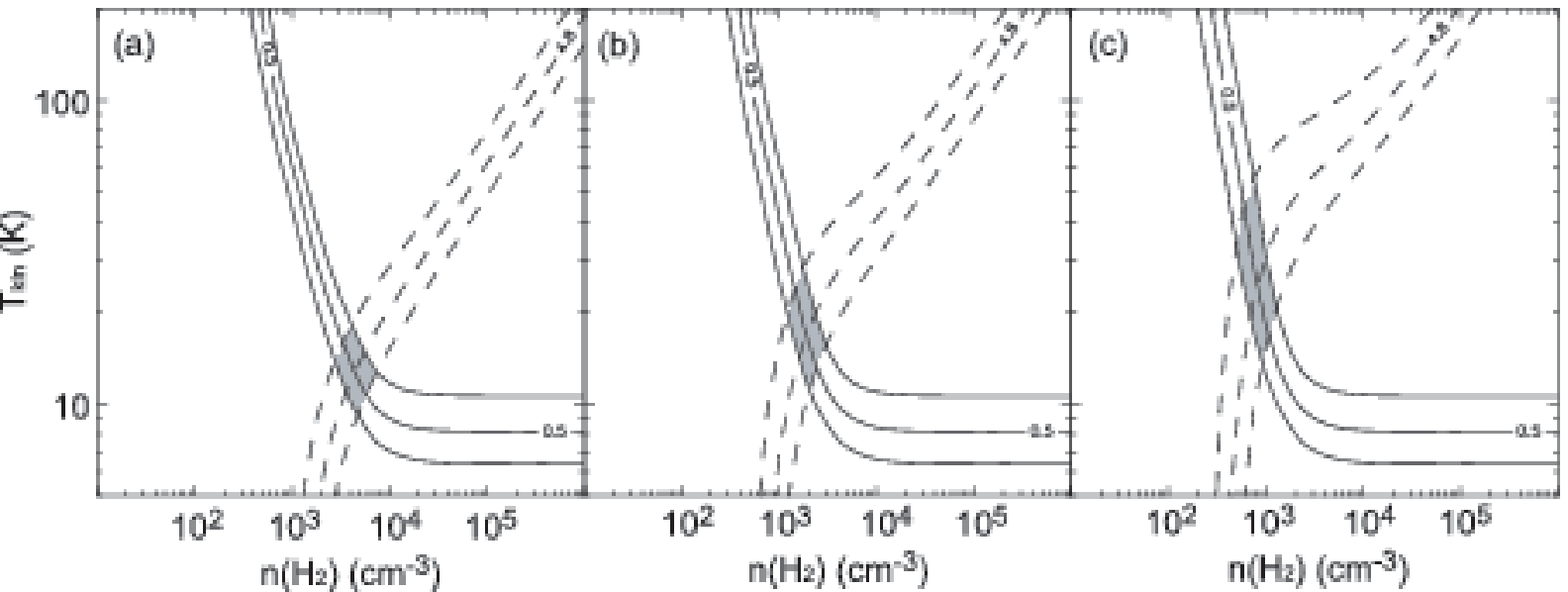}
\caption{Contour plots of LVG analysis of N 206D No.1. The vertical axis is kinetic temperature, \Tkin, and the horizontal axis is molecular hydrogen density, $n$(H$_2$). Solid lines are $R_{3-2/1-0, clump}$, and dashed lines are $R_{12/13}$. Hatched areas are the regions in which these two ratios oberlap within intensity calibration errors and uncertainty due to a possible variation of $^{12}$CO/$^{13}$CO abandance ratio. $X$(CO) = (a)1$\times$10$^{-6}$, (b)3$\times$10$^{-6}$, and (c)1$\times$10$^{-5}$. \label{fig35}}
\end{figure}

\begin{figure}
% \figurenum{E12}
\epsscale{.80}
\plotone{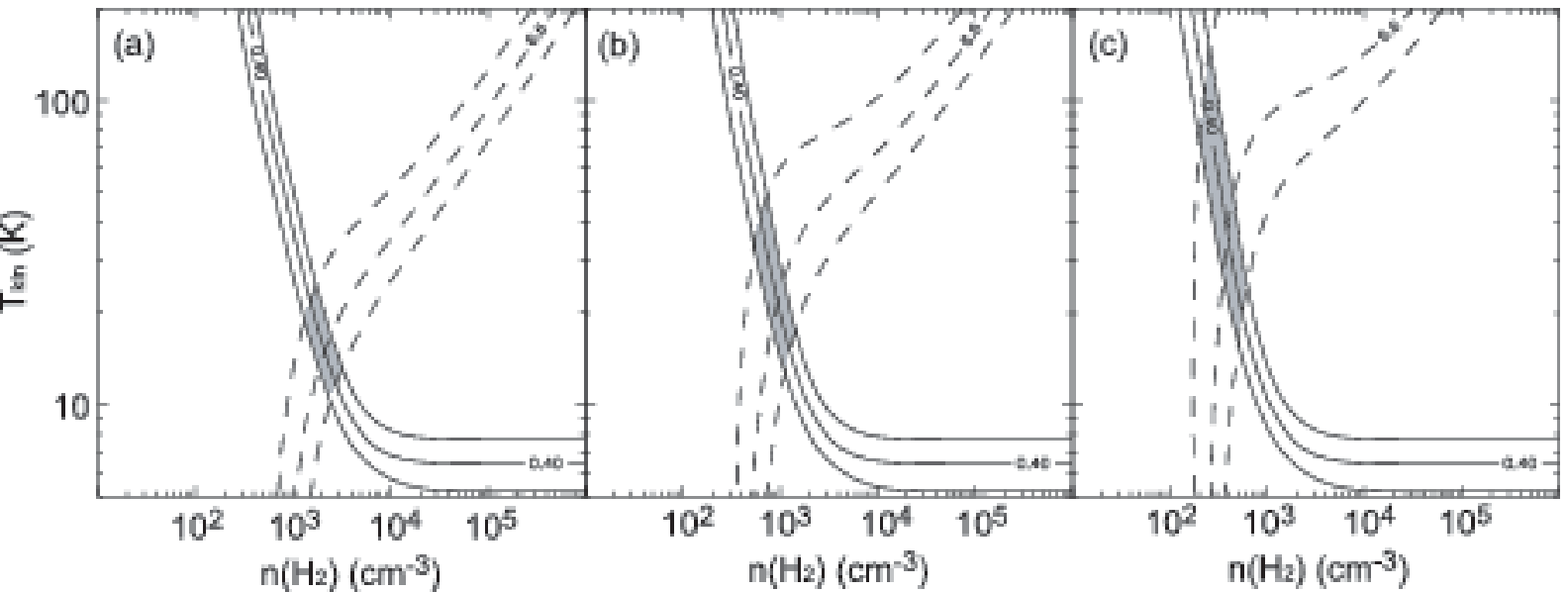}
\caption{Contour plots of LVG analysis of GMC 225 No.1. The vertical axis is kinetic temperature, \Tkin, and the horizontal axis is molecular hydrogen density, $n$(H$_2$). Solid lines are $R_{3-2/1-0, clump}$, and dashed lines are $R_{12/13}$. Hatched areas are the regions in which these two ratios oberlap within intensity calibration errors and uncertainty due to a possible variation of $^{12}$CO/$^{13}$CO abandance ratio. $X$(CO) = (a)1$\times$10$^{-6}$, (b)3$\times$10$^{-6}$, and (c)1$\times$10$^{-5}$. \label{fig36}}
\end{figure}

\begin{figure}
% \figurenum{E13}
\epsscale{.80}
\plotone{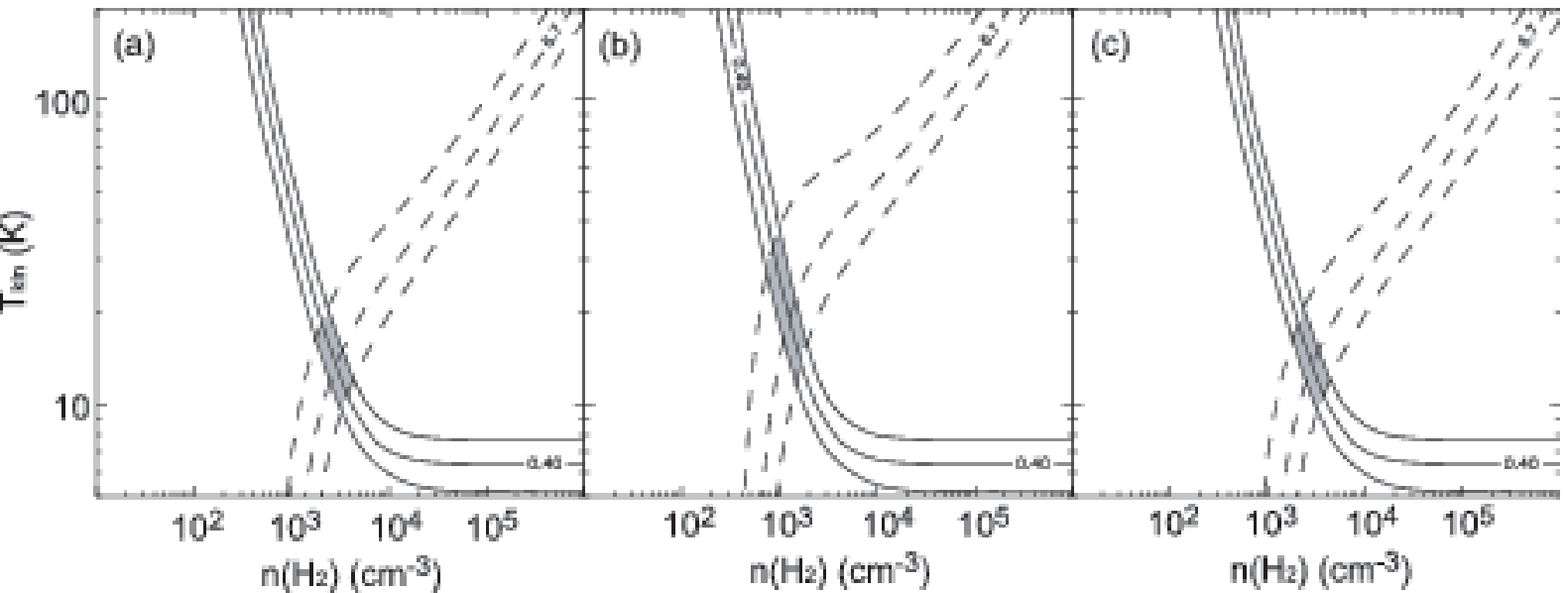}
\caption{Contour plots of LVG analysis of GMC 225 No.3. The vertical axis is kinetic temperature, \Tkin, and the horizontal axis is molecular hydrogen density, $n$(H$_2$). Solid lines are $R_{3-2/1-0, clump}$, and dashed lines are $R_{12/13}$. Hatched areas are the regions in which these two ratios oberlap within intensity calibration errors and uncertainty due to a possible variation of $^{12}$CO/$^{13}$CO abandance ratio. $X$(CO) = (a)1$\times$10$^{-6}$, (b)3$\times$10$^{-6}$, and (c)1$\times$10$^{-5}$. \label{fig37}}
\end{figure}

Figures \ref{fig25} -- \ref{fig37} show the complete LVG results described in section 4.2. We present all cases: $X$(CO) of (a)1$\times$10$^{-6}$, (b)3$\times$10$^{-6}$, and (c)1$\times$10$^{-5}$ for 13 clumps.

\section{\Halpha flux}

\begin{figure}
%\figurenum{A1}
\epsscale{.80}
\plotone{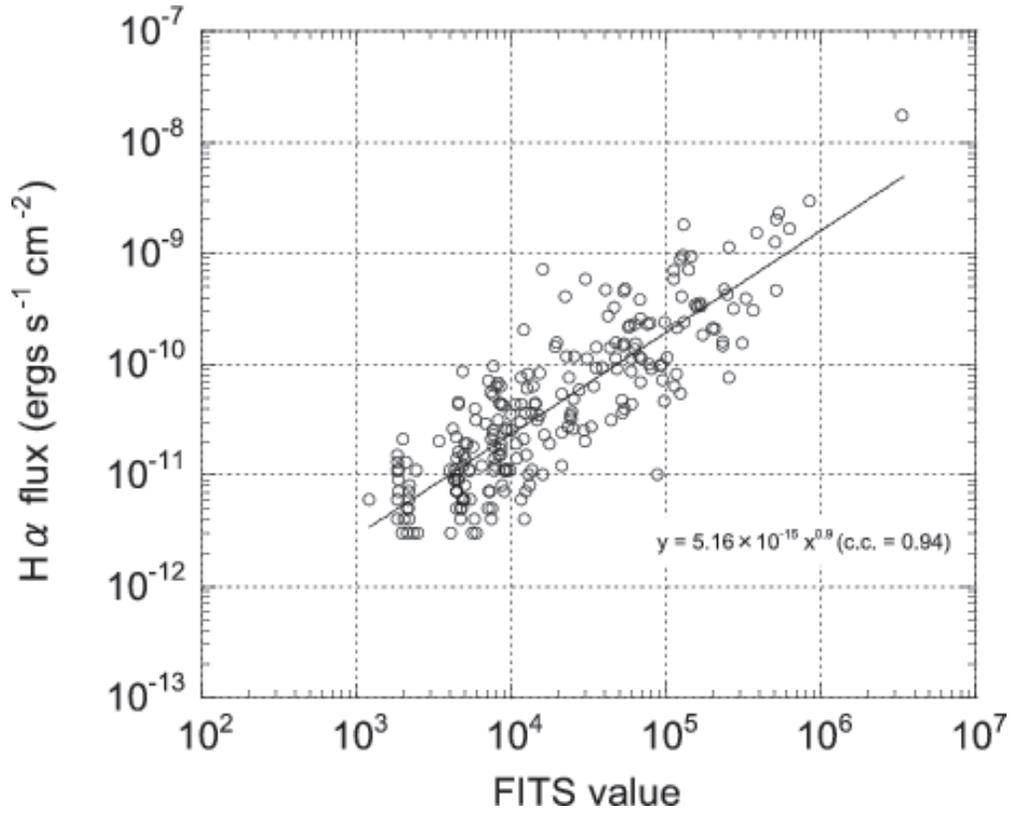}
\caption{Plot of relation between integrated \Halpha flux and FITS value. The solid line represents the least-squares fit. \label{fig38}}
\end{figure}

We shall here describe the method of scaling data values in FITS cube (\Halpha image in \citet{Kim1999}; hereafter, "Kim's FITS").
%%%%%%%%%%
%\textcolor{green}{%
The data values in Kim's FITS is not flux scale, and the calibration is needed for quantitative comparison with CO clouds and $R_{3-2/1-0}$.
%}%
%%%%%%
The procedure is as follows. 
1) Sum up the data values of Kim's FITS inside apertures which are listed in \citet{Kennicutt1986} for each listed HII region.
2) Plot cataloged values of \Halpha flux in \citet{Kennicutt1986} as a function of summed values derived in 1). They are well fitted by a power function of $y = 5.16\times10^{-15} x^{0.9}$ (c.c. = 0.94) (Figure \ref{fig38}).
3) Convert data values of Kim's FITS to \Halpha flux scale (ergs s$^{-1}$ cm$^{-2}$) using the function derived above.

\section{LVG results in the other planes}

\begin{figure}
% \figurenum{G01}
\epsscale{.80}
\plotone{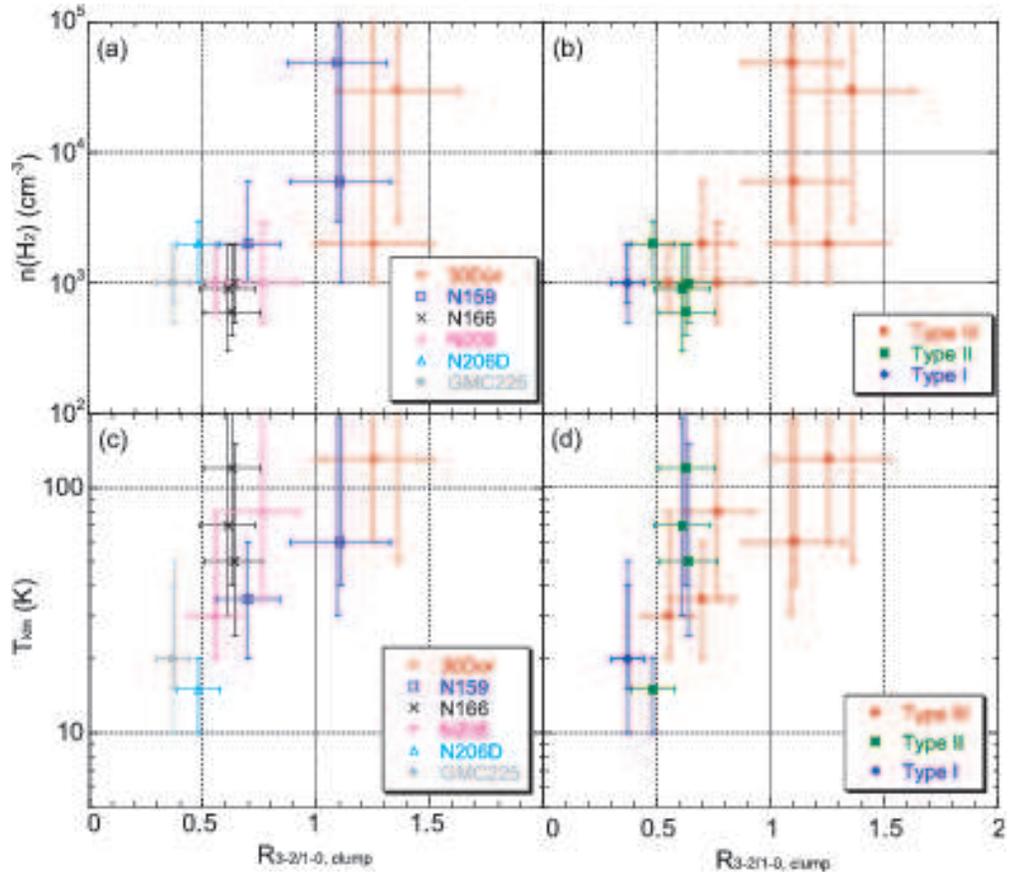}
\caption{Plots of physical properties as a function of $R_{3-2/1-0, clump}$. (a) $n$(H$_2$) by region. (b) $n$(H$_2$) by GMC type. (c) \Tkin by region. (d) \Tkin by GMC type.\label{fig39}}
\end{figure}

\subsection{Physical properties -- $R_{3-2/1-0,clump}$ (Figure \ref{fig39})}

The density plots (Figure \ref{fig39}(a) and (b)) show that higher $R_{3-2/1-0,clump}$ (\textgreater 1.0) correspond to higher densities of 10$^3$ to 10$^5$ cm$^{-3}$, while lower $R_{3-2/1-0,clump}$ (\textless 1.0) correspond to lower densities of around 10$^3$ cm$^{-3}$. The temperature plots (Figure \ref{fig39}(c) and (d)) show that higher $R_{3-2/1-0,clump}$ (\textgreater 0.5) correspond to higher temperatures of \textgreater 30K, while lower $R_{3-2/1-0,clump}$ (\textless 0.5) correspond to lower temperatures of \textless 30K. 

Then, roughly speaking, we can say that clumps with $R_{3-2/1-0,clump}$ lower than 0.5 have lower densities of around 10$^3$ cm$^{-3}$ and lower temperatures of \textless 30K, clumps with $R_{3-2/1-0,clump}$ of 0.5 to 1.0 have lower density around 10$^3$ cm$^{-3}$ and higher temperatures of \textgreater 30K, and clumps with $R_{3-2/1-0,clump}$ higher than 1.0 have higher densities of 10$^3$ to 10$^5$ cm$^{-3}$ and higher temperatures of \textgreater 30K, although ratios, densities, and temperatures are with large error bars.

Of course, there are great benefits to using $R_{12/13}$ in LVG analyses. We could not obtain the above results with only $R_{3-2/1-0}$, as mentioned in section 4.

\begin{figure}
% \figurenum{G02}
\epsscale{.80}
\plotone{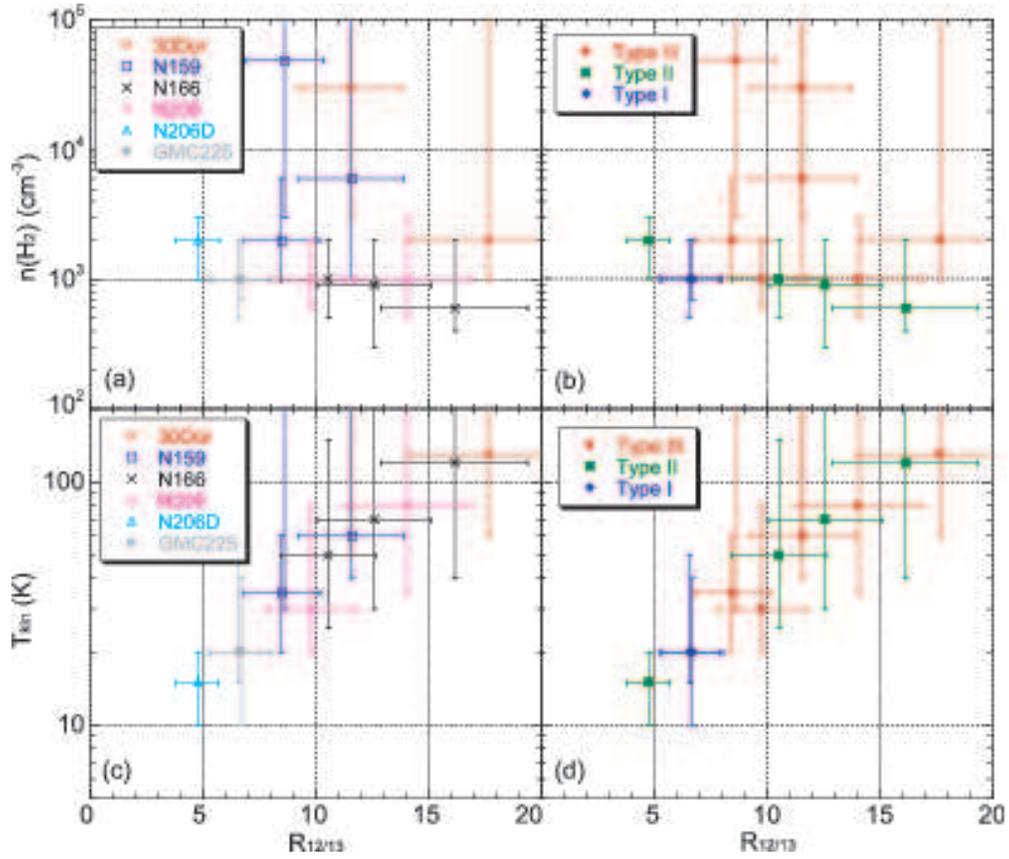}
\caption{Plots of physical properties as a function of $R_{12/13}$. (a) $n$(H$_2$) by region. (b) $n$(H$_2$) by GMC type. (c) \Tkin by region. (d) \Tkin by GMC type.\label{fig40}}
\end{figure}

\subsection{Physical properties -- $R_{12/13}$ (Figure \ref{fig40})}

The density (Figure \ref{fig40}(a) and (b)) does not show a significant correlation with $R_{12/13}$. The temperature (Figure \ref{fig40}(c) and (d)) shows a good correlation with $R_{12/13}$, that is, higher ratios indicate higher temperatures.
Usually, $R_{12/13}$ correspond to density, but in this case, due to the LVG analysis using both $R_{3-2/1-0}$ and $R_{12/13}$, larger $R_{12/13}$ indicate lower density and lower density tends to higher temperature.

\begin{figure}
% \figurenum{G03}
\epsscale{.80}
\plotone{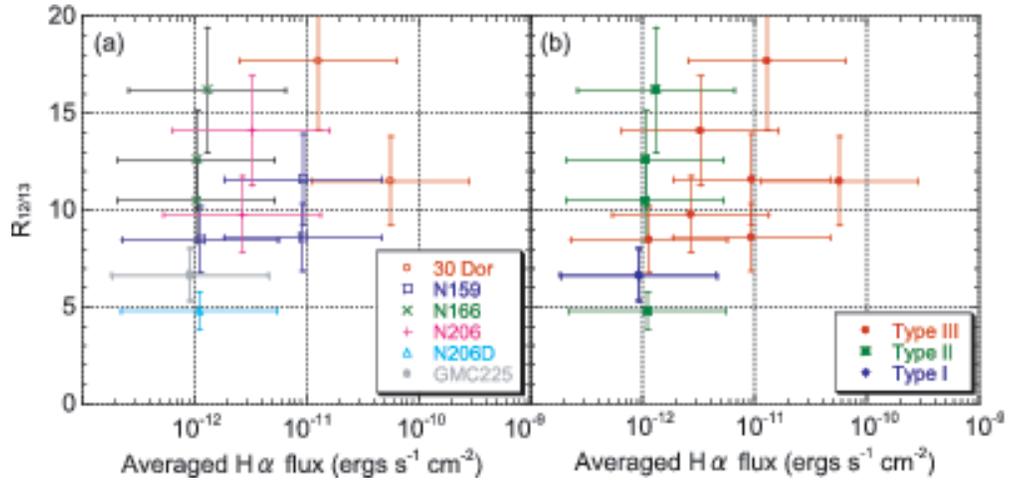}
\caption{Plots of $R_{12/13}$ as a function of clump averaged H$\alpha$ flux (a) by region and (b) by GMC type. \label{fig41}}
\end{figure}

\subsection{$R_{12/13}$ -- \Halpha flux (Figure \ref{fig41})}

There is not significant relation between $R_{12/13}$ and \Halpha flux, contrary to $R_{3-2/1-0}$ ratio.




%% ApJ = \apj
%% ApJL = \apjl
%% ApJS = \apjs
%% A&A = \aap
%% AJ = \aj
%% PASA = \pasa
%% PASJ = \pasj
%% PASP = \pasp

% \input{reference.tex}

\clearpage

%% the file, uncomment the \plotone, \plottwo, and
%% If you need a layout that cannot be achieved with \plotone or
%% \plottwo, you can invoke the graphicx package directly with the
%% \includegraphics command or use \plotfiddle. For more information,
%% please see the tutorial on "Using Electronic Art with AASTeX" in the
%% documentation section at the AASTeX Web site,
%% http://www.journals.uchicago.edu/AAS/AASTeX.
%%
%% The examples below also include sample markup for submission of
%% supplemental electronic materials. As always, be sure to check
%% the instructions to authors for the journal you are submitting to
%% for specific submissions guidelines as they vary from

\clearpage

\clearpage

\clearpage

\clearpage

\clearpage

\clearpage

\clearpage

\clearpage

\clearpage

\clearpage

\clearpage

\clearpage

\clearpage

\clearpage

\clearpage

\clearpage

\clearpage

\clearpage

\clearpage

\clearpage

\clearpage

\clearpage

\clearpage

\clearpage

%% Figures for Appendix

%% LVG results fig25 -- 37

\clearpage

\clearpage

\clearpage

\clearpage

\clearpage

\clearpage

\clearpage

\clearpage

\clearpage

\clearpage

\clearpage

\clearpage

\clearpage

\clearpage

\clearpage

\clearpage

\clearpage

%% This example uses \plotone to include an EPS file scaled to
%% 80% of its natural size with \epsscale. Its caption
%% has been written to indicate that additional figure parts will be
%% available in the electronic journal.

\begin{deluxetable}{ccccccccccc}
%% \tabletypesize{\small}		%% 11pt
%% \tabletypesize{\footnotesize}	%% 10pt
\tabletypesize{\scriptsize}	%%  8pt
\rotate
\tablewidth{0pt}
%% \tablenum{}
%% \tablecolumns{}
\tablecaption{List of observed GMCs and transitions\label{tbl01}}

\tablehead{
\multicolumn{3}{c}{GMC} & \colhead{} & \multicolumn{3}{c}{Position \tablenotemark{a}} & 
\colhead{} & \multicolumn{3}{c}{Telescope}\\
\cline{1-3} \cline{5-7} \cline{9-11}
\colhead{Num. \tablenotemark{1}} & \colhead{Name \tablenotemark{1}} & \colhead{Type \tablenotemark{2}} & 
\colhead{Region Name} & \colhead{$\alpha$(1950)} & \colhead{$\delta$(1950)} & \colhead{Ref.\tablenotemark{b}} & 
\colhead{} & \colhead{$^{12}$CO(J=3-2)} & \colhead{$^{12}$CO(J=1-0)} & \colhead{$^{13}$CO(J=1-0)}\\
\colhead{} & \colhead{} & \colhead{} & \colhead{} & \colhead{(h m s)} & \colhead{(\arcdeg\phn\arcmin\phn\arcsec)} & 
\colhead{} & \colhead{} & \colhead{} & \colhead{} & \colhead{}
}

\startdata
186 & LMC N J0538-6904 & III & 30 Dor & 5 39 \phn0.2 & -69 \phn8 \phn0.0 & 3 & & ASTE & MOPRA & SEST \tablenotemark{3} \\
197 & LMC N J0540-7008 & III & N 159 & 5 40 18.2 & -69 47 \phn0.0 & 3 & & ASTE & MOPRA & SEST \tablenotemark{3} \\
 & & & N 171 & 5 40 24.1 & -70 \phn8 \phn0.0 & 4 & & ASTE & MOPRA & \nodata \\
216 & LMC N J0544-6923 & II & N 166 & 5 44 52.5 & -69 26 39.1 & 5 & & ASTE & SEST \tablenotemark{5} & SEST \tablenotemark{5} \\
153 & LMC N J0530-7106 & III & N 206 & 5 31 33.9 & -71 10 \phn0.0 & 6 & & ASTE & SEST & SEST \\
156 & LMC N J0532-7114 & II & N 206D & 5 32 52.2 & -71 16 \phn0.0 & 6 & & ASTE & SEST & SEST \\
225 & LMC N J0547-7041 & I & GMC 225 & 5 48 35.7 & -70 40 \phn0.0 & 6 & & ASTE & SEST & SEST \\
\enddata
\tablecomments{
Column (1): Running number of GMC used in Table 1 in Fukui et al.(2006b). 
Column (2): Name of GMC. 
Column (3): Type of GMC. 
Column (4): Region name used in this paper. 
Column (5)-(7): Coordinates used as reference position in each region for these $^{12}$CO($J=3-2$) observations. 
Column (8)-(10): Telescope used for each observation.
}	%% Notes
%% \tablenotemark{}
\tablenotetext{a}{
These coordinates are used as reference positions in each regions for this $^{12}$CO($J=3-2$) observations
}

\tablenotetext{b}{References for the positions.}

% \tablenotetext{c}{These names are based on Mizuno et al. 2001}

\tablerefs{
(1) Fukui et al. 2006b; 
(2) Kawamura et al. 2006; 
(3) Johansson et al. 1998; 
(4) Kutner et al. 1997; 
(5) Garay et al. 2002; 
(6) Mizuno et al. 2001
}

\end{deluxetable}

\clearpage

\begin{deluxetable}{lccccccccc}
%% \tabletypesize{\small}		%% 11pt
%% \tabletypesize{\footnotesize}	%% 10pt
\tabletypesize{\scriptsize}	%%  8pt
%% \rotate
\tablewidth{0pt}
%% \tablenum{}
%% \tablecolumns{}
\tablecaption{Observed properties of $^{12}$CO($J=3-2$) clumps \label{tbl02}}
\tablehead{
\colhead{} & \colhead{} & 
\multicolumn{2}{c}{Position} & \colhead{} & 
\multicolumn{4}{c}{Peak properties} & 
\colhead{} \\

\cline{3-4} \cline{6-9}

\colhead{Region} & \colhead{No.} & 
\colhead{$\alpha$(1950)} & \colhead{$\delta$(1950)} & \colhead{} & 
\colhead{$T_{\mathrm{mb}}$} & \colhead{$V_{\mathrm{LSR}}$} & \colhead{$\Delta V$} & \colhead{$I.I.$} & 
\colhead{Another Ident.} \\

\colhead{} & \colhead{} & 
\colhead{(h m s)} & \colhead{(\arcdeg\phn\arcmin\phn\arcsec)} & \colhead{} & 
\colhead{(K)} & \colhead{(km s$^{-1}$)} & \colhead{(km s$^{-1}$)} & \colhead{(K km s$^{-1}$)} & 
\colhead{} 
}

\startdata
30 Dor & \phn1 & 5 39 \phn8.6 & -69 \phn6 15 &  & \phn5.2\phn & 250.4 & 8.4 & \phn49.5 & 30Dor-10 \tablenotemark{1} \\
 & \phn2 & 5 38 54.6 & -69 \phn8 \phn0 &  & \phn4.3\phn & 246.9 & 6.9 & \phn31.7 & 30Dor-12 \tablenotemark{1} \\
 & \phn3 & 5 38 49.0 & -69 \phn4 30 &  & \phn3.1\phn & 251.9 & 5.4 & \phn22.5 & \\
 & \phn4 & 5 38 49.0 & -69 \phn3 30 &  & \phn2.6\phn & 248.2 & 5.0 & \phn21.1 & 30Dor-06 \tablenotemark{1} \\
 & \phn5 & 5 38 54.6 & -69 \phn9 \phn0 &  & \phn1.2\phn & 247.4 & 4.5 & \phm{00}5.3 & \\ 
\hline
N 159 & \phn1 & 5 40 \phn3.7 & -69 47 \phn0 &  & 12.3\phn & 237.3 & 8.1 & 106.6 & N159-W \tablenotemark{1} \\
 & \phn2 & 5 40 35.5 & -69 46 \phn0 &  & \phn7.6\phn & 233.2 & 7.0 & \phn55.5 & N159-E \tablenotemark{1} \\
 & \phn3 & 5 40 47.1 & -69 46 15 &  & \phn5.4\phn & 234.6 & 6.7 & \phn46.0 \\
 & \phn4 & 5 40 32.7 & -69 52 \phn0 &  & \phn3.5\phn & 234.4 & 9.7 & \phn35.4 & N159-S \tablenotemark{1} \\
 & \phn5 & 5 39 49.3 & -69 47 \phn0 &  & \phn3.0\phn & 233.5 & 9.5 & \phn29.7 & \\
 & \phn6 & 5 40 47.2 & -69 51 30 &  & \phn3.6\phn & 235.6 & 5.3 & \phn19.8 & \\
 & \phn7 & 5 39 55.1 & -69 45 \phn0 &  & \phn4.5\phn & 238.3 & 3.4 & \phn16.6 & \\
 & \phn8 & 5 40 29.8 & -69 50 \phn0 &  & \phn0.91 & 234.6 & 6.9 & \phn10.5 & \\
 & \phn9 & 5 40 41.4 & -69 51 \phn0 &  & \phn0.87 & 235.7 & 8.0 & \phm{00}9.3 & \\
 & 10 & 5 39 37.7 & -69 45 30 &  & \phn0.89 & 238.6 & 7.9 & \phm{00}8.3 & \\ 
\hline
N 171 & \phn1 & 5 40 \phn0.5 & -70 12 30 &  & \phn2.5\phn & 224.3 & 5.2 & \phn15.4 & 30DOR-CENTER-04 \tablenotemark{2} \\
 & \phn2 & 5 40 \phn0.5 & -70 10 30 &  & \phn1.3\phn & 227.3 & 8.3 & \phn12.7 & \\
 & \phn3 & 5 40 24.1 & -70 \phn9 45 &  & \phn1.3\phn & 230.9 & 9.2 & \phn18.2 & \\
 & \phn4 & 5 40 24.1 & -70 \phn8 30 &  & \phn2.3\phn & 231.0 & 6.7 & \phn16.1 & 30DOR-CENTER-03 \tablenotemark{2} \\
 & \phn5 & 5 40 \phn6.4 & -70 \phn9 \phn0 &  & \phn1.7\phn & 240.6 & 4.9 & \phn11.5 & \\
 & \phn6 & 5 40 18.2 & -70 \phn9 30 &  & \phn1.9\phn & 240.7 & 5.1 & \phn10.4 & \\ 
\hline
N 166 & \phn1 & 5 44 58.2 & -69 26 39 &  & \phn4.5\phn & 227.6 & 4.5 & \phn22.1 & Cloud-C \tablenotemark{3} \\
 & \phn2 & 5 44 46.8 & -69 27 24 &  & \phn2.9\phn & 229.4 & 6.2 & \phn20.3 & \\
 & \phn3 & 5 45 \phn3.9 & -69 29 \phn9 &  & \phn2.2\phn & 227.9 & 5.5 & \phn14.6 & Cloud-D \tablenotemark{3} \\
 & \phn4 & 5 44 32.6 & -69 23 39 &  & \phn1.0\phn & 236.1 & 9.2 & \phn10.1 & Cloud-B \tablenotemark{3} \\
 & \phn5 & 5 45 \phn9.6 & -69 30 \phn9 &  & \phn0.82 & 231.6 & 4.5 & \phm{00}9.6 & \\ 
\hline
N 206 & \phn1 & 5 31 29.8 & -71 \phn9 40 &  & \phn5.1\phn & 229.0 & 4.6 & \phn26.2 & \\
 & \phn2 & 5 31 46.3 & -71 \phn8 20 &  & \phn2.3\phn & 231.2 & 4.6 & \phn13.5 & \\ 
\hline
N 206D & \phn1 & 5 32 58.4 & -71 15 20 &  & \phn4.5\phn & 224.6 & 4.0 & \phn18.9 & \\ 
\hline
GMC 225 & \phn1 & 5 47 51.3 & -70 41 20 &  & \phn1.8\phn & 216.5 & 3.9 & \phm{00}9.0 & \\
 & \phn2 & 5 48 25.6 & -70 41 20 &  & \phn1.2\phn & 216.2 & 6.2 & \phm{00}8.6 & \\
 & \phn3 & 5 48 53.9 & -70 40 30 &  & \phn1.8\phn & 216.4 & 3.8 & \phm{00}7.9 & \\
\enddata
\tablecomments{
Column (1): Region.
Column (2): Running number in each region.
Column (3)-(4): Positions of observed points with maximum $^{12}$CO($J=3-2$) integrated intensities within the $^{12}$CO($J=3-2$) clumps are given in equatorial coordinates. 
Column (5)-(8): Observed properties of the $^{12}$CO($J=3-2$) spectra obtained at the peak positions of the $^{12}$CO($J=3-2$) clumps. 
The peak main-beam temperature $T_{\mathrm{mb}}$, $V_{\mathrm{LSR}}$, and the FWHM line width $\Delta V$ are derived from a single Gaussian fitting, and are given in columns 5, 6, and 7, respectively. In column 8, the $^{12}$CO($J=3-2$) integrated intensities at the peak positions of the $^{12}$CO($J=3-2$) clumps are shown.
Column (9): Another indentifications based on $^{12}$CO($J=1-0$) observations with SEST.
}	%% Notes

%% \tablenotemark{a}

\tablerefs{
(1) Johansson et al. 1998; 
(2) Kutner et al. 1997; 
(3) Garay et al. 2002
}

\end{deluxetable}

\clearpage

\begin{deluxetable}{lcccc}
%% \tabletypesize{\small}		%% 11pt
%% \tabletypesize{\footnotesize}	%% 10pt
\tabletypesize{\scriptsize}	%%  8pt
%% \rotate
\tablewidth{0pt}
%% \tablenum{}
%% \tablecolumns{}
\tablecaption{Physical properties of $^{12}$CO($J=3-2$) clumps \label{tbl03}}
\tablehead{
\colhead{} & \colhead{} & 
\multicolumn{3}{c}{Clump properties}\\

\cline{3-5}

\colhead{Region} & \colhead{No.} & 
\colhead{$\Delta V_{\mathrm{clump}}$} & \colhead{$R_{\mathrm{deconv}}$} & \colhead{$M_{\mathrm{vir}}$}\\

\colhead{} & \colhead{} & 
\colhead{(km s$^{-1}$)} & \colhead{(pc)} & \colhead{($\times$10$^4$ M$_{\sun}$)}
}

\startdata
30 Dor & \phn1 & 12.8 & \phn7.2 & 22\phd\phn\phn \\
 & \phn2 & \phn6.9 & \phn3.1 & \phn2.9\phn \\
 & \phn3 & \phn5.8 & \phn3.1 & \phn2.0\phn \\
 & \phn4 & \phn4.0 & \phn3.7 & \phn1.1\phn \\
 & \phn5 & \phn4.6 & \phn1.1 & \phn0.46 \\ 
\hline
N 159 & \phn1 & \phn8.7 & \phn4.7 & \phn6.8\phn \\
 & \phn2 & \phn7.3 & \phn6.9 & \phn6.9\phn \\
 & \phn3 & \phn7.6 & \phn4.7 & \phn5.2\phn \\
 & \phn4 & \phn8.5 & 11.3 & 15\phd\phn\phn \\
 & \phn5 & 10.2 & \phn4.7 & \phn9.4\phn \\
 & \phn6 & \phn5.2 & \phn4.3 & \phn2.2\phn \\
 & \phn7 & \phn4.1 & \phn2.3 & \phn0.73 \\
 & \phn8 & \phn6.3 & \phn6.3 & \phn4.7\phn \\
 & \phn9 & \phn7.1 & \phn2.3 & \phn2.2\phn \\
 & 10 & \phn7.7 & \phn2.3 & \phn2.7\phn \\ 
\hline
N 171 & \phn1 & \phn6.5 & \phn6.6 & \phn5.2\phn \\
 & \phn2 & \phn6.9 & \phn8.3 & \phn7.4\phn \\
 & \phn3 & \phn7.2 & \phn7.8 & \phn7.7\phn \\
 & \phn4 & \phn6.2 & \phn5.9 & \phn4.3\phn \\
 & \phn5 & \phn6.1 & \phn7.5 & \phn5.3\phn \\
 & \phn6 & \phn5.6 & \phn5.9 & \phn3.5\phn \\
\hline
N 166 & \phn1 & \phn6.1 & \phn9.9 & \phn6.9\phn \\
 & \phn2 & \phn6.5 & \phn9.5 & \phn7.6\phn \\
 & \phn3 & \phn6.3 & 10.3 & \phn7.7\phn \\
 & \phn4 & \phn9.2 & 12.4 & 20\phd\phn\phn \\
 & \phn5 & \phn9.5 & \phn4.7 & \phn8.1\phn \\ 
\hline
N 206 & \phn1 & \phn6.6 & \phn7.1 & \phn5.9\phn \\
 & \phn2 & \phn5.2 & \phn6.3 & \phn3.2\phn \\ 
\hline
N 206D & \phn1 & \phn4.0 & \phn7.3 & \phn2.2\phn \\ 
\hline
GMC 225 & \phn1 & \phn4.2 & \phn8.8 & \phn3.1\phn \\
 & \phn2 & \phn6.6 & \phn5.5 & \phn4.6\phn \\
 & \phn3 & \phn4.9 & \phn8.8 & \phn3.9\phn \\
\enddata
\tablecomments{
Column (1): Region.
Column (2): Running number in each region.
Column (3): Line width, $\Delta V_{\mathrm{clump}}$, derived by using a single Gaussian fitting for a spectrum obtained by averaging all the spectra within a single clump.
Column (4): Deconvolved effective radii, $R_{\mathrm{deconv}}$, defined as $[R_{\mathrm{nodeconv}}^2 - (\theta_{\mathrm{HPBW}}/2)^2]^{0.5}$. $R_{\mathrm{nodeconv}}$ is effective radii defined as ($A/\pi$)$^{0.5}$, where A is the total cloud surface area.
Column (5): Virial masses of the $^{12}$CO($J=3-2$) clumps, a density profile of $\rho \propto$ r$^{-1}$ was assumed; the virial mass is written as $M_{\mathrm{vir}}$ / M$_{\sun}$ = 190 $\times$ [$\Delta V_{\mathrm{clump}}$(km s$^{-1}$)]$^2$ $\times$ $R_{\mathrm{deconv}}$(pc) (MacLaren et al. 1998)
}	%% Notes

%% \tablenotemark{a}

%% \tablerefs{}

\end{deluxetable}

\clearpage

\begin{deluxetable}{cccccc}
%% \tabletypesize{\small}		%% 11pt
%% \tabletypesize{\footnotesize}	%% 10pt
\tabletypesize{\scriptsize}	%%  8pt
%% \rotate
\tablewidth{0pt}
%% \tablenum{}
%% \tablecolumns{}
\tablecaption{$^{12}$CO($J=3-2$)/$^{12}$CO($J=1-0$) ratio and H$\alpha$ flux \label{tbl04}}
\tablehead{
\colhead{Region} & \colhead{No.} & 
\colhead{$T_{\mathrm{mb}}$(3-2)} & \colhead{$T_{\mathrm{mb}}$(1-0)} & \colhead{$R_{3-2/1-0,clump}$} & 
\colhead{Averaged H$\alpha$ flux}\\

\colhead{} & \colhead{} & 
\colhead{(K)} & \colhead{(K)} & \colhead{} & 
\colhead{($\times$10$^{-12}$ ergs s$^{-1}$ cm$^{-2}$)}
}
\startdata
30 Dor & \phantom{0}1 & 1.63 & 1.20 & 1.4 & 56.49 \\
 & \phantom{0}2 & 1.76 & 1.12 & 1.6 & 73.37 \\
 & \phantom{0}3 & 1.56 & 1.13 & 1.2 & 24.09 \\
 & \phantom{0}4 & 1.98 & 1.58 & 1.3 & 12.83 \\
 & \phantom{0}5 & 0.73 & 0.87 & 0.8 & 39.71 \\ 
\hline
N 159 & \phantom{0}1 & 5.50 & 5.04 & 1.1 & \phantom{0}9.28 \\
 & \phantom{0}2 & 3.69 & 3.34 & 1.1 & \phantom{0}9.37 \\
 & \phantom{0}3 & 2.92 & 3.25 & 0.9 & \phantom{0}6.36 \\
 & \phantom{0}4 & 2.33 & 3.34 & 0.7 & \phantom{0}1.14 \\
 & \phantom{0}5 & 1.84 & 2.19 & 0.8 & \phantom{0}1.93 \\
 & \phantom{0}6 & 1.61 & 1.57 & 1.0 & \phantom{0}1.42 \\
 & \phantom{0}7 & 1.67 & 1.98 & 0.9 & \phantom{0}3.14 \\
 & \phantom{0}8 & 0.78 & 0.57 & 1.4 & \phantom{0}1.52 \\
 & \phantom{0}9 & 0.80 & 0.99 & 0.8 & \phantom{0}1.26 \\
 & 10 & 0.57 & 0.84 & 0.7 & \phantom{0}1.42 \\ 
\hline
N 171 & \phantom{0}1 & 1.04 & 2.20 & 0.5 & \phantom{0}1.00 \\
 & \phantom{0}2 & 0.69 & 1.55 & 0.4 & \phantom{0}0.93 \\
 & \phantom{0}3 & 1.02 & 3.09 & 0.3 & \phantom{0}0.97 \\
 & \phantom{0}4 & 1.34 & 3.89 & 0.3 & \phantom{0}0.98 \\
 & \phantom{0}5 & 0.88 & 1.92 & 0.4 & \phantom{0}0.95 \\
 & \phantom{0}6 & 0.87 & 2.72 & 0.3 & \phantom{0}0.97 \\ 
\hline
N 166 & \phantom{0}1 & 1.97 & 3.09 & 0.6 & \phantom{0}1.05 \\
 & \phantom{0}2 & 1.59 & 2.60 & 0.6 & \phantom{0}1.01 \\
 & \phantom{0}3 & 1.17 & 1.92 & 0.6 & \phantom{0}1.06 \\
 & \phantom{0}4 & 0.79 & 1.26 & 0.6 & \phantom{0}1.31 \\
 & \phantom{0}5 & 0.62 & 0.92 & 0.7 & \phantom{0}1.06 \\ 
\hline
N 206 & \phantom{0}1 & 1.83 & 2.39 & 0.8 & \phantom{0}3.26 \\
 & \phantom{0}2 & 1.34 & 2.44 & 0.5 & \phantom{0}2.68 \\ 
\hline
N 206D & \phantom{0}1 & 2.26 & 4.70 & 0.5 & \phantom{0}1.11 \\ 
\hline
GMC 225 & \phantom{0}1 & 1.04 & 2.81 & 0.4 & \phantom{0}0.93 \\
 & \phantom{0}2 & 0.55 & 1.62 & 0.3 & \phantom{0}0.91 \\
 & \phantom{0}3 & 0.77 & 2.09 & 0.4 & \phantom{0}0.92 \\
\enddata
\tablecomments{
Column (1): Region.
Column (2): Running number in each region.
Column (3): The peak main-beam temperature, $T_{\mathrm{mb}}$, of the $^{12}$CO($J=3-2$) spectra derived by using a single Gaussian fitting for a spectrum obtained by averaging all the spectra to the beam within a single clump. The intensities refer to the $^{12}$CO($J=1-0$) beam size (45\arcsec).
Column (4): The peak main-beam temperature, $T_{\mathrm{mb}}$, of the $^{12}$CO($J=1-0$) spectra derived by using a single Gaussian fitting for a spectrum obtained by averaging all the spectra within a single clump.
Column (5): The ratios of $T_{\mathrm{mb}}$(3-2) to $T_{\mathrm{mb}}$(1-0).
Column (6): The \Halpha flux obtained by averaging within a single clump.
}	%% Notes
%% \tablenotemark{}
%% \tablenotetext{}{}
%% \tablerefs{}

\end{deluxetable}

\clearpage

\begin{deluxetable}{cccccccccccccc}
%% \tabletypesize{\small}		%% 11pt
%% \tabletypesize{\footnotesize}	%% 10pt
\tabletypesize{\scriptsize}	%%  8pt
\rotate
\tablewidth{0pt}
%% \tablenum{}
%% \tablecolumns{}
\tablecaption{Summary of LVG and MPE analyses \label{tbl05}}
\tablehead{
\colhead{} & \colhead{} & 
\colhead{} & \colhead{} & \colhead{} & \colhead{} & \colhead{} & 
\colhead{$n$(H$_2$) (cm$^{-3}$)} & 
\colhead{} & \colhead{} & \colhead{} & \colhead{} & \colhead{} & \colhead{}\\

\colhead{Region} & \colhead{No.} & 
\colhead{d$v$/d$r$} & \colhead{$R_{3-2/1-0,clump}$} & \multicolumn{2}{c}{$R_{12/13}$} & \colhead{} & 
\colhead{$T_{\mathrm{kin}}$} & 
\multicolumn{6}{c}{(Previous studies)} \\
% \colhead{} & \colhead{} & \colhead{} & \colhead{} & \colhead{} & \colhead{}\\

\cline{5-6} \cline{8-14}

\colhead{} & \colhead{} & 
\colhead{(km s$^{-1}$ pc$^{-1}$)} & \colhead{} & \colhead{Value} & \colhead{Ref.} & \colhead{} & 
\colhead{This work} & 
% \multicolumn{6}{c}{Previous studies} \\

% \cline{9-14}

% \colhead{} & \colhead{} & 
% \colhead{} & \colhead{} & \colhead{} & \colhead{} & \colhead{} & 
% \colhead{} & 
\colhead{(1)} & \colhead{(4)} & \colhead{(5)} & \colhead{(6)} & \colhead{(7)} & \colhead{(8)}
}

\startdata
30 Dor & 1 & 0.9 & 1.4 & 11.5 & (1) &  & 3$\times$10$^{3}$ - 3$\times$10$^{5}$ & 
10$^{4.5}$ & 1$\times$10$^{5}$ &  &  &  & 10$^{4.3}$ \\
 &  &  &  &  &  &  & \textgreater 50 & 
40 - 80 & 50 &  &  &  & 100 \\ 
\cline{2-14}
 & 4 & 0.5 & 1.3 & 17.7 & (1) &  & 1$\times$10$^{3}$ - 1$\times$10$^{5}$ & 
10$^{4.5}$ &  &  &  &  &  \\
 &  &  &  &  &  &  & \textgreater 60 & 
\textgreater 20 &  &  &  &  &  \\ 
\hline
N 159 & 1 & 0.9 & 1.1 & \phantom{0}8.6 & (1) &  & 3$\times$10$^{3}$ - 8$\times$10$^{5}$ & 
10$^{4.5}$ & 3$\times$10$^{5}$ & 10$^{3}$ & 10$^{5}$ & 10$^{2}$ &  \\
 &  &  &  &  &  &  & \textgreater 30 & 
16 - 23 & 25 & 150 & 20 & 100 &  \\ 
\cline{2-14}
 & 2 & 0.5 & 1.1 & 11.6 & (1) &  & 1$\times$10$^{3}$ - 3$\times$10$^{5}$ & 
 &  &  &  &  &  \\
 &  &  &  &  &  &  & \textgreater 40 & 
 &  &  &  &  &  \\ 
\cline{2-14}
 & 4 & 0.4 & 0.7 & \phantom{0}8.5 & (1) &  & 1$\times$10$^{3}$ - 6$\times$10$^{3}$ & 
 & 1$\times$10$^{5}$ &  &  &  &  \\
 &  &  &  &  &  &  & 20 - 60 & 
 & 10 &  &  &  &  \\ 
\hline
N 166 & 1 & 0.3 & 0.6 & 10.5 & (2) &  & 5$\times$10$^{2}$ - 2$\times$10$^{3}$ & 
 &  &  &  &  &  \\
 &  &  &  &  &  &  & 25 - 150 & 
 &  &  &  &  &  \\ 
\cline{2-14}
 & 3 & 0.3 & 0.6 & 12.6 & (2) &  & 3$\times$10$^{2}$ - 2$\times$10$^{3}$ & 
 &  &  &  &  &  \\
 &  &  &  &  &  &  & \textgreater 30 & 
 &  &  &  &  &  \\ 
\cline{2-14}
 & 4 & 0.4 & 0.6 & 16.2 & (2) &  & 4$\times$10$^{2}$ - 2$\times$10$^{3}$ & 
 &  &  & 10$^{4}$ & 10$^{2}$ - 10$^{3}$ &  \\
 &  &  &  &  &  &  & \textgreater 40 & 
 &  &  & 20 & 30 - 60 &  \\ 
\hline
N 206 & 1 & 0.5 & 0.8 & 14.1 & (3) &  & 5$\times$10$^{2}$ - 3$\times$10$^{3}$ & 
 &  &  &  &  &  \\
 &  &  &  &  &  &  & \textgreater 35 & 
 &  &  &  &  &  \\ 
\cline{2-14}
 & 2 & 0.4 & 0.5 & \phantom{0}9.8 & (3) &  & 6$\times$10$^{2}$ - 2$\times$10$^{3}$ & 
 &  &  &  &  &  \\
 &  &  &  &  &  &  & 20 - 80 & 
 &  &  &  &  &  \\ 
\hline
N 206D & 1 & 0.3 & 0.5 & \phantom{0}4.8 & (3) &  & 1$\times$10$^{3}$ - 3$\times$10$^{3}$ & 
 &  &  &  &  &  \\
 &  &  &  &  &  &  & 10 - 20 & 
 &  &  &  &  &  \\ 
\hline
GMC 225 & 1 & 0.2 & 0.4 & \phantom{0}6.6 & (3) &  & 5$\times$10$^{2}$ - 2$\times$10$^{3}$ & 
 &  &  &  &  &  \\
 &  &  &  &  &  &  & 15 - 20 & 
 &  &  &  &  &  \\ 
\cline{2-14}
 & 3 & 0.3 & 0.4 & \phantom{0}6.7 & (3) &  & 7$\times$10$^{2}$ - 2$\times$10$^{3}$ & 
 &  &  &  &  &  \\
 &  &  &  &  &  &  & 10 - 40 & 
 &  &  &  &  &  
\enddata
\tablecomments{
% Column (1): Region. 
% Column (2): Running number in each region. 
% Column (3): The velocity gradient of the clump. 
% Column (4): Intensity ratio of $^{12}$CO($J=3-2$) to $^{12}$CO($J=1-0$). 
% Column (5)-(6): Intensity ratio of $^{12}$CO($J=1-0$) to $^{12}$CO($J=1-0$) and their references. 
Column (7): Results of LVG analysis 
Column (8)-(13): Results of previous studies, including both LVG and MEP analyses.
}	%% Notes
%% \tablenotemark{}
%% \tablenotetext{}{}
\tablerefs{
(1) Johansson et al. 1998; 
(2) Garay et al. 2002; 
(3) This work; 
(4) Heikkil\"{a} et al. 1999; 
(5) "Single-component fit" in Bolatto et al. 2005; 
(6) "Cold Dense Component" and (7) "Hot Tenuous Component" of the "Dual-component fit" in Bolatto et al. 2005; 
(8) Kim 2006
}

\end{deluxetable}

\clearpage

\begin{deluxetable}{lcccc}
%% \tabletypesize{\small}		%% 11pt
%% \tabletypesize{\footnotesize}	%% 10pt
\tabletypesize{\scriptsize}	%%  8pt
%% \rotate
\tablewidth{0pt}
%% \tablenum{}
%% \tablecolumns{}
\tablecaption{Effect of $X$(CO) \label{tbl06}}
\tablehead{
\colhead{} & \colhead{} & \colhead{30 Dor No.1} & \colhead{N 159 No.1} & \colhead{N 159 No.4}
%% \\ \colhead{} & \colhead{(30Dor-10)} & \colhead{(N159W)} & \colhead{(N159S)}
}
\startdata
\multicolumn{2}{l}{This work} &  &  & \\
(Uniform $X$(CO)) & \multicolumn{1}{r}{$X$(CO)} & 3$\times$10$^{-6}$ & 3$\times$10$^{-6}$ & 3$\times$10$^{-6}$ \\
 & \multicolumn{1}{r}{$n$(H$_2$) (cm$^{-3}$)} & 3$\times$10$^{3}$ - 3$\times$10$^{5}$ & 3$\times$10$^{3}$ - 8$\times$10$^{5}$ & 1$\times$10$^{3}$ - 6$\times$10$^{3}$ \\
 & \multicolumn{1}{r}{$T_{\mathrm{kin}}$ (K)} & \textgreater 50 & \textgreater 30 & 20 - 60 \\
%% &  &  &  &  \\ 
\cline{2-5}
(Different $X$(CO)) & \multicolumn{1}{r}{$X$(CO)} & 1$\times$10$^{-6}$ & 1$\times$10$^{-5}$ & 3$\times$10$^{-6}$ \\
 & \multicolumn{1}{r}{$n$(H$_2$) (cm$^{-3}$)} & 6$\times$10$^{3}$ - 1$\times$10$^{6}$ & 2$\times$10$^{3}$ - 2$\times$10$^{5}$ & 1$\times$10$^{3}$ - 6$\times$10$^{3}$ \\
 & \multicolumn{1}{r}{$T_{\mathrm{kin}}$ (K)} & \textgreater 35 & \textgreater 45 & 20 - 60 \\
%% &  &  &  \\ 
\hline
\multicolumn{2}{l}{Heikkil\"{a} et al. 1999} &  &  & \\
 & \multicolumn{1}{r}{$X$(CO)} & 1.4$\times$10$^{-6}$ & 1.2$\times$10$^{-5}$ & 4.6$\times$10$^{-6}$ \\
 & \multicolumn{1}{r}{$n$(H$_2$) (cm$^{-3}$)} & 1$\times$10$^{5}$ & 3$\times$10$^{5}$ & 1$\times$10$^{5}$ \\
 & \multicolumn{1}{r}{$T_{\mathrm{kin}}$ (K)} & 50 & 25 & 10 \\
%% &  &  &  
\enddata
%% \tablecomments{}	%% Notes
%% \tablenotemark{}
%% \tablenotetext{}{}
%% \tablerefs{}

\end{deluxetable}

\clearpage

\begin{deluxetable}{ccccl}
%% \tabletypesize{\small}		%% 11pt
%% \tabletypesize{\footnotesize}	%% 10pt
\tabletypesize{\scriptsize}	%%  8pt
%% \rotate
\tablewidth{0pt}
%% \tablenum{}
%% \tablecolumns{}
\tablecaption{Estimated PDR surface temperatures \label{tbl07}}
\tablehead{
\colhead{Region} & \colhead{$n$} & \colhead{$G_0$} & \colhead{$T_s$ \tablenotemark{a}} & \colhead{References}\\

\colhead{} & \colhead{(cm$^{-3}$)} & \colhead{} & \colhead{(K)} & \colhead{}
}
\startdata
30 Dor & 10$^4$ & 3500 & 300 & 1, 2, 3, 4 \\
N 159 & 10$^4$ & 300 & 100 & 1, 4, 5 \\
\enddata

\tablecomments{
Column(1): Regions. 
Column(2): Gas density. 
Column(3): FUV flux in units of local interstellar value; 1.6$\times$10$^{-3}$ ergs cm$^{-2}$ s$^{-1}$. 
Column(4): Derived PDR surface temperature of the atomic gas. 
Column(5): References of FUV flux. 
}	%% Notes
%% \tablenotemark{}
\tablenotetext{a}{Estimation is done by using Figure 1 of Kaufman et al. 1999}

\tablerefs{
(1) Bolatto et al. 1999; 
(2) Poglitsch et al. 1995; 
(3) Werner et al. 1978; 
(4) Israel \& Koornneef 1979;
(5) Israel et al. 1996
}

\end{deluxetable}

\clearpage

%% The following command ends your manuscript. LaTeX will ignore any text

\end{document}